\documentclass{cpbtex}
\usepackage{authblk}
\usepackage{hyperref}

\usepackage{float}
\usepackage{graphicx}
\usepackage{xcolor}
\usepackage{lineno}

\usepackage{rotating}
\usepackage{booktabs}  
\usepackage{makecell}
\usepackage{CJKutf8}

\usepackage{natbib}
\usepackage{xcolor}

\newcommand{\REV}[1] {\textbf{#1}}

\newcommand{\AFF}[1]{$^{\foreach\d[count=\ni]in{#1}{\ifnum\ni=1\ref{\d}\else,\ref{\d}\fi}}$}

\def \arcsec {''}

\usepackage{titlesec}

\titleformat{\section}[runin]
  {\normalfont\itshape}    
  {\thesection }
  {1em}
  {}

\titlespacing*{\section}
  {0pt}                    
  {0.5\baselineskip}      
  {0.5em}  
  
\titleformat{\subsection}[runin]
  {\normalfont\itshape}    
  {\thesubsection}
  {1em}
  {}

\titlespacing*{\subsubsection}
  {0pt}                    
  {0.3\baselineskip}      
  {0.3em}                 
  
  \titleformat{\subsubsection}[runin]
  {\normalfont\itshape}    
  {\thesubsubsection}
  {1em}
  {}

\titlespacing*{\subsection}
  {0pt}                    
  {0.3\baselineskip}      
  {0.3em}                 

\title{To understand the radiative processes of pulsars and fast radio bursts with the FAST}

\begin{document}

\definecolor{blue}{RGB}{51,87,255}


\author{
\begin{CJK*}{UTF8}{gbsn}
Wei-Yang Wang(王维扬)\textsuperscript{1}, %
Shunshun Cao(曹顺顺)\textsuperscript{2}, %
Zhipeng Huang(黄志鹏)\textsuperscript{3,4}, %
Jiguang Lu(卢吉光)\textsuperscript{5,6}, %
Yunpeng Men(门云鹏)\textsuperscript{7}, %
Lingqi Meng(孟令祺)\textsuperscript{7}, %
Jiarui Niu(牛佳瑞)\textsuperscript{5}, %
Zhichen Pan(潘之辰)\textsuperscript{5,6}, %
Pengfei Wang(王鹏飞)\textsuperscript{5}, %
Dejiang Zhou(周德江)\textsuperscript{5}, %
Yi Feng(冯毅)\textsuperscript{8,9}, %
Jinlin Han(韩金林)\textsuperscript{5}, 
Jinchen Jiang(姜金辰)\textsuperscript{7}, %
Bin Liu(刘彬)\textsuperscript{5}, %
Rui Luo(罗睿)\textsuperscript{10}, %
Honguang Wang(王洪光)\textsuperscript{10}, %
Shuangqiang Wang(王双强)\textsuperscript{11}, %
Tao Wang(王涛)\textsuperscript{5}, %
Zhengli Wang(王正理)\textsuperscript{12}, %
Heng Xu(胥恒)\textsuperscript{5}, %
Jiangwei Xu(徐江伟)\textsuperscript{5}, %
Renxin Xu(徐仁新)\textsuperscript{2}, 
Yonghua Xu(徐永华)\textsuperscript{13}, %
Yi Yan(颜一)\textsuperscript{5}, %
Zhen Yan(闫振)\textsuperscript{14}, %
Yukai Zhou(周宇凯)\textsuperscript{2}, %
Siyuan Chen\textsuperscript{14}, %
Yinfeng Dai(戴印丰)\textsuperscript{15,16}, %
Mingyu Ge(葛明玉)\textsuperscript{17}, %
Zejun Jiang(姜泽军)\textsuperscript{18}, %
Kejia Lee(李柯伽)\textsuperscript{2}, %
Yujie Lian(连禹杰)\textsuperscript{15,7}, %
Kuo Liu(刘阔)\textsuperscript{14}, %
Lei Qian(钱磊)\textsuperscript{5,6}, %
Hao Tong(仝号)\textsuperscript{10}, %
Lin Wang(王琳)\textsuperscript{14},%
Yujie Wang(王宇杰)\textsuperscript{1,14}
Zihao Xu(徐子浩)\textsuperscript{2}, %
Jumei Yao(姚菊枚)\textsuperscript{11}, %
Dejiang Yin(尹德江)\textsuperscript{16}, %
Li Zhang(张力)\textsuperscript{18}, %
Weiwei Zhu(朱炜玮)\textsuperscript{5}%
\end{CJK*}
}

\affil{\textsuperscript{1}{ \textcolor{black}{University of Chinese Academy of Sciences, Chinese Academy of Sciences, Beijing 100049, China}},\\
\textsuperscript{2}{ \textcolor{black}{School of Physics, Peking University, Beijing 100871, China; {\tt r.x.xu@pku.edu.cn}}},\\
\textsuperscript{3}{ \textcolor{black}{School of Physics and Mechanical Electrical \& Engineering, Hubei University of Education, Wuhan 430205, China}},\\
\textsuperscript{4}{ \textcolor{black}{Institute of Astronomy and High Energy Physics, Hubei University of Education, Wuhan 430205, China}},\\
\textsuperscript{5}{ \textcolor{black}{National Astronomical Observatories, Chinese Academy of Sciences, Beijing 100012, China; {\tt hjl@nao.cas.cn}}},\\
\textsuperscript{6}{ \textcolor{black}{Guizhou Radio Astronomical Observatory, Guiyang 550025, China}},\\
\textsuperscript{7}{ \textcolor{black}{Max-Planck-Institut fur Radioastronomie, Aufdem H\"{u}gel 69, Bonn, D-53121, Germany}},\\
\textsuperscript{8}{ \textcolor{black}{Research Center for Astronomical Computing, Zhejiang Laboratory, Hangzhou 311100, China}},\\
\textsuperscript{9}{ \textcolor{black}{Institute for Astronomy, School of Physics, Zhejiang University, Hangzhou 310027, China}},\\
\textsuperscript{10}{ \textcolor{black}{Department of Astronomy, School of Physics and Materials Science, Guangzhou University, Guangzhou 510006, China}},\\
\textsuperscript{11}{ \textcolor{black}{Xinjiang Astronomical Observatory, Chinese Academy of Sciences, 150 Science 1-Street, Urumqi 830011, China}},\\
\textsuperscript{12}{ \textcolor{black}{Guangxi Key Laboratory for Relativistic Astrophysics, School of Physical Science and Technology, Guangxi University, Nanning 530004, China}},\\
\textsuperscript{13}{ \textcolor{black}{Yunnan Observatories, Chinese Academy of Sciences, Kunming 650216, China}},\\
\textsuperscript{14}{ \textcolor{black}{Shanghai Astronomical Observatory, CAS, Shanghai 200030, China}},\\
\textsuperscript{15}{ \textcolor{black}{School of Physics and Astronomy, Beijing Normal University, Beijing 100875, China}},\\
\textsuperscript{16}{ \textcolor{black}{College of Physics, Guizhou University, Guiyang 550025, China}},\\
\textsuperscript{17}{ \textcolor{black}{Department of Physics, Faculty of Arts and Sciences, Beijing Normal University, Zhuhai 519087, China}},\\
\textsuperscript{18}{ \textcolor{black}{State Key Laboratory of Particle Astrophysics, Institute of High Energy Physics, Chinese Academy of Sciences, Beijing 100049, China}},\\
\textsuperscript{19}{ \textcolor{black}{Department of Physics, Yunnan University, Kunming, 650091 China}}\\

}

\renewcommand*{\Affilfont}{\small\it}

\maketitle

\begin{abstract}
The radiative mechanism of coherent radio emission has remained an enigma since the discovery of pulsars, even the emergence of fast radio bursts (FRBs), which exhibit similarities to the single-pulse behavior of pulsars and have opened a new view for deciphering the long-standing mystery. Besides tremendous efforts in modelling, advanced facilities matter for solving the problem. The authors review the observational breakthroughs from the Five-hundred-meter Aperture Spherical radio Telescope (FAST), which are providing pivotal insights to unravel the underlying physics of pulsars and FRBs. This study offers a novel perspective in the era when pulsars meet FRBs, and further investigations are encouraged to utilize the highly sensitive telescope, the FAST.

\end{abstract}

\hspace{2em}\textbf{Keywords:} Pulsar, Fast radio burst, Radiative mechanism, FAST


\section{~Introduction.}

New discoveries can be categorised as either predicted or unpredicted, and the discovery of the pulsar belongs to the latter. %
Indeed, as noted on the flyleaf of ``{\em Pulsars}''~\cite{PSR77}, without Jocelyn Bell’s attention to detail and perseverance in her youth, we may not have been able to appreciate the wonders of pulsars and to enjoy studying them for so long$^*$.\footnote{$^*$The authors (R. Manchester and J. Taylor) wrote~\cite{PSR77}: ``The book is dedicated to Jocelyn Bell, without whose perceptiveness and persistence we might not yet have had the pleasure of studying pulsars.''
} %
Unfortunately, even after nearly sixty years, we still do not fully understand the radiative mechanism of cosmic coherent radio emission, which is probably relevant to the nature of pulsars, particularly with regard to their surfaces. %
Besides theoretical efforts, advanced facilities help, and China's FAST (the five-hundred-meter aperture spherical radio telescope) is therefore focused in this paper. %

\par 
FAST was completed with its main structure installed on September 25, 2016.$^\dag$\footnote{$^\dag$%
FAST was initially proposed as China's contribution to the international Large Telescope (LT) project in 1994, with the aim of building a radio telescope array comprising large dishes in small numbers. LT was renamed the Square Kilometre Array (SKA) in 1999. The book of ``{\it The FAST Project}'' by Bo Peng collects a lot of architectural and anecdotal stories.~\cite{FAST21}
} %
It mainly consists of the active reflector system, the feed support system and the high precision measurement system, and the organic coordination among all systems ensures the reliable operation of the telescope~\cite{jiang2019}. So far, the sky coverage of FAST can reach up to zenith angle (ZA) 40$^\circ$. With ZA less than 26.4$^\circ$, the effective reflector is a 300-m diameter paraboloid, and the reflection area reduces by about one-third when ZA reaches 40$^\circ$. Given the primarily used 19-beam 1.05-1.45 GHz receiver, the telescope efficiency, pointing accuracy, and system noise temperature were measured as 63\%, 7.9$^{\prime\prime}$ and less than 24\,K with ZA less than 26.4$^\circ$ during the commissioning test~\cite{jiang2020}. At present, FAST has already conducted comprehensive pulsar observations, and will continue to enhance its capabilities and broaden its application potential in the future.

\par
Without a doubt, pulsars are among the best targets for FAST, and many research achievements have already been published since 2019. In order to promote the research into coherent radio emission from pulsar magnetospheres and the relevant surface conditions, it is now timely to review these previous FAST results, as well as to set the focus for future efforts. Meanwhile, fast radio bursts (FRBs), discovered in 2007, poses a similar challenge in terms of exploring the underlying physics of radio coherence, and therefore we have also included them in this article. In fact, this is our motivation for organizing the dialogue at the Dream Field near the FAST site (https://psr.pku.edu.cn/conference/psrfrb/) in order to promote collaborative researches between pulsars and FRBs, and the review here is also an extended summary of the meeting to some extent.

\par
Although FAST's observations have yielded remarkable achievements, three major questions remain that require significant efforts to answer.
\begin{enumerate}
    \item What is the physical nature of a {\em pulsar's surface}, and how does this impact its magnetospheric electrodynamics?
    \item What is the {\em coherent radiative mechanism} responsible for radio emission in pulsars and FRBs?
    \item What is the underlying physics that governs the {\em coupling of waves} in highly magnetised plasma?
\end{enumerate}
The first question relates to the challenge problem: the equation of state for cold supra-nuclear matter, as the surface/crust of a compact star is essentially determined by its interior. The second and the third questions are interesting in both astrophysics and plasma physics. Many derivative questions are entailed from the three questions: (1) How can the true geometry of pulsar emission be derived observationally? (2) Are there any overall differences in the radiation properties of normal and millisecond pulsars? (3) Can nulling pulsars tell us more about the emission processes? (4) Besides the difference in powering mechanism, what can we learn from the analogy between the signals of  FRB and pulsar? Nevertheless, we would elucidate our attempts on answering these questions using FAST observations, and certainly, newly discovered pulsars may reveal surprising answers.

\par
The structure of this article is as follows. In Section 2, we outline several key observational results from FAST in pulsar astronomy, as well as those for FRBs in Section 3. We discuss the underlying physical mechanisms of these pulsar and FRB observations in Section 4. In Section 5, we introduce future prospects. Finally, we present a summary in Section 6.

\section{~Pulsars.}

The discovery of pulsars in 1967 is meaningful not only in fundamental sciences (e.g., checking the interactions of gravitational and strong forces, probing the interstellar/galactic medium, and detecting nano-Hertz gravitational waves), but also triggering technological developments (e.g., digital techniques, wideband radio receivers and X/$\gamma$-ray telescopes in space). Nevertheless, after more than half a century, the central and simple question remains: {\em How does the emission radiate}?

\par
Admittedly, the pulsar emission process is still poorly understood even now, and it is, however, generally accepted that primary pairs are produced and accelerated in regions (gaps) with a strong electric field along the magnetic line, while more secondary pairs (with multiplicity of $10^2-10^5$) are created, \REV{and} instability may develop in order to produce coherent radio emission. Although numerous models have been suggested concerning gap acceleration, there is a consensus that the acceleration process may depend on the condition of the pulsar's surface: either a vacuum gap~\cite{RS75} with sufficient binding energy to hold particles on the surface, or a space charge-limited flow~\cite{Arons79} with negligible binding. 

\par
Recently, numerical simulations in magnetohydrodynamic and particle-in-cell (PIC) approaches have made great progress in pulsar emission physics. A novel coherent radiation mechanism for radio pulsars has been proposed in PIC simulation~\cite{Philippov2020}. Simulations above the polar cap have also challenged some of our conventional ideas on spark discharge~\cite{ Chernoglazov2024} and polarization~\cite{Benacek2025}. Coherent emission at different regions of pulsar magnetospheres has been explored as well \cite {Bransgrove2023}. However, our knowledge on real pulsar emission processes is still manifestly limited. Surface conditions and propagation effects are not well studied, and we have almost no idea about the mechanisms behind some time-varying phenomena, like the mode switches. We are expecting advanced facilities, e.g., the FAST could help. Combining observations with theories more tightly may push us forward in understanding pulsar emission.

\par
First of all, let's use FAST to discover new pulsars, so that we can learn more about other populations.

\subsection{~Pulsar searching.}

It is certainly motivated to search pulsars and to target FRBs with the FAST, and this study is in fact focused by a few groups.


\subsubsection{~The FAST Galactic Plane Pulsar Snapshot survey.}

\par 
The GPPS survey represents the most sensitive systematic search for pulsars in the Galactic plane to date. Utilizing the L-band 19-beam receiver with a system temperature of approximately 22 K, the survey employs a specialized ``snapshot'' strategy. This mode covers a hexagonal sky region of 0.1575 square degrees via four pointings, with an integration time of 5 minutes per pointing. This strategy allows for a high survey speed while maintaining a sensitivity capable of detecting flux densities down to the magnitude of a few $\mu$Jy \cite{HanJL2021RAA....21..107H}. As of the latest release, the survey has covered approximately one-quarter of the planned area within $\pm 10^{\circ}$ of the Galactic latitude, resulting in the discovery 
over 800 pulsars \cite{HanJL2025RAA....25a4001H}.

\par
\begin{figure*}
  \centering
  \includegraphics[height=0.45\textwidth]{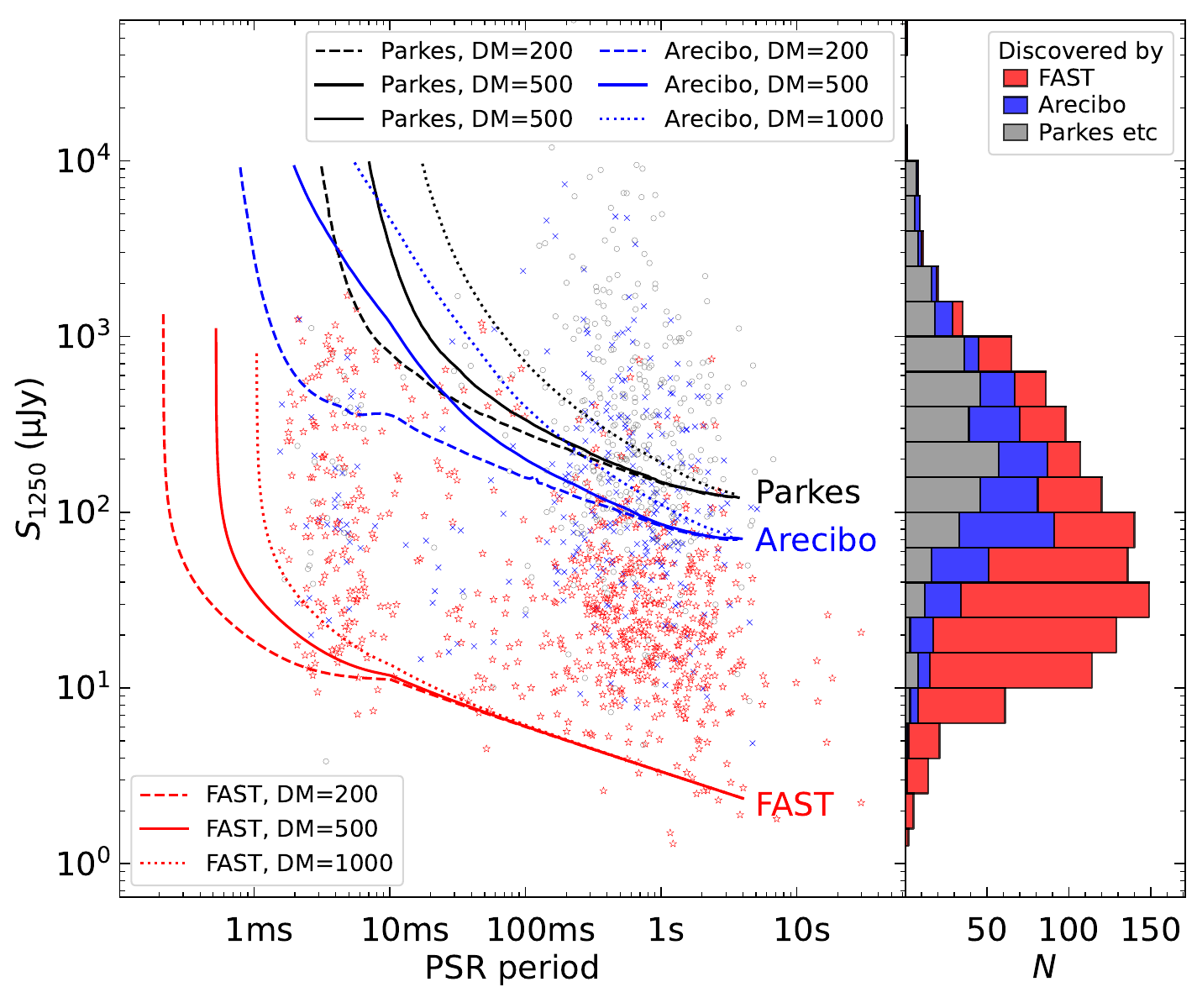}
  \includegraphics[height=0.45\textwidth]{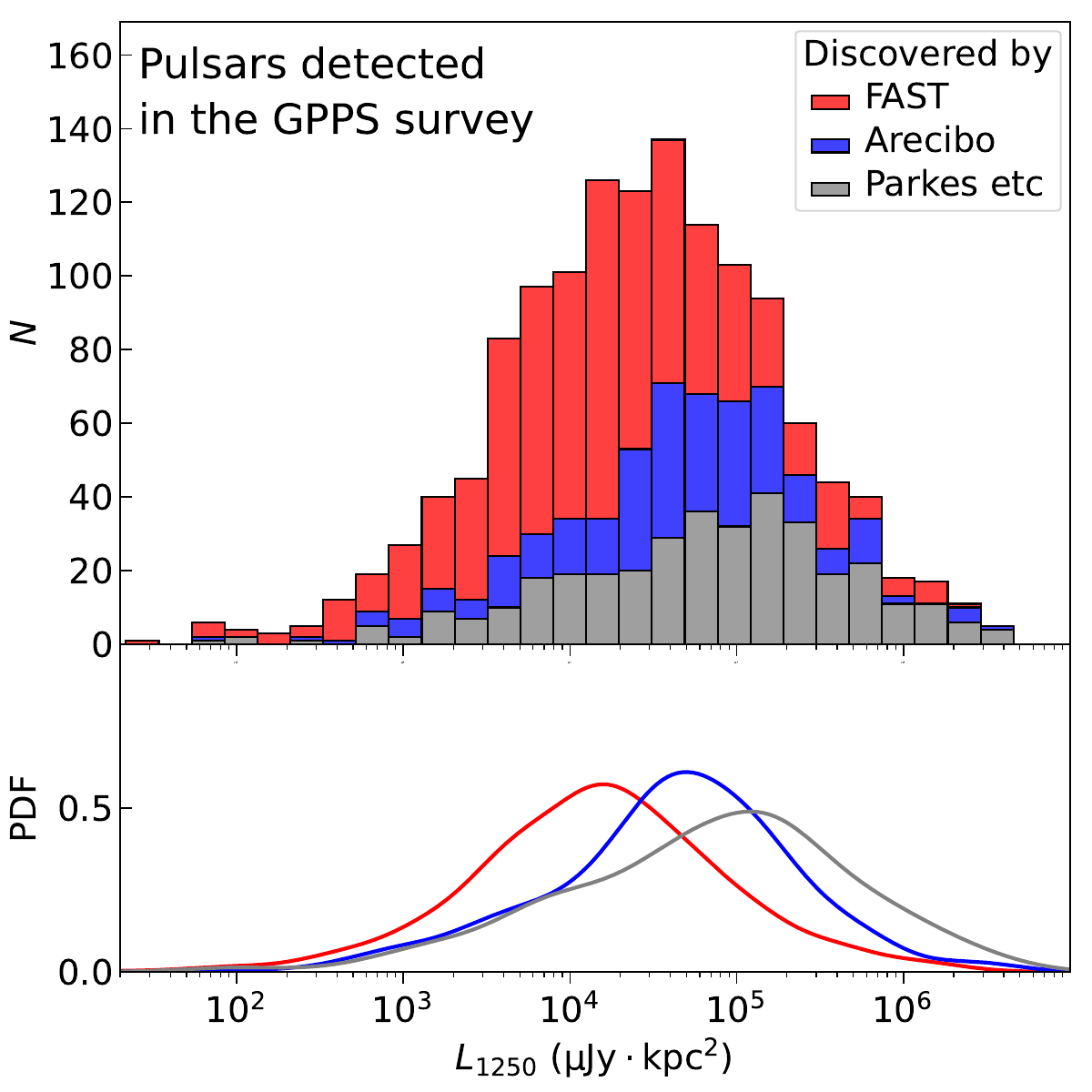}
  \caption{The distributions of flux densities and luminosities of pulsars observed by FAST. In the left panel, the sensitivity curves for different DMs of the FAST GPPS survey are given, compared to those for the Arecibo and Parkes surveys \cite{CordesJM2006ApJ...637..446C,ManchesterRN2001MNRAS.328...17M}, together with the histogram of the three surveys. The right panel shows that the FAST GPPS pulsars have much lower flux densities and dominate at the lower end of luminosity distributions. The probability distribution functions (PDFs) are smoothed curves for the normalized fraction distribution (original from \cite{HanJL2025RAA....25a4001H}). }
  \label{GPPS:flux_lum}
\end{figure*}

 
A defining achievement of the GPPS survey is its probing of the faint end of the pulsar luminosity function. The newly discovered population is dominated by extremely weak sources, with flux densities reaches $\mu$Jy, which falls below the detection thresholds of previous major surveys like those conducted at Parkes or Arecibo (Figure \ref{GPPS:flux_lum}). Consequently, the GPPS results have significantly reshaped the known distribution of pulsar luminosities, providing a decisive contribution to the lower end of the distribution. Furthermore, the survey has revealed a substantial number of distant pulsars with high dispersion measures (DMs). Many of these sources exhibit DMs exceeding the maximum limits predicted by current Galactic electron density models (NE2001 and YMW16), particularly in the tangential directions of spiral arms, suggesting large unmodeled electron densities in the scutum and Sagittarius arms.

\par 
In the time domain, the GPPS survey has utilized a sensitive single-pulse search module to uncover 76 new Rotating Radio Transients (RRATs) \cite{ZhouDJ2023RAA....23j4001Z}. Detailed follow-up observations and polarization analyses of these sources, alongside 59 previously known RRATs, have provided critical insights into the ``RRAT enigma''. The high sensitivity of FAST revealed that the majority of these objects are not a distinct class of neutron stars but rather ordinary pulsars that are either extremely nulling or intrinsically weak with sporadic strong pulses. The polarization angle swings of the sporadic bursts from these sources trace the same geometric curves as their integrated profiles, confirming that the emission geometry remains consistent between the transient and persistent phases.

\par 
The survey has also significantly expanded the sample of exotic pulsar populations. The GPPS discoveries include 177 millisecond pulsars (MSPs), many of which are in binary systems \cite{WangPF2025RAA....25a4003W}, including eclipsing binaries and spider systems. Notably, 34 of these new MSPs possess timing precision better than 3 $\mu$s, making them viable candidates for Pulsar Timing Arrays. Conversely, the survey has been effective in finding long-period pulsars, identifying eight sources with rotation periods greater than 10 seconds, including one with a period of approximately 30 seconds. Additionally, the survey serendipitously detected fast radio bursts (FRBs), distinguished from Galactic sources by their excessive DMs, and pulsars exhibiting significant scattering tails, which allow for the characterization of the turbulent interstellar medium in the distant Galactic disk.

In summary, the FAST GPPS survey has not only vastly increased the census of Galactic pulsars—particularly faint, high-DM, long-period, and binary systems—but has also provided high-quality polarization, timing, and single-pulse data essential for probing pulsar emission physics, neutron star demographics, and the structure of the Galactic interstellar medium. Its continued observations are expected to further delineate the distribution and evolution of neutron stars in the Milky Way.

\subsubsection{~Pulsar Searching in CRAFTS.}

\par The Commensal Radio Astronomy FAST Survey (CRAFTS) established as a remarkable project in pulsar astronomy by pioneering a highly efficient commensal drift-scan technique that simultaneously acquires neutral hydrogen and pulsar data.
As of late 2024, CRAFTS has contributed over 200 pulsars to FAST’s total of more than 1,000 discoveries, with this sample including at least more than 70 MSPs, more than 100 non-recycled pulsars, and several RRATs.
Single-pulse studies of long-period pulsars have revealed complex phenomena including the coexistence of nulling, regular, and bright pulses within individual pulse sequences, each characterized by distinct abundance and duration distributions \cite{tedila2022}.
Timing measurements of PSR J0535--0231 have yielded accurate determinations of its position, rotational parameters, and DM. The integrated pulse profile reveals at least three distinct emission components, leading, central, and trailing, with the leading component exhibiting an anticorrelation with the trailing component.
The time-varying subpulse modulation including drifting modulation, phase-stationary amplitude modulation, and diffuse modulation, can not explained by the traditional carousel model \cite{cai2025}.
Furthermore, CRAFTS-discovered pulsar exhibits DM exceeding the predictions of the widely used NE2001 and YMW16 Galactic electron density models, highlighting the need for refined and improved models of the interstellar medium \cite{blackmon2025}.

\subsubsection{~Pulsar Searching in Globular Clusters.}
\par Pulsars with their companions such as low mass white dwarfs, main sequence stars, 
neutron stars, or even black holes are labs in the universe to test the theories of
star formation and evolution, general relativity, gravity waves, state equation, etc.
Besides the galactic plane, 
Globular Clusters are another target for searching for the binary pulsars 
due to their billion year evolutionary ages 
and at least 10 times higher stellar density in the core region than the Milky Way.
Binary pulsar in the extremely eccentric orbit \cite{2015ApJ...807L..23D}, 
double neutron stars (M71D for the massive binary star evolution and M15C for post keplarian parameters measurement and gravity tests),
possible the mini-PTA \cite{2025PhRvR...7c3300C},
fastest spining pulsar (Terzan 5 ad, \cite{2006Sci...311.1901H})
were all found in Globular Clusters.
Till the January of 2026, 350 pulsars discovered in 46 Globular Clusters, 
while more than 60 are from FAST discovery.

\par
\begin{figure*}
  \includegraphics[height=0.8\textwidth]{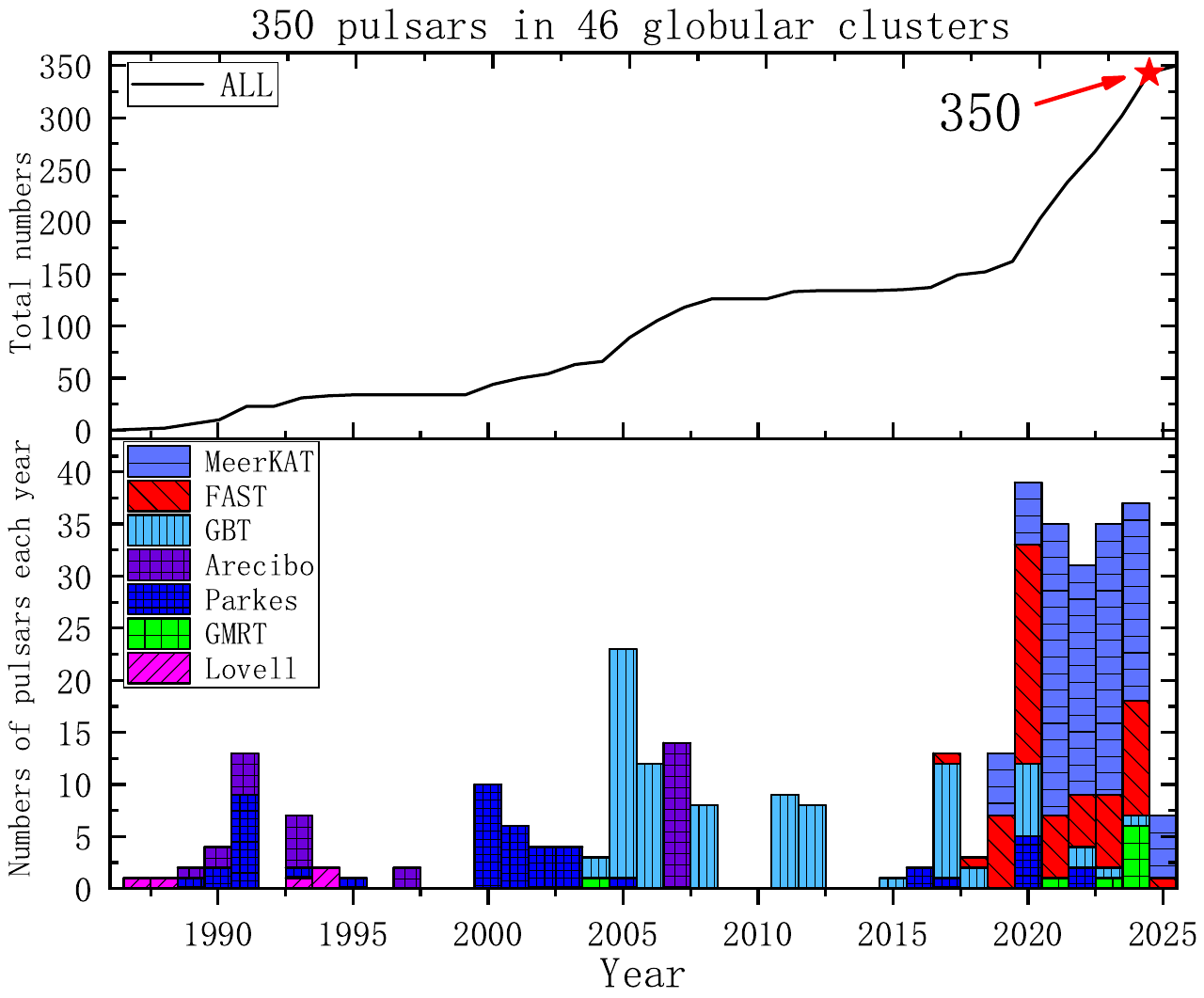}
  \caption{GC pulsar discoveries from Paulo's GC pulsar catalog (https://www3.mpifr-bonn.mpg.de/staff/pfreire/GCpsr.html) and references therein. }
  \label{GCpsr}
\end{figure*}

 Among all the $\rm \sim$160 Globular Clusters, 45 are in FAST sky. Before FAST, $\rm \sim$30 pulsars were discovered in these 45 Globular Clusters (e.g., \cite{2007ApJ...670..363H}). The Globular Cluster pulsar search with FAST (e.g., GC FANS, \cite{2025ApJS..279...51L,2020ApJ...892...43W}) resulted in more than 60 discoveries, 
more than doubling the number of Globular Cluster pulsar in FAST sky. It is no doubt that the pulsar discoveries come after the built of large radio telescopes (Fig \ref{GCpsr}). The Arecibo plus Parkes around 1990, GBT around 2005, and MeerKAT plus FAST currently, are three waves of GC pulsar disocveries.

\par The earliest Globular Cluster pulsar search attempts were done with drift scan mode on the M15,
with the detection of M15A and many related harmonics.
Till now, 7 pulsars from M15 were discovered by FAST.
None of them are bright enough to be found in drift scan, 
indicating that finding new and faint pulsars still needs tracking and long intergration. 
The first set of Globular Cluster pulsars from FAST,
came from October of 2017, can be M92A \cite{2020ApJ...892L...6P} and M13F\cite{{2020ApJ...892...43W}}.
Both of them were discovered with the ultra-wide-band receiver and at $\rm \sim$500~MHz, even if the low frequency is not so suitable for searching these MSPs.
In addition, after J0203-0150 (discovered in August, 2017\cite{2023MNRAS.518.1672M}), 
these two MSPs were also the second and third MSPs discovered by FAST.
Since the Autumn of 2020, the 19-beam L-band receiver of FAST was ready, 
the data center was ready, 
and the pipeline for searching for the exotic pulsars was ready, 
the GC pulsar discovery jackpot came, resulting in more than 20 pulsars discovered within 2 months.
The biased data reduction focused on faint signals, long spin period signals, short orbit binaries, or identifying candidates were then used, 
resulted in many discovered.

\par Among the 60 pulsars discovered by GC FANS, there are 5 spider pulsars, inclucing 2 redbacks \cite{2020ApJ...892L...6P}(M92A; \cite{WanginprepM14E}, M14E) and 3 black widows (\cite{WanginprepM14E}, M14A), NGC 6712A\cite{2021ApJ...921..120Y}, M2G\cite{2025ApJ...991...44L}), and one inter-stage spider M71E\cite{2023Natur.620..961P}.
In addition, M3A was discovered previously and confirmed to be a black widow \cite{2024ApJ...972...43L}.
Two long spin period pulsars from M15 (M15K and L, \cite{2024ApJ...974L..23W}) were from using Fast Folding Algorithm, may indicating a new neutron star forming routine\cite{2024ApJ...977L..42K}.
With a widely used spectra stacking method (e.g., \cite{2016MNRAS.459L..26P}), 
6 faint pulsars from M15 and NGC 6517 were discovered.
These faint signals can't be found by typical frequency domain pulsar search pipeline.
Besides, viewing the spectra was a useful method to identify new signals from the forest of known pulsar harmonics.
The pulsar M15O, which has a very similar (less than 1\%) to the 10$^{th}$ harmonic of M15A,
was discovered \cite{2025RAA....25g1001D}.
In addition, in the pulsar identification, 
the on-off observing mode was used to create one key feature for candidate classification \cite{2021RAA....21..185Q}.
This is a new idea for the machine learning and may be used to break the bottleneck in sample feature selection.

\par Currently, the GC FANS project were focus on three aspects.

\par \textbf{NGC 6517 and the possible intermediate mass black hole near the center:  }
Before FAST, there are 4 pulsars in NGC 6517.
With FAST data, 17 more were found \cite{2021RAA....21..143P,2024ApJ...969L...7Y}.
Among all the 21 pulsars, 17 were timed.
Their positions are near a unknown faint x-ray object near the GC center.
As a very dense and core-collapse GC, the gathering of pulsars may be cause by the gravitational potential and the N-body simulation agrees, too.
Thus, we suggest that there should be an IMBH near the center of NGC 6517 (Lian et al. in prep).

\par \textbf{M10B as the result of stellar object inter-action:  }
M10B is a binary pulsar with the spin period of 7.35~ms.
The orbital eccentricity is 0.985.
From the OMDOT, the total mass of this binary system is 1.9 solar mass,
smaller than the total mass when assuming the pulsar mass is 1.4 solar mass and the orbit is edge-on.
The accretion is the key for MSP forming and makes the orbit being circular.
So, such a very quick MSP should be formed earlier than forming the current M10B system.
Thus, M10B is the result of inter-action of the objects in GCs, 
proving that the interaction tends to form eccentric binaries.

\par \textbf{The mysterious M12B:  }
M12B was discovered by so-called cross matching method. 
It is a binary in a $\rm \sim$0.5 day orbit, however the timing solution was not obtained till now.
By folding the data with a partial timing solution, it seems that M12B's orbit was changed periodically, suggesting that maybe there is a third object around it and affect the orbit.
The monitoring observing is on-going, and currently no more evidence were found.

\subsection{~Mean Pulse profile.}%

\par 
Pulsars are highly polarized radio sources. Their mean pulse profiles exhibit diverse polarization characteristics, including linear polarization, circular polarization, and position angle swings. These features have long been studied to deduce the geometry and mechanisms of pulsar emission. Databases of polarized mean pulse profiles covering approximately 1,500 pulsars have been established using data from telescopes such as the Lovell Telescope, Arecibo, Parkes Telescope, and MeerKAT, etc. However, these databases are limited in sample size, particularly for faint pulsars, due to constraints on observational sensitivity. The FAST addresses this gap with its unprecedented sensitivity and 19-beam L-band receiver, enabling the detection of polarization signals from 460 previously unobserved pulsars and construction of a database for 682 pulsars \cite{WangPengfei2023}. 

\begin{figure*}
    \centering
    \includegraphics[width=0.33\textwidth]{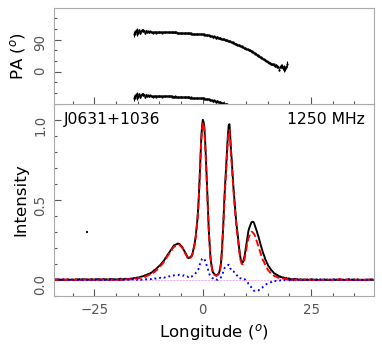}
    \includegraphics[width=0.33\textwidth]{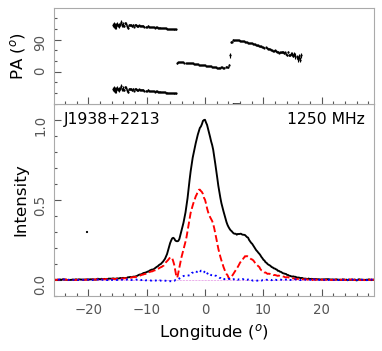}
    \includegraphics[width=0.33\textwidth]{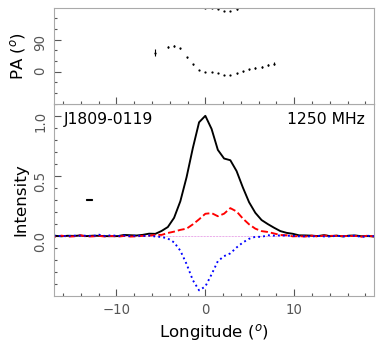}\\
    \includegraphics[width=0.33\textwidth]{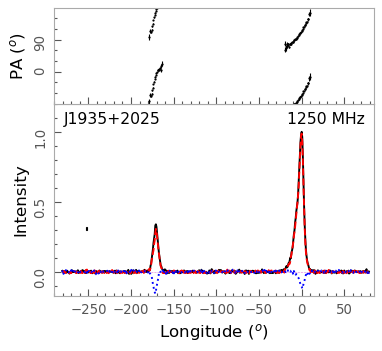}
    \includegraphics[width=0.33\textwidth]{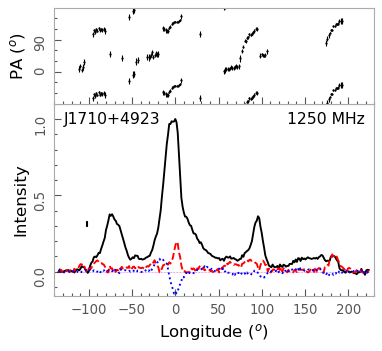}
    \includegraphics[width=0.33\textwidth]{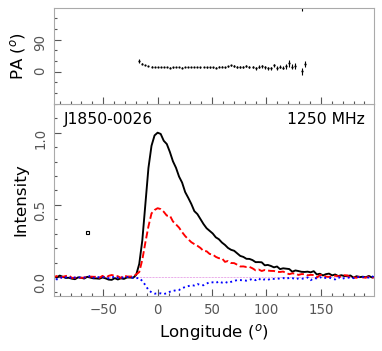}
    \caption{Mean pulse profiles with diverse polarization. In the bottom panels, the black solid, red dashed and blue dotted lines are for the total intensity, linear and circular polarization intensities. The position angle of linear polarization are represented by black dots in the top panels.  
    }
    \label{fig:Mean_Prof}
\end{figure*}

\par 
This sample of mean pulse profiles exhibits diverse emission features, which are categorized into eight main types. \textbf{(1) S-Shaped Polarization Position Angle (PA) Curves.} Approximately 29\% of the pulsars exhibit a monotonic ``S"-shaped swing in PA with rotation phase, which indicates emission originating from a dipole magnetosphere. Notably, some millisecond pulsars (MSPs, e.g., PSR J0605+3757) also displayed S-shaped PA curves, demonstrating that MSP polarization behavior is not fundamentally different from normal pulsars. These curves are critical for deriving pulsar geometry via the Rotating Vector Model (RVM). \textbf{(2) Orthogonal Modes.} 25\% of the pulsars show sudden $90^\circ$ jumps in PA, which is often accompanied by reduced linear polarization intensity. Most have single jumps, some have double jumps or multiple jumps. These jumps are caused by change of the dominance of ordinary (O) modes and extraordinary (X) modes within the magnetosphere \cite{wwh14}. Even MSPs (e.g., PSR J0340+4130) exhibit this feature, which suggests similar magnetosphere conditions for emission and wave propagation across pulsar types. Non-$90^\circ$ jumps are also detected (e.g., PSR J1946+2244), which is attributed to mixed emission from different magnetosphere heights. \textbf{(3) Highly Linearly Polarized Emission.} 11\% of the pulsars have linear polarization fractions $>70\%$, either across the entire profile or in specific components. This high linear polarization aligns with theoretical predictions of curvature radiation from relativistic particles streaming along curved magnetic field lines \cite{wh16}. It is also confirmed that pulsars with higher energy loss rates are more likely to exhibit strong linear polarization. \textbf{(4) Highly Circularly Polarized Emission.} Only some pulsars show circular polarization fractions $>30\%$, with either left-handed or right-handed circular polarized. This feature might be caused by curvature radiation models, where density gradients of relativistic particles and line-of-sight geometry produce net circular polarization \cite{wwh12}. Non-symmetric cyclotron absorption in the magnetosphere may also contribute. \textbf{(5) Conal Double Pulsars.} Some pulsars have two distinct components, likely from conal emission beams. The study verified a previously reported correlation that increasing PA swing correlates with right-handed circular polarization, while decreasing swing correlates with left-handed polarization. While some pulsars, like PSR J1919+1745, deviate from this trend, which suggests complex beam structures. \textbf{(6) Interpulse Emission.} Approximately 4\% of the pulsars exhibit an "interpulse" component offset by roughly $180^\circ$ from the main pulse, indicating emission from both magnetic poles of the pulsar. \textbf{(7) Very Wide Profiles.} 3\% of the pulsars have profiles spanning $>180^\circ$ of rotation phase, with some ones covering nearly the entire rotation cycle. Most of them are MSPs, but some ones are normal pulsars, e.g., PSR J1932+1059 (B1929+10), as caused by small inclination angles that keeps the line of sight within the emission beam or large polar caps. \textbf{(8) Scattering Profiles.} 3\% of the pulsars show extended tails due to interstellar scattering, accompanying with flat PA curves in the tail region. 

\par Pulsars are known as the ``lighthouse'' of the universe. As mentioned above, most of them behave as periodic pulses are usually narrow with a duty cycle of $\sim 10 \%$. Meanwhile, the mean pulse profiles of the pulsars are crucial for understanding their radiation mechanisms, as they result from the line of sight repeatedly cutting the radio emission beam. Thanks to the extremely high sensitivity of FAST, we find two whole pulse phase radio emission pulsars, B0950$+$08 and B1929$+$10\cite{2022MNRAS_Wang, 2025A&A_Wang}. After observing for approximately 160 minutes with FAST, we identify, for the first time, a weak emission component in the mean pulse profile of the pulsar B0950$+$08, its observed flux density is about $10^{-5}$ of the magnitude of the peak radio emission. To investigate and estimate the relative intensity of the extremely weak emission component further even, we propose a function, $\Theta(n)$, to identify the intrinsic radio emission from this pulsar \cite{2022MNRAS_Wang}. The $\Theta(n) $ function is defined as the following,
\begin{equation*}
    \Theta(n) = \sum_{k=1}^{N_{\mathrm{period}}} { 2(I_{k,n-1} - I_{k,n+1})^2 - [(I_{k,n}-I_{k,n-1})^2 + (I_{k,n}-I_{k,n+1})] },
    \label{func_theta}
\end{equation*}
where $I_{k,n}$ denotes the radio signal intensity in the $k$th period and its $n$the pulse phase bin, and $N_{\mathrm{period}}$ corresponds to the total rotation periods.
\par
\begin{figure*}
    \centering
    \includegraphics[width=1\linewidth]{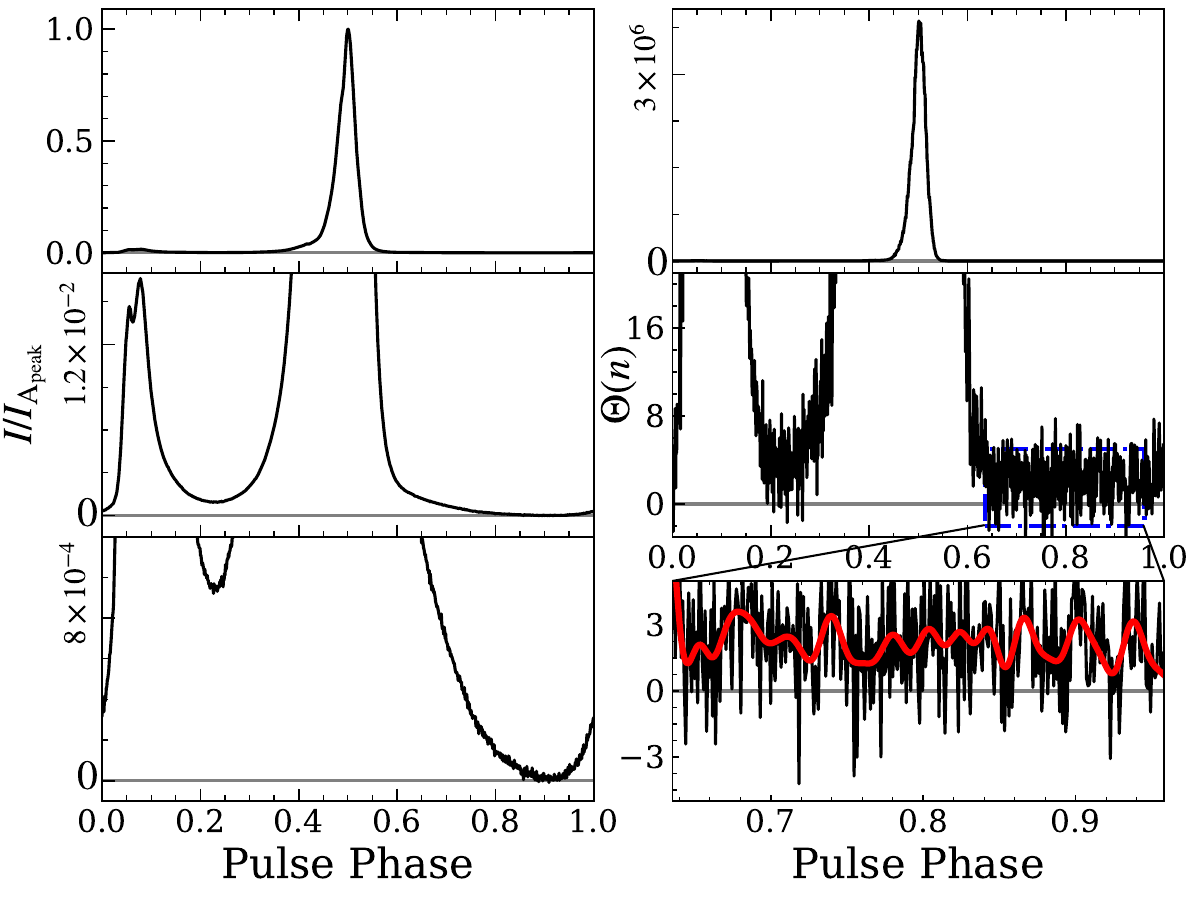}
    \caption{Left: the mean pulse profile of PSR B0950$+$08 over the number of about $36000$ individual pulses based on the conventional baseline subtraction. To reveal the weak emission components and its whole pulse profile feature, the corresponding vertical expanded views in $\times 60$ and $\times 900$ are included in the middle and bottom panels. The intensity of the region between the pulse phase $0.88$ and $0.92$ is believed to be the baseline intensity. The $I_{\mathrm{A_{peak}}}$ denotes the peak radio emission. Right: The $\Theta(n)$ function for the whole pulse phase is shown in the top panel, and a vertically zoomed-in view is plotted in the middle panel. Moreover, a detailed view of the region in the dashed blue box in included in the bottom panel to display the relative intensity of the weak emission component. The red curve corresponds to the smooth value of the function $\Theta(n)$.}
    \label{B0950}
\end{figure*}

 The radio emission feature of pulsar B0950$+$08 is shown in Figure \ref{B0950}, one can see that the mean pulse profile of the region between 0.65 and 0.95 displays a ``valley'' structure, and the observed flux density of the bottom of the ``valley'' structure is about $10^{-6}$ of the magnitude of the peak radio emission. The right panel of Figure \ref{B0950} implies that the intrinsic radio emission of this pulsar over its whole pulse phase, demonstrating that pulsar B0950$+$08 is a whole pulse phase radio emission pulsar.
\par
\begin{figure*}
    \centering
    \includegraphics[width=1\linewidth]{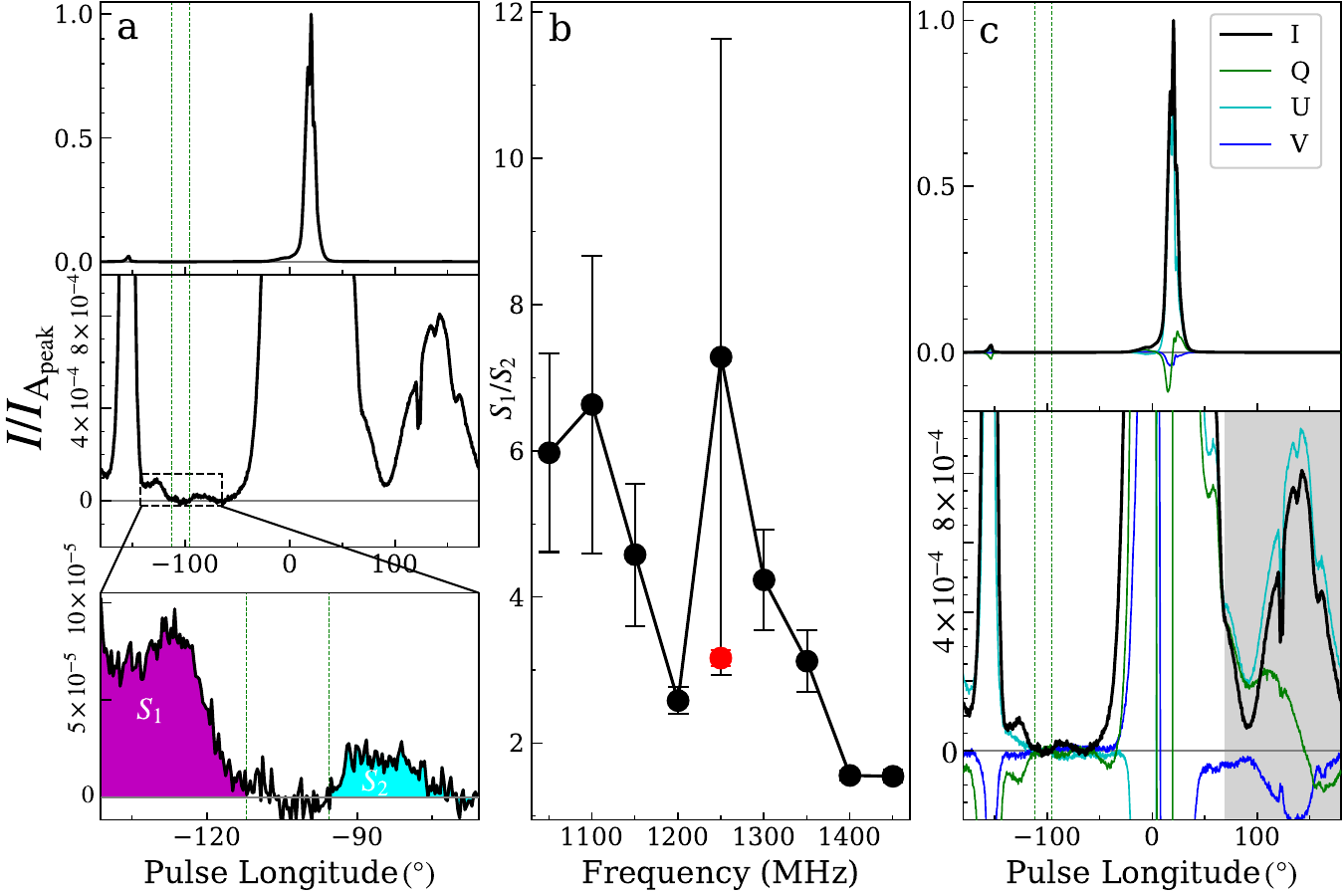}
    \caption{Radio emission feature of PSR B1929$+$10 based on the conventional baseline subtraction. (a) The averaged pulse profile is included in the top panel, and the corresponding $\times 1000$ expanded scale view is plotted in the middle panel. To unravel the profile longitudes with extremely weak emission at the pulse longitude range $-150^{\circ}$ to $-70^{\circ}$ even further, a detailed view of the rectangle region in the dashed-dotted box is shown in the bottom panel. The labeled $S_1$ and $S_2$ correspond to the areas of the first and second new pulse components in the left (magenta region) and the right (cyan region), respectively. The region between the two vertical green dashed lines is believed to be the baseline position, and the solid gray line represents the baseline position. The $I_{\mathrm{A_{peak}}}$ denotes the peak radio emission of the average pulse. (b) To identify whether the two new pulse components are still visible, which are similar to the results shown in the bottom panel (a) at nine narrow bands, we calculate the ratio between the $S_1$ and $S_2$, $S_1/S_2$, at nine narrow bands and plot them in this subpanel. The red dot corresponds to the ratio $S_1/S_2$ for the two new pulse components of the average pulse profile. (c) The observed profiles of Stokes parameters for PSR B1929$+$10 based on the conventional baseline subtraction. We plot the total intensity (the Stokes $I$ in the black), the Stokes $Q$ in the green, the Stokes $U$ in the cyan, and the circular polarization intensity (the Stokes $V$ in the blue) in the top panel. To reveal the observed profiles in the weak emission region in more detail, the $ \times 1000$ expanded scale view is included in the bottom panel. The vertical axis is the same as the vertical axis of panel (a). Same as panel (a), the region between the two vertical green dashed lines is believed to be the baseline position and is then subtracted. The horizontal gray line represents the baseline position. And the vertical gray shadow shows that an unphysical phenomenon is that the Stokes $U$ is greater than the total intensity Stokes $I$, due to the conventional baseline subtraction, which is not suitable for this pulsar.}
    \label{B1929l}
\end{figure*}

 The radio emission of PSR B1929$+$10 emits a similar feature as the PSR B0950$+$08, both of them have high brightness temperature, and are located in a nearby location. To measure the radio emission characteristics and polarization behaviors of the normal and bright pulsar PSR B1929+10, we carried out a long-term observation to track this pulsar. We find, for the first time, two new emission components with an extremely weak observed flux density of about $10^{‑4}$ of the magnitude of the peak radio emission of PSR B1929$+$10\cite{2025A&A_Wang}. Our results show that the intrinsic radio emission of PSR B1929$+$10 covers the $360^{\circ}$ of longitude, demonstrating that this pulsar is a whole $360^{\circ}$ of longitude radio emission pulsar. The radio emission feature of PSR B1929$+$10 is shown in Figure \ref{B1929l}.

\par
\begin{figure*}
\centering
\includegraphics[width=0.33\textwidth]{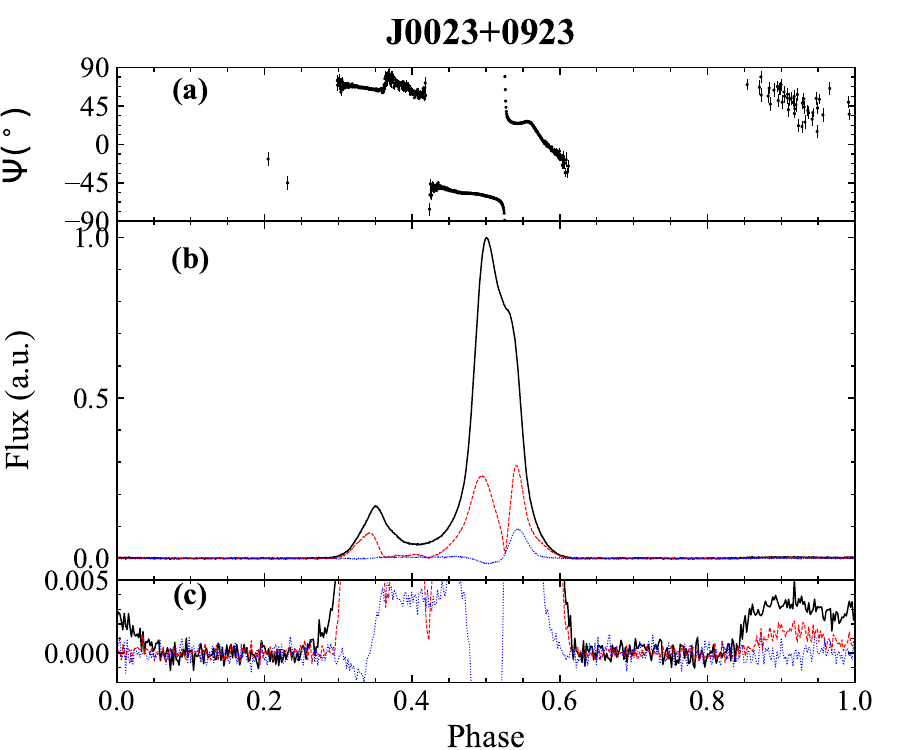}  
\includegraphics[width=0.33\textwidth]{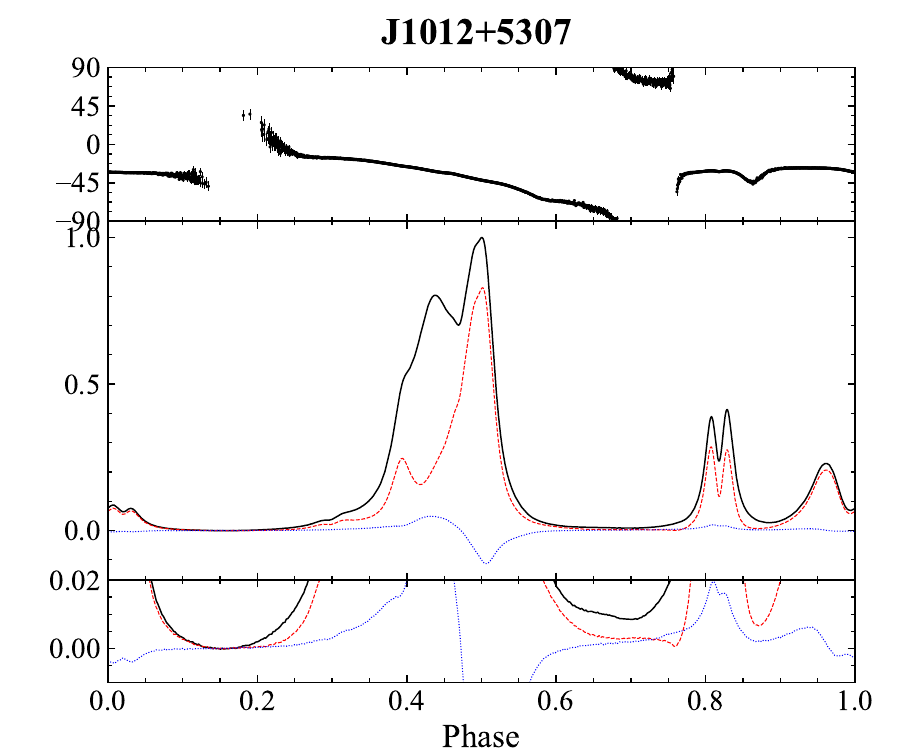}
\includegraphics[width=0.33\textwidth]{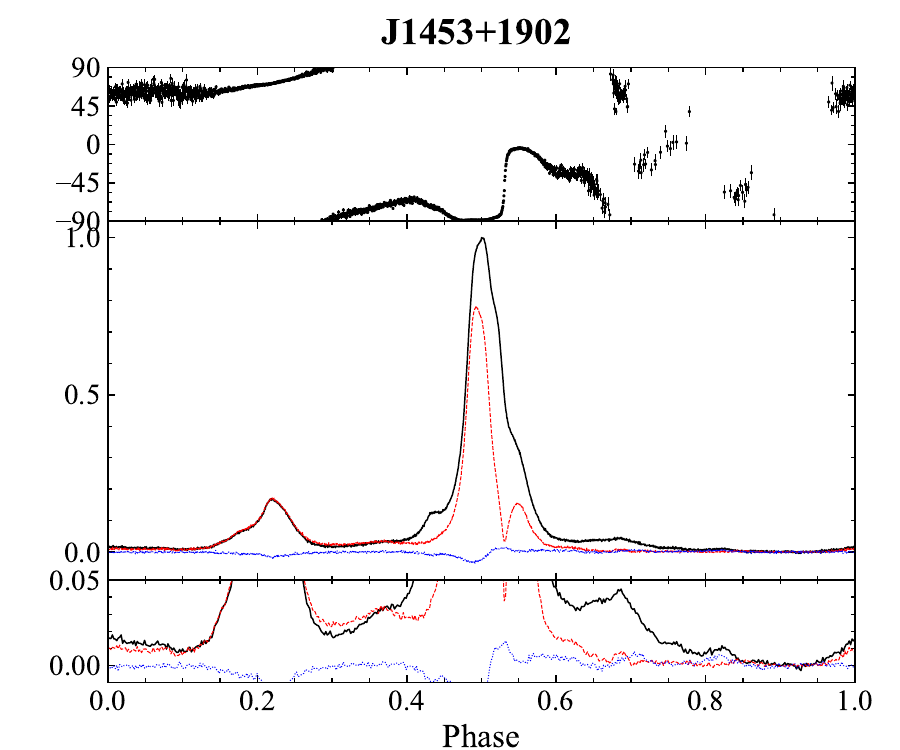}\\
\includegraphics[width=0.33\textwidth]{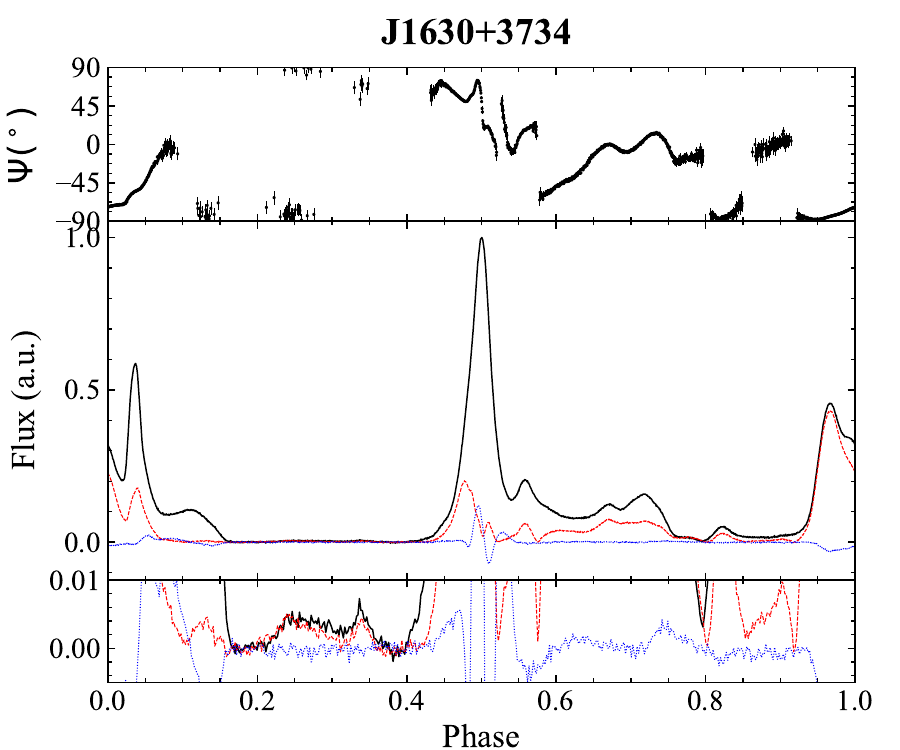}
\includegraphics[width=0.33\textwidth]{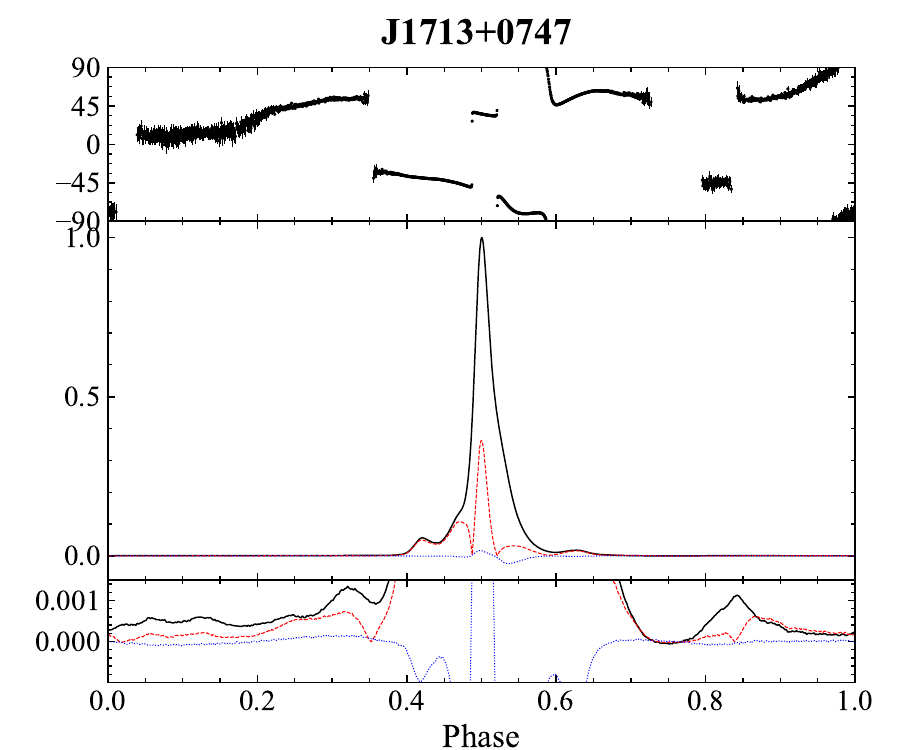}  
\includegraphics[width=0.33\textwidth]{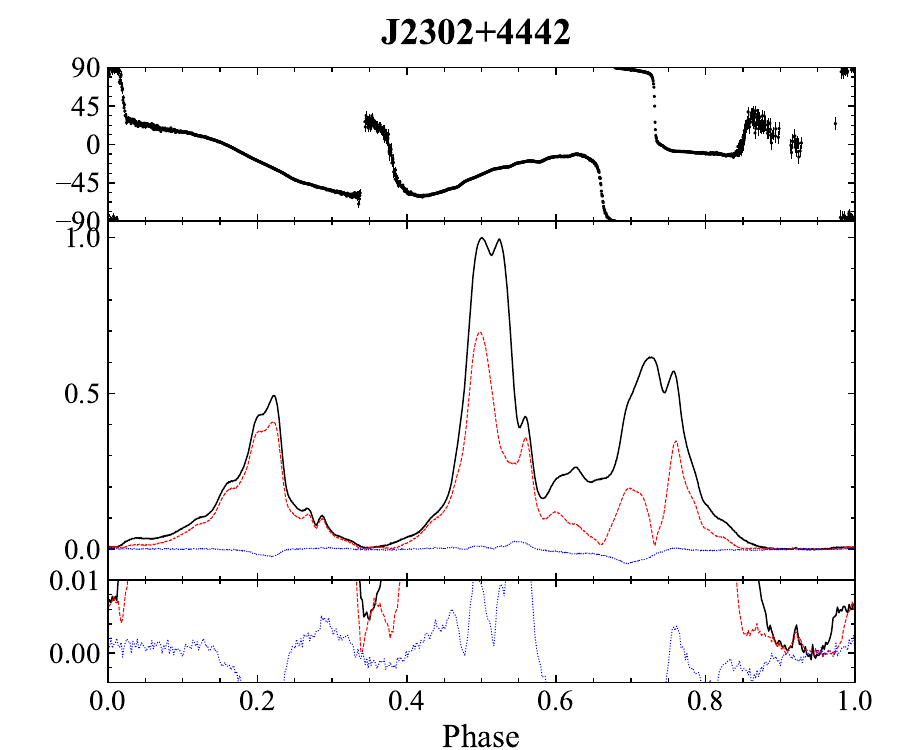} 
\caption{Integrated polarization profiles of some millisecond pulsars. Panel (a) presents the position angle curve, and panel (b) gives the total intensity (black), linear polarization (blue) and circular polarization curves (red). The panel (c) concentrates on smaller intensity range to more explicitly illustrate weak emission components. \label{fig:cpta_pol}}
\end{figure*}

 
Compared with normal pulsars, the profiles of millisecond pulsars (MSPs) generally occupy broader pulse phase\cite{WangPengfei2023,2024MNRAS_Karastergiou}. The correlation between profile width and spin period exhibits a spectral index of roughly -0.3 from some large pulsar sample studies\cite{2019MNRAS_Johnston,2024MNRAS_Karastergiou}. It is reasonable since the emission beam size is inversely correlated with the spin period under the hollow cone assumption\cite{1970Nature_Komesaroff}. Owing to the long term monitoring on some MSPs, the pulsar timing array (PTA) projects provide a substantial sample of profiles with unprecedented signal-to-noise ratio (S/N)\cite{2004ApJ_Manchester,2011MNRAS_Yan,2015MNRAS_Dai,2018ApJ_Gentile,2022PASA_Spiewak,2022ApJ_Wahl}. Some very weak emission components were detected for the first time in a subset of MSPs\cite{2018ApJ_Gentile,2022ApJ_Wahl}. Their appearances demonstrate that MSPs sustain emission over much broader rotation phases than previously thought. Recently, the Chinese pulsar timing array (CPTA), which mainly utilizes the FAST telescope, delivers highly delicated polarization profiles of 56 MSPs\cite{2025A&A_Xu} as depicted in Fig.\ref{fig:cpta_pol}. Nearly 80\% of them were detected to contain weak emission below 3\% of the peak flux. These weak emission components present as pulse-like structure such as J0023$+$0923, J1630$+$3734, J1713$+$0747 and J2302$+$4442, or bridge-like emission as shown in J1012$+$5307 and J1713$+$0747. In addition, we also found some whole pulse phase radio emission candidates as noted above. The duty cycles of six pulsars are larger than 90\% such as J1012$+$5307, J1713$+$0747 and J2302$+$4442, which illustrates that they may sustain radiation over the whole spin period. Another 12 pulsars are also potential whole-phase radiators, although their duty cycles are slightly smaller. Some of them are limited by insufficient S/Ns like J1630$+$3734, where the weak emission components are obscured by noise. They can be further checked with longer integration time accumulated. Some pulsars exhibit "polarization excess" as happened around phase 0.3 for J1453$+$1902, where the linear intensity is little higher than the total intensity. The total intensity is therefore underestimated, due to the incorrect removal of underlying base emission.  With interferometer observations in the future like Square Kilometre Array (SKA) or FAST core array, we expect to better determine their baselines. 
\par 
These whole pulse phase radio emission pulsars challenge the typical picture of pulsar radiation, which are supposed to concentrate within narrow windows. The pulsar emission region can be broadly distributed within the magnetosphere. From the geometry side, a nearly aligned rotator is a simple and reasonable explanation. It can also explain that the majority of whole-phase radiators are MSPs, since the magnetic inclination angle may decrease with time\cite{2017MNRAS_Johnston,2023MNRAS_Johnston}. However, we notice that some pulsars contain roughly 180$^\circ$ separated inter-pulse like B1929$+$10 and J1630$+$3734. For this case, some radiation components should come from higher emission height where the magnetic field line has larger inclination angle with magnetic axis. As denoted in B1929$+$10\cite{2025A&A_Wang}, they found the region responsible for weak emission components locate around the light cylinder. A recent simulation on static dipole magnetosphere under the assumption of curvature radiation has also approved both geometric configurations\cite{2025ApJ_Wang}. Moreover, they found that the persistent radiation, although much weaker than the main pulse, could be generated at high emission height. We may therefore expect more pulsars become whole-phase radiators with enough integration time in the future.

\subsection{~Single dwarf pulses.}

\par
\begin{figure*}[!thp]
    \centering
    \includegraphics[width=\columnwidth]{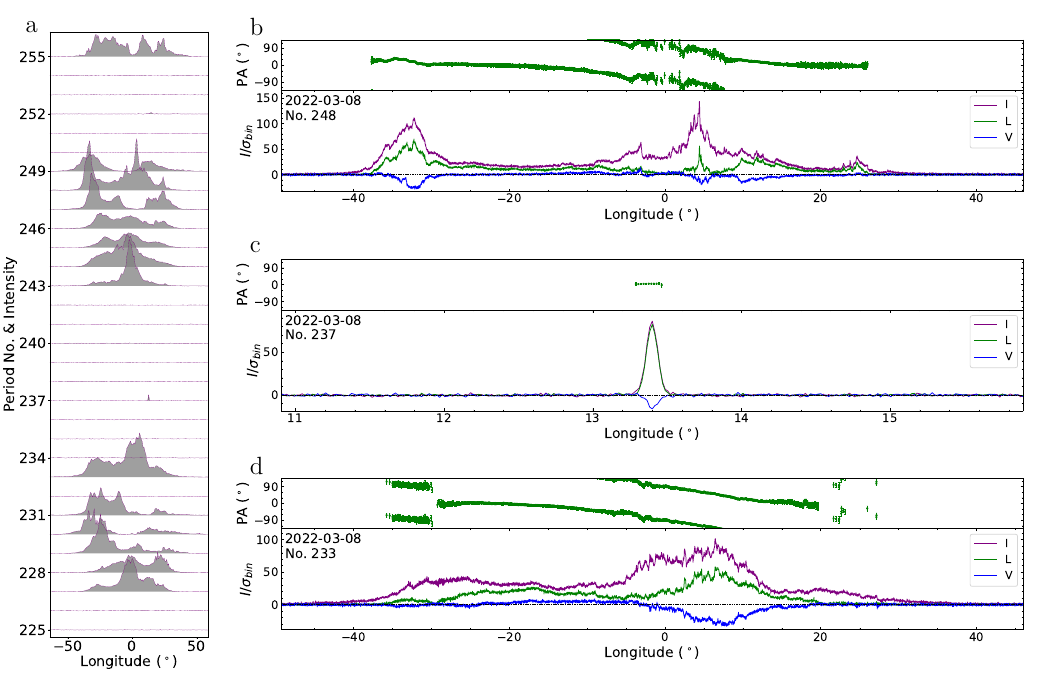}
    \caption{ FAST detection of a dwarf pulse in a series pulses of B2111+46. (a) A segment of pulse trains of PSR B2111+46 observed in the session of 2022-03-08 by the FAST, showing some emission and nulling periods; (b-d): polarization profiles of 3 individual pulses. In the lower sub-panels the total intensity $I$, linear polarization $L$ and circular polarization $V$ (with positive values for the left hand sense) are plotted in the original time resolution (49.152~$\mu$s) of FAST observations, and the polarization angles PA are plotted in the upper sub-panels. The dwarf pulse in the period No.237 has only one resolved emission cell, almost fully linearly polarized with a width of about $0.1^{\circ}$. Many notches of the other two pulse profiles are sensitive significant detection of real intensity fluctuations caused by emission cells with different strength. The error bar for PA is  $\pm1\sigma$. The intensity is scaled with the off-pulse fluctuations expressed by $\sigma_{\rm bin}$ (original from \cite{Chen2023}).}
    \label{B2111+46:DP-pulse}
\end{figure*}

\begin{figure*}[!thp]
    \centering
    \includegraphics[width=\columnwidth]{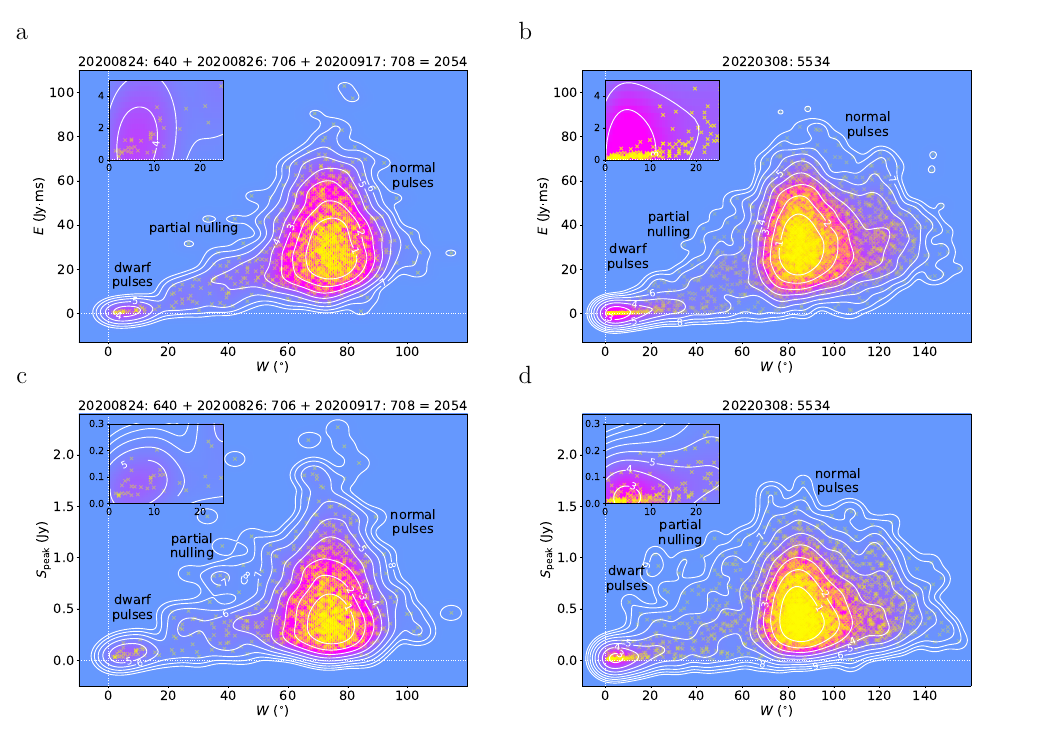}
    \caption{Dwarf pulses of PSR B2111+46 as a distinct population from the partial nulling and normal pulses. Pulse width is measured at the most outer profile at the 3$\sigma$ detection level. {\it Upper panels}:  Pulse fluence integrated over each pulse, $E$, against pulse width, $W$, and the density distribution of data are presented in color and also in contours at levels of 1/$2^{-n}$ of the maximum density ($n$ = 1 $-$ 8). More sensitive observations on 2022-03-08 gives larger widths for normal pulses. No mention of about 20\% of nulling period, normal pulses are concentrated around the main peak, with a fluence in the range from 10 to about 50~Jy$\cdot$ms and a pulse width $60^{\circ}<W<100^{\circ}$. The dwarf pulses are concentrated on another peak, with a fluence less than 1 Jy$\cdot$ms and a pulse width less than $15^{\circ}$ (i.e. 40 ms). In between are partially nulling pulses (see Method: Individual pulses). {\it Lower panels}: The same as {\it the upper panel} but for peak flux density $S$ against the pulse width. The {\it left panels} are made for individual pulses obtained the three sessions in 2020, the {\it right panels} for pulses detected in the longer verification observation session on 2022-03-08  (original from \cite{Chen2023}).}
    \label{B2111+46:DP-EW}
\end{figure*}

 
Using the high-sensitivity observations of FAST, a number of narrow and weak pulses was detected in the ordinary nulling state of pulsar B2111+46 \cite{Chen2023}, and later for PSR J2323+1214 \cite{Tedila2025}, PSR B1931+24 \cite{Rusul2025arXiv} and other 10 pulsars \cite{Yan2024}. Utilizing the unprecedented sensitivity of FAST, a distinct population of pulsar radio emission known as ``dwarf pulses" has been identified. In the detailed study of the nulling pulsar B2111+46, FAST revealed a large number of sporadic, weak, and narrow pulses appearing during phases previously categorized as ``nulling" states by Chen et al \cite{Chen2023} (see Figure \ref{B2111+46:DP-pulse}). Unlike giant pulses, which can exceed the average pulse energy by orders of magnitude, dwarf pulses occupy the lower extreme of the energy distribution. They exhibit a peak flux density significantly lower than that of normal pulses and appear as a distinct island in the two-dimensional distribution of pulse width and intensity (see Figure \ref{B2111+46:DP-EW}). These pulses are typically composed of only one or a few elementary emission cells with widths narrower than $15^{\circ}$, distinguishing them from the broader ``thunderstorm-like" emission of normal pulses.

\par 
Following the initial discovery in B2111+46, the phenomenon was further established as a common feature among nulling pulsars. A systematic search of FAST archival and survey data identified dwarf pulses in ten additional sources, including well-known nulling pulsars such as PSRs B0525+21, B1237+25, and B1944+17 \cite{Yan2024}. For pulsars with sufficient data, such as B1237+25, the dwarf pulses clearly separate from the normal pulse population in width-intensity space. The energy histograms of these sources often display a bimodal distribution, where the dwarf pulses reside within the low-energy peak traditionally associated with nulls. This suggests that for many pulsars, the "nulling" state is not a complete cessation of radio emission but rather a state of extremely weak radiation detectable only by high-sensitivity instruments like FAST.

\par 
Polarization analysis of these dwarf pulses has provided critical constraints on the magnetospheric geometry during the weak emission state. For both B2111+46 and the sample of ten pulsars reported by Yan et al. \cite{Yan2024}, the PAs of the dwarf pulses generally follow the same S-shaped curve or the orthogonal polarization modes defined by the average profile of normal pulses. This consistency implies that the magnetic field structure and the emission geometry remain stable during the transition from the normal state to the dwarf-pulse state. The observation of dwarf pulses appearing at various longitudes, including both core and cone components, further supports that the geometric configuration of the magnetosphere is preserved even when the emission intensity drops drastically.

\par 
The physical origin of dwarf pulses provides insight into the pair production processes in aging pulsars. All pulsars currently known to exhibit dwarf pulses are located in the ``death valley'' of the $P-\dot{P}$ diagram. The emission of normal pulses is interpreted as resulting from a ``thunderstorm'' of particles produced by copious discharges in a regular gap. In contrast, dwarf pulses are likely produced by one or a few ``raindrops'' of particles generated in a fragile gap where the electric potential is barely sufficient to ignite electron-positron discharges. The detection of these pulses suggests that the gap activity in older pulsars is non-stationary and fragile, occasionally producing sparse particle streams that result in weak, narrow, but highly polarized radio emission.

\subsection{~Polarization with single pulses.}

\par 
The emission mechanism sets an initial polarization state of the signal, and propagation effects in the pulsar magnetospheric plasma shape the polarization. Therefore, the polarization of radio pulsar signals contains much information, and is related to several basic questions in pulsar radiation theories. The pulsar polarization is also used for probing the interstellar medium. FAST provides high-quality polarization data of pulsars, and we will introduce some researches done with FAST data in the following paragraphs.

\subsubsection{~Measurements of the radiation geometry.}

\par 
The PAs of pulsar pulse profiles are associated with the radiation geometry (inclination angle $\alpha$, impact angle $\beta$, longitude of fiducial plane $\phi_{0}$, ...) through the RVM \cite{RVM1969} and its modifications. The radiation geometry is important for understanding the magnetospheric structures, and for further analysis of radiative processes. However, RVM fitting of PA curves is always ineffective~\cite{Everett2001}. One important reason is that pulsar signals usually have small duty-cycles, which introduces large uncertainties in estimating geometrical parameters. FAST performs better in measuring radiation geometry, because its high sensitivity could make wider on-pulse regions of integrated pulse profiles, and smaller measure errors of PA values.

\par 
Some large-sample works on measuring pulsar radiation geometry with FAST data have been done, like those in~\cite{WangPengfei2023} (190 pulsars) and in~\cite{SunShengnan2025} (16 interpulse pulsars). Some researches focusing on single sources also achieve good RVM fitting. For example, \cite{CaoShunshun2024} measured the radiation geometry of a frequently studied mode-switching pulsar B0943$+$10. This updated geometry, different from previous ones, is significantly oblique, and performs better in explaining the X-ray pulsation of B0943$+$10. Great sensitivity of FAST also helps measuring the geometry of some sources that have never been measured, like the work done in~\cite{DangShijun2024}.

\par 
Of course, RVM has its limitations, as is already proposed by former theories and observations~\cite{BP2012,Johnston2024}. There are many pulsars with PA curve deviating from RVM curves~\cite{WangPengfei2023}. With higher precision in polarimetry, we may establish a better model for describing radiation geometry in the future.

\subsubsection{~Study of the propagation effects in magnetosphere.}
\par 
The pulsar polarization records what has happened in the pulsar magnetosphere. The key physics is related to the normal wave modes in the pulsar magnetospheric plasmas, about their generation, conversion, and absorption. The basic reflections of wave modes in observations are the orthogonal polarized modes (OPMs). We can decode the propagation effects and diagnose the magnetospheric plasma by analyzing the polarization data. Such magnetospheric study might reveal even more interesting physics when combined with polarized single pulses provided by FAST observation, which present a dynamical magnetosphere.

\par
\begin{figure}
		\centering
		\includegraphics[scale=0.5]{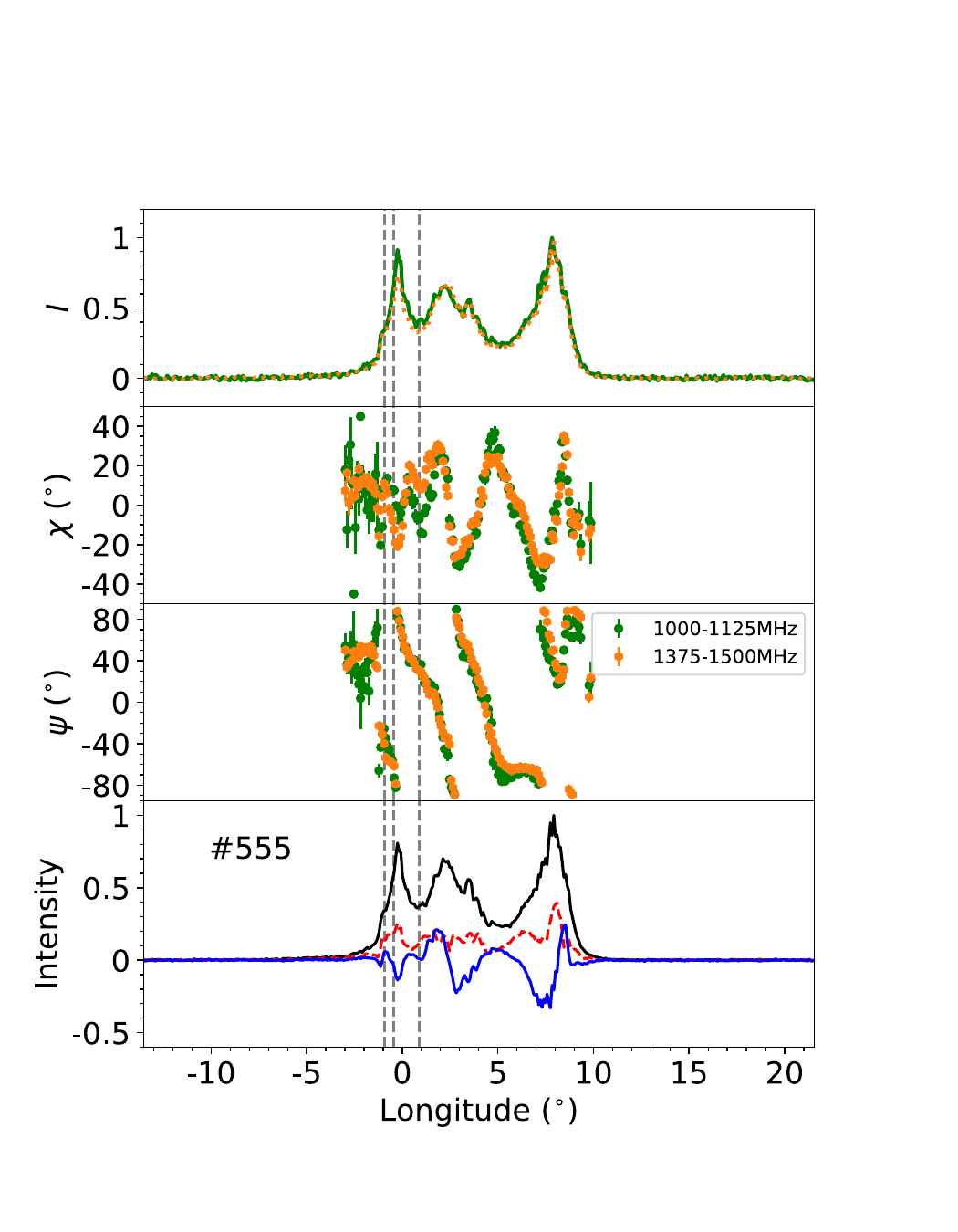}
	\caption{A single pulse of B1919$+$21 from \cite{CaoShunshun2025}, with PA ($\psi$), EA ($\chi$) and total intensity $I$ plotted in two frequency subbands (1000--1125 MHz in green and 1375--1500 MHz in orange). In the $I$ panel, the intensity curves are normalized to the maximum values separately for two subbands. Black line---total intensity ($I$); red dashed line---linear polarization intensity ($L=\sqrt{Q^{2}+U^{2}}$); blue line---circular polarization intensity ($V$). $I$, $L$ and $V$ are normalized by the maximum total intensity of the respective profiles.\label{fig:B1919}}
\end{figure}

 
One example is the analysis and modeling of the single pulses polarization of the first pulsar B1919$+$21 \cite{CaoShunshun2025}. The integrated profile of B1919$+$21 has a distorted PA curve, with almost no sign for RVM. Such complexity comes from its single pulses, where PA could change rapidly along the pulse longitude. An interesting phenomenon is that in some pulses, the PAs rotate quasi-monotonically along pulse longitudes over large angles ($>180^{\circ}$), where the largest quasi-monotonic rotation approaches 720$^{\circ}$. An example is given in Figure~\ref{fig:B1919}. Such large rotation could not be produced by RVM, so the authors attributed it to the coherent summation of orthogonal polarization modes. When the phase lag between two wave modes changes along pulse longitude, both the polarization orientation and the ellipticity of the added polarization will vary accordingly. The phase lag is determined by parameters like plasma density and Lorentz factor.

\par 
Since there has been various kinds of analytical models developed for describing propagation effects in the pulsar magnetosphere (e.g., \cite{Petrova2000,WangChen2010}), we can use FAST data for testing those models, which could be complementary to numerical simulations nowadays. Some primary work has been done with FAST released data:  Cao et al.~\cite{CaoShunshun2025b} finds that in the limiting polarization region of pulsar magnetospheres, the wave mode coupling effect could explain the diversity of observed single-pulse circular polarization spectra. By assuming that emission process itself does not generate significant circular polarization component, the observed circular polarization could come from the coupling of OPMs in the limiting polarization region~\cite{LP1998}. Cao et al. \cite{CaoShunshun2025b} also try to constrain plasma multiplicity $\kappa$ and Lorentz factor $\gamma$ by fitting circular polarization spectra. The results are $\kappa\sim 10^{0}$-$10^{2}$ and $\gamma\sim 10^{0.5}$-$10^{2}$, for three pulsars. They demonstrate that the plasma kinetic energy density account for radio emission is only a very small part of total spin down energy loss.

\subsection{~Pulsar Zoo.}
\subsubsection{~Mode changing.}

\par

\begin{figure*}
\centering
\includegraphics[width=0.3\linewidth]{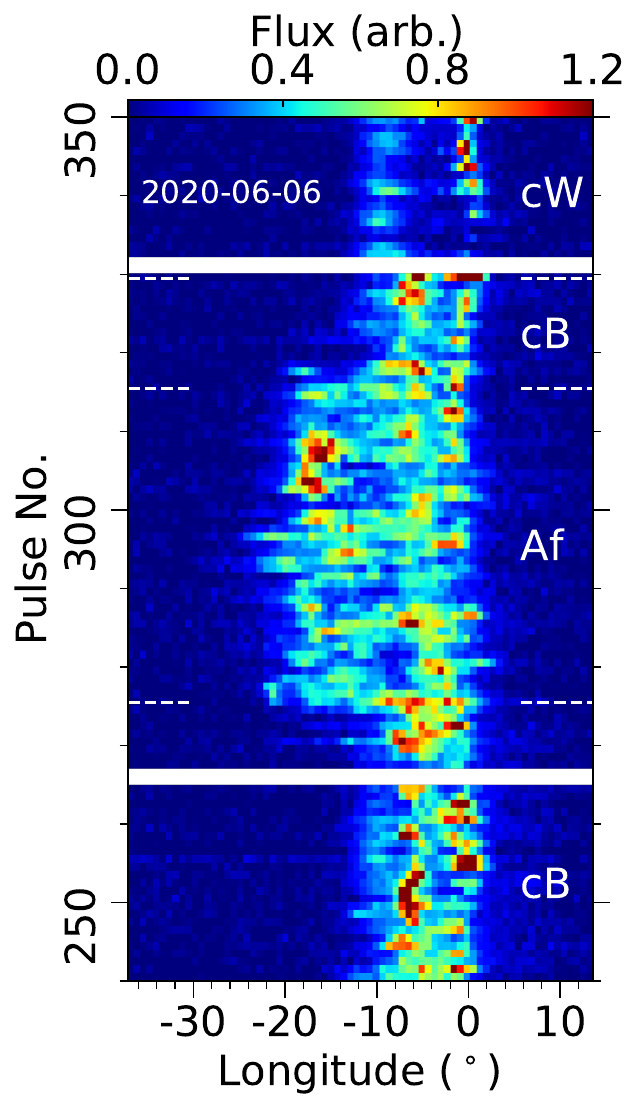}
\includegraphics[width=0.3\linewidth]{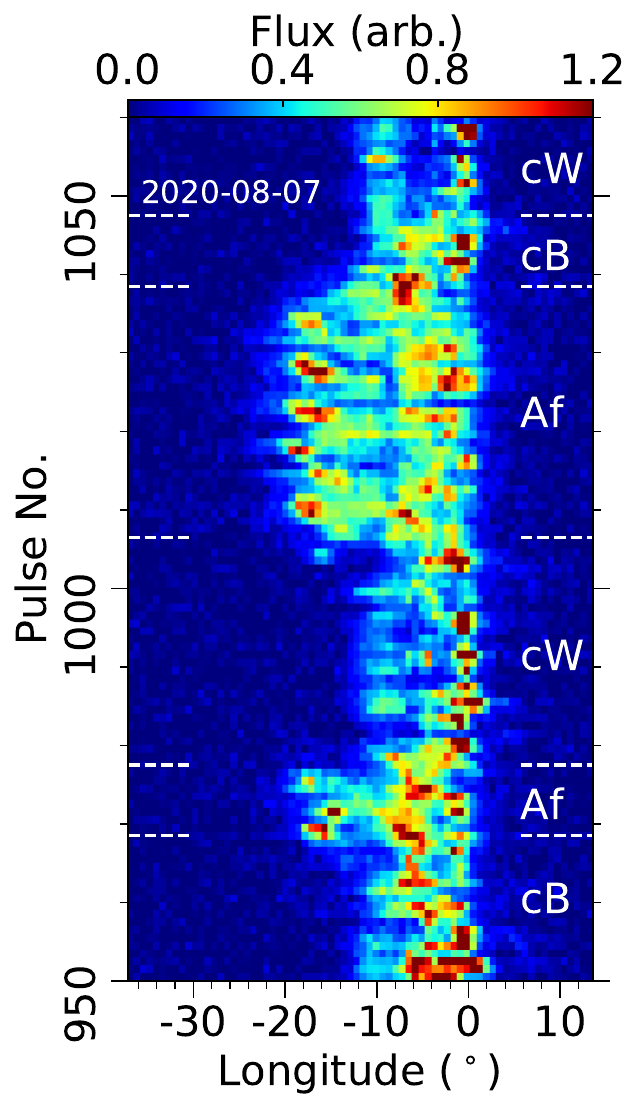}
\includegraphics[width=0.3\linewidth]{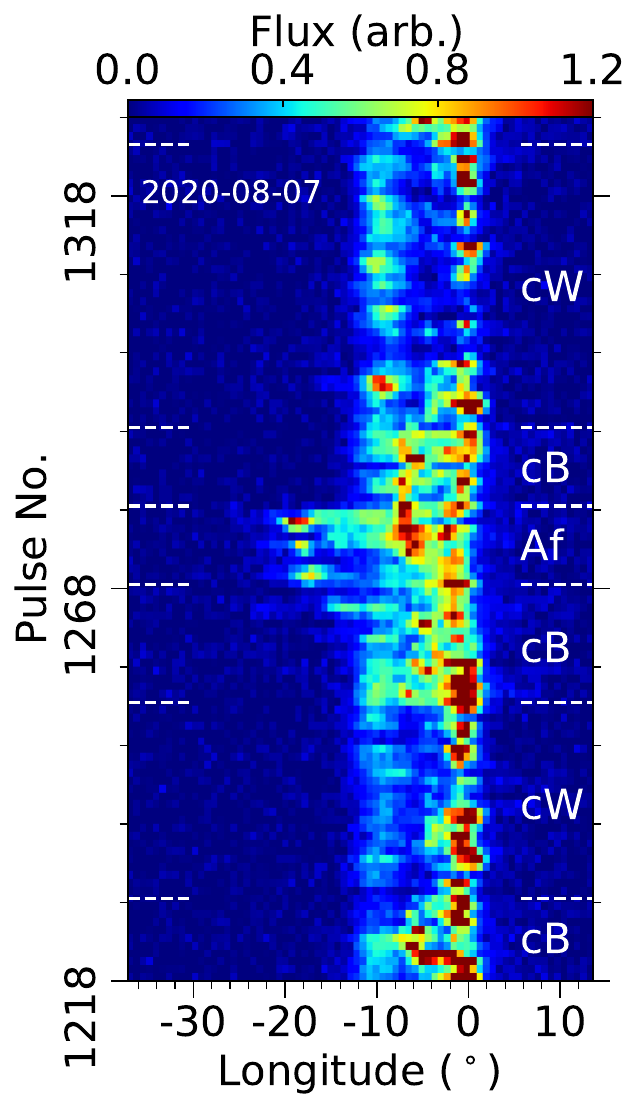}
\caption{Three examples for the anomalous-filled mode, which combines the new leading component in the abnormal mode and the slightly squished components of the core-burst and core-weak modes \cite{wwh+2023}. The data with radio-frequency interference are eliminated as marked with white lines.}
\label{swooshType}
\end{figure*}

\par 
The pulse energy of a pulsar fluctuates significantly from pulse to pulse; however, when a large number of individual pulses are averaged, the resulting mean profile remains remarkably stable~\cite{1975ApJ...195..513T}. The single-pulse energy distribution of most pulsars typically follows a log-normal or Gaussian form~\cite{2018MNRAS.479.5413M}. Nevertheless, some pulsars display more complex forms of intensity modulation, characterized by transitions between two or more quasi-stable emission states.

\par Pulse nulling refers to the sudden cessation of pulsed emission, followed by its later reappearance~\cite{1970Natur.228...42B}. The duration of these nulls varies widely, from a few pulse periods to several minutes or even hours~\cite{2007MNRAS.377.1383W}. In rare cases, pulsars enter a prolonged inactive phase lasting days or years, known as the intermittent pulsar state~\cite{2006Sci...312..549K}. Multi-frequency observations demonstrate that the nulling phenomenon is frequency-independent~\cite{2014ApJ...797...18G}.

\par Mode changing is another distinct form of emission variability, in which the average pulse profile alternates between two or more stable configurations, with each mode persisting for duration ranging from a few pulses to several hours~\cite{1982ApJ...258..776B,2007MNRAS.377.1383W,2024ApJ...964....6W}. The observational similarities between nulling and mode changing suggest that they may share a common physical origin~\cite{2007MNRAS.377.1383W}. Mode changing was first identified in PSR~B1237+25~\cite{1970Natur.228.1297B}, and subsequent studies have found this phenomenon in dozens of pulsars.These observations offer valuable insights into the underlying mechanisms of pulsar radio emission.

\par Several theoretical models have been proposed to explain mode changing, but its physical origin remains uncertain. Early studies suggested that mode transitions arise from abrupt changes in magnetosphere current flows~\cite{1982ApJ...258..776B}. Pulsar magnetosphere may occupy multiple quasi-stable states, each characterized by different current distributions or magnetic geometries, with mode changes reflecting transitions between these states \cite{2010MNRAS.408L..41T}. The correlations between pulse shape variations and changes in spin-down rate in six pulsars, linking emission state transitions to magnetosphere torque variations \cite{2010Sci...329..408L}. This strongly supports the idea that mode changing is driven by reconfiguration of magnetosphere currents, which changes the shape of active emission regions.

\par However, it is still unclear whether the emission heights of different modes are identical or distinct, a key question for understanding the physical origin of mode changing. Measuring the emission height of pulsars is challenging, since both the delay–radius and geometrical methods rely on the assumption that the emission beam is fully filled. In many pulsars, however, the beam is likely underfilled, meaning that estimates based directly on the observed pulse profiles only provide lower limits on the true emission height.

\begin{figure}
    \centering
    \includegraphics[width=0.5\textwidth]{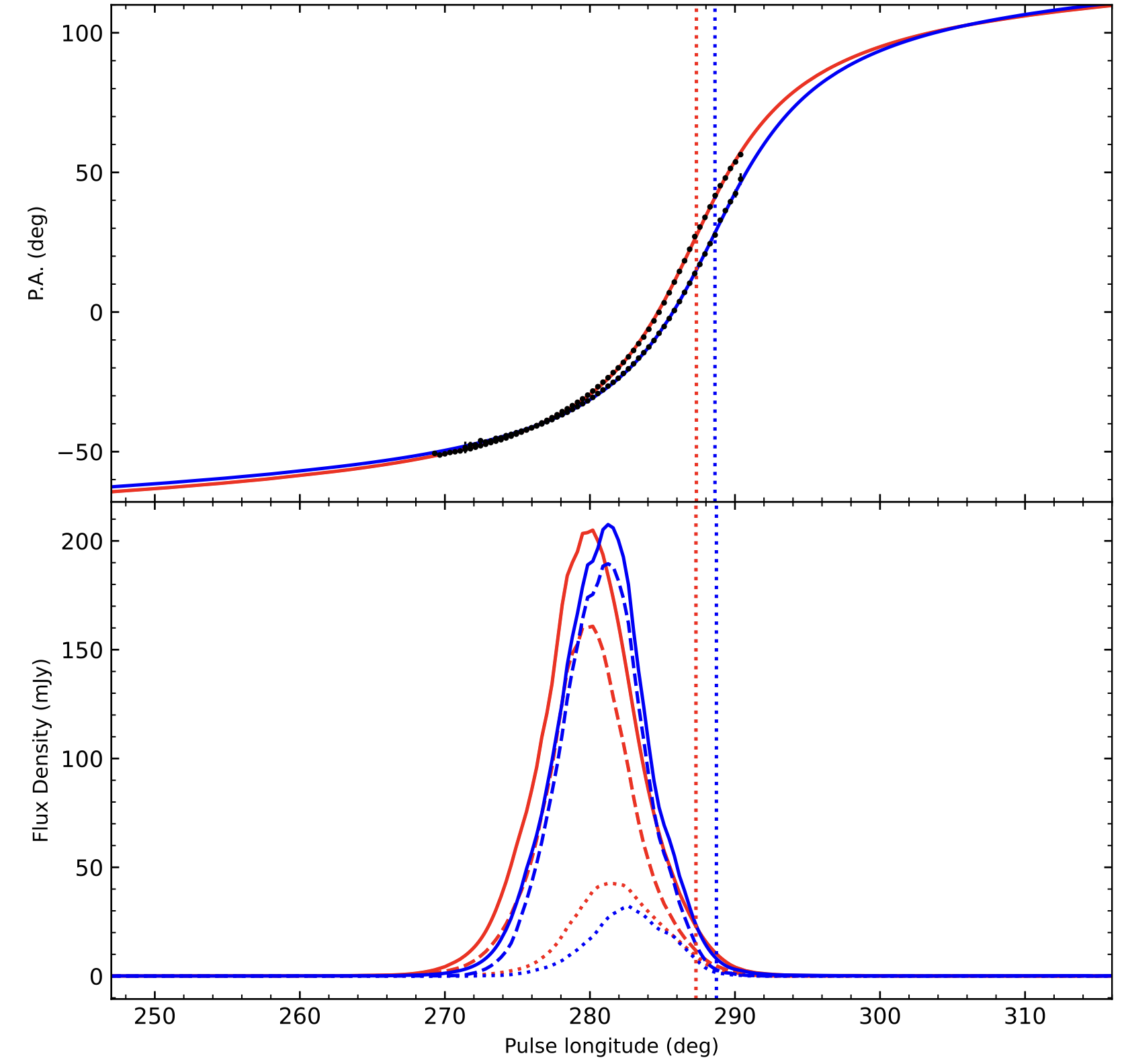}
    \caption{Average polarization properties of mode A (red) and mode B (blue). Solid lines show the total intensity, while dashed and dotted lines indicate the linear and circular polarization components, respectively. In the upper panel, the S-shaped solid curves correspond to the best-fit RVM for the PA swings. Vertical dotted lines mark the phases of the steepest gradient in the RVM fits.}
    \label{fig:J0614}
\end{figure}

\par With its high sensitivity and exceptional polarization capability, FAST enables more detailed studies of pulsar mode changing. 
We observed the well-known mode changing pulsar PSR~J0614+2229 using FAST and found that the PA swings differ significantly between modes (see Figure~\ref{fig:J0614}) \cite{2022ApJ...934...57S}. 
This provides strong evidence that the two modes originate at different altitudes within the pulsar magnetosphere. These results further support the idea that magnetospheric current reconfigurations can alter not only the geometry of the active emission region, but also the height at which radio emission is generated.

\begin{figure}
    \centering
    \includegraphics[width=0.8\textwidth]{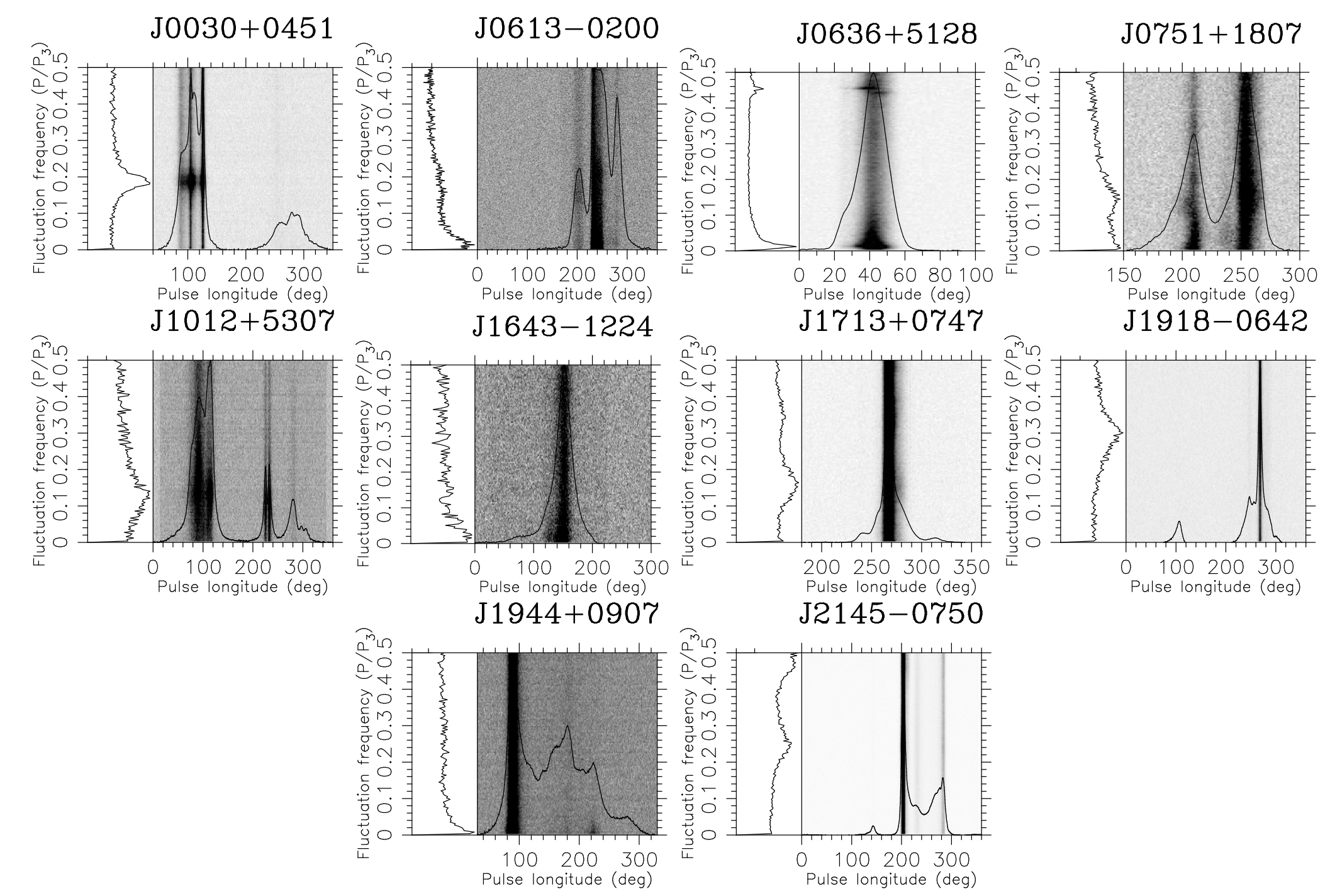}
    \caption{Longitude-resolved fluctuation spectra (LRFS) for 10 MSPs, with the average pulse profile overlaid for reference. Side panels show the horizontally integrated power of each LRFS.}
    \label{fig:msp-sp}
\end{figure}

\par Moreover, FAST offers a substantial advantage in the study of millisecond pulsars (MSPs), which are generally faint and have previously been difficult to observe in detail. Mode changing has mainly been investigated in normal pulsars, and prior to FAST’s operation, only two MSPs (PSRs  B1957+20 and J1909$-$3744) were known to exhibit this phenomenon~\cite{2018ApJ...867L...2M,2022MNRAS.510.5908M}. Using FAST, we observed a large sample of MSPs and investigated their emission variability~\cite{2024ApJ...964....6W}. We discovered a new MSP (PSR J0030+0451) showing clear periodic mode changing, and more than ten MSPs exhibiting periodic intensity modulations similar to this behavior—most of which are reported for the first time (Figure~\ref{fig:msp-sp}). Unlike normal pulsars, in MSPs these periodic modulations often occur only across part of the pulse profile, and no phase-locked relationship between the main pulse and interpulse emission is observed. These distinct characteristics may reflect the presence of multiple emission regions within MSP magnetospheres, where different pulse components originate from separate locations.

\par The emission swoosh phenomenon refers to a peculiar behavior observed in certain pulsars, where the emission profile shifts toward earlier rotational phases for several consecutive pulses, gradually returning to the normal phase. This phenomenon was initially identified in PSRs B0919+06 and B1859+07 \cite{rrw+2006},  and characterized by a displacement of the emission window, with a new bright leading component and a weakened or absent trailing component. The phenomenon was later recognized as quasi-periodic, with intervals of recurrence ranging from 150 to 700 rotation periods \cite{wor+2016,wyk+2022}. Recent findings from FAST also highlight that the duration of emission shifts can vary from a single rotation period to several tens of periods \cite{wyk+2022}. 

\par 
Early studies observed the swoosh as a gradual phase shift, with no abrupt transitions typical of mode changesy \cite{rrw+2006}. Subsequent research revealed different types of swooshes \cite{wor+2016,hhp+2016}, such as abrupt shifts ($\alpha$-type), gradual shifts (M-type), and triangular shifts, each with distinct temporal profiles and emission morphologies. These shifts typically last tens of pulses and often exhibit frequency-dependent characteristics, with the profile separation between normal and anomalous modes becoming smaller at higher frequencies for B1859+07. However, another pulsar B0919+06 exhibits the converse frequency-dependence \cite{rps+2021}. Additionally, the relation between swoosh phenomenon and spin frequency derivatives had been explored, but no certain correlation was reported \cite{psw+2016}.

\par 
Recent observations with the FAST provided a significant advancement by identifying a new emission mode in PSR B1859+07, termed the anomalous-filled (Af) mode \cite[see Figure~\ref{swooshType}]{wwh+2023}. In this mode, both the leading component from the anomalous mode and the trailing component from the normal mode coexist, though the trailing component is slightly compressed in longitude. The Af mode differs from the typical swoosh behavior by retaining the trailing components of the normal emission, rather than completely eliminating them. This suggests that the emission swooshes are not caused by a rigid phase shift, but rather by varying intensities of different emission components within a largely static emission beam. 

\par 
In conclusion, the emission swoosh phenomenon is best understood as the dynamic modulation of the pulsar’s emission beam, with varying intensities of different components. The identification of the Af mode has resolved the ambiguity surrounding the origin of swooshes, ruling out models involving phase shifts or orbital interactions. These results enhance our understanding of pulsar magnetospheres and the complex mechanisms that govern their emission.

\par In the future, with the ever-expanding sample of pulsars observed by FAST, we will be able to construct a comprehensive statistical database of pulsar emission properties. Such a dataset will provide crucial input for theoretical and numerical simulations, advancing our understanding of the pulsar emission mechanism and the magnetospheric processes that drive its variability.

\subsubsection{~Millisecond pulsars.}

\par 
Generally, the spin period of MSPs is smaller than 100 ms and the first derivative is less than 10$^{-17} {\rm s\,s^{-1}}$. Up to now, the number of MSPs is roughly 900, among which FAST has detected more than 200 MSPs\cite{HanJL2025RAA....25a4001H,2025ApJS_Lian}. The MSPs are supposed to form through accreting materials from their companion stars, as confirmed by the discovery of transitional MSP binaries\cite{2009Sci_Archibald,2013Natur_Papitto,2015ApJ_Roy}. The accreting process will distort the configuration of magnetosphere, which may have effect on the emission properties. 

\par
\begin{figure*}
\begin{minipage}[t]{0.33\columnwidth}
    \centering
    \includegraphics[width=1\columnwidth]{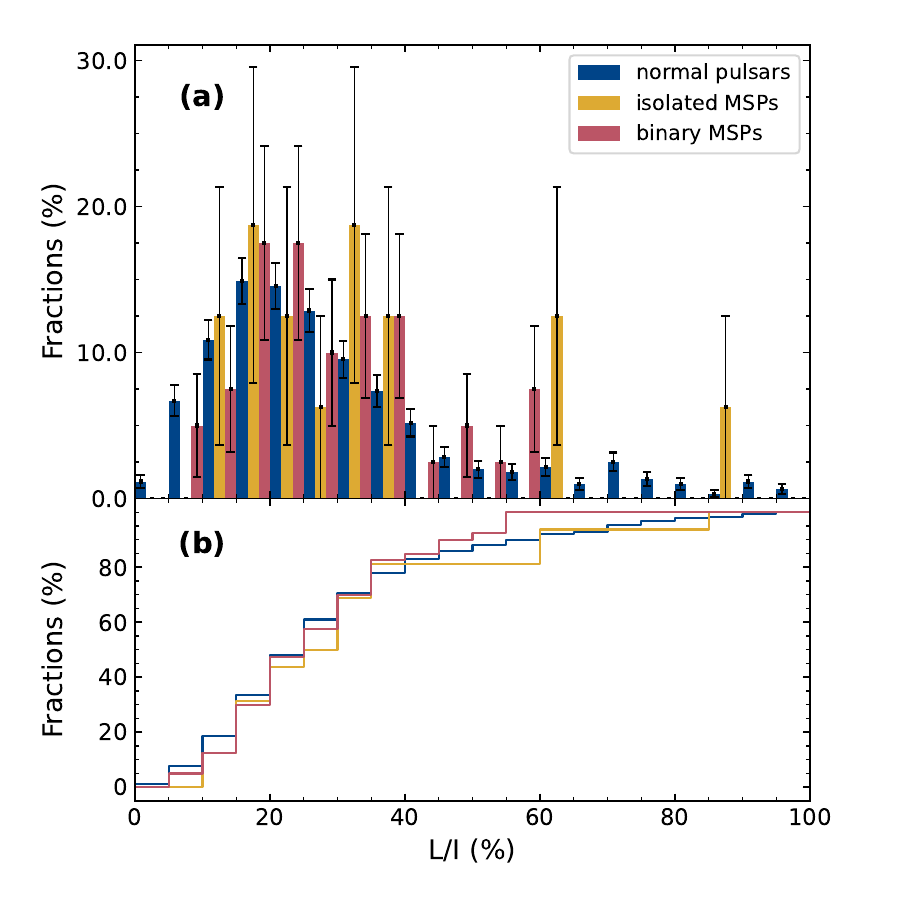}  
\end{minipage}
\begin{minipage}[t]{0.33\columnwidth}
    \centering
    \includegraphics[width=1\columnwidth]{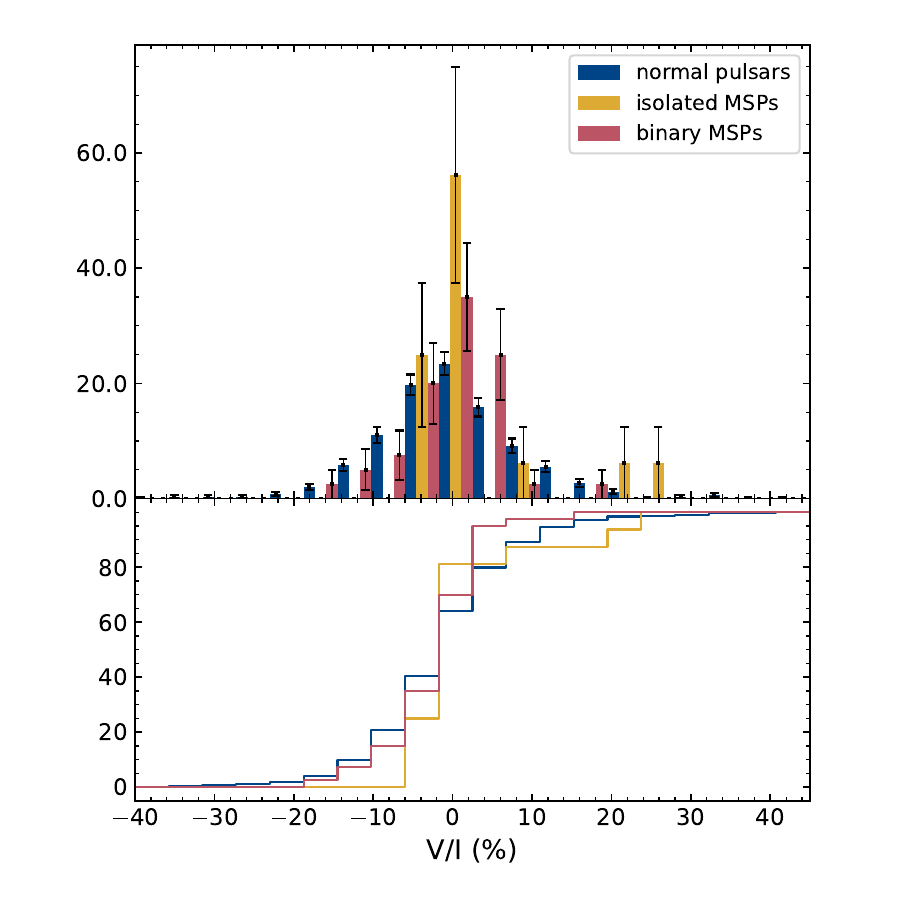}   
\end{minipage}
\begin{minipage}[t]{0.33\columnwidth}
    \centering
    \includegraphics[width=1\columnwidth]{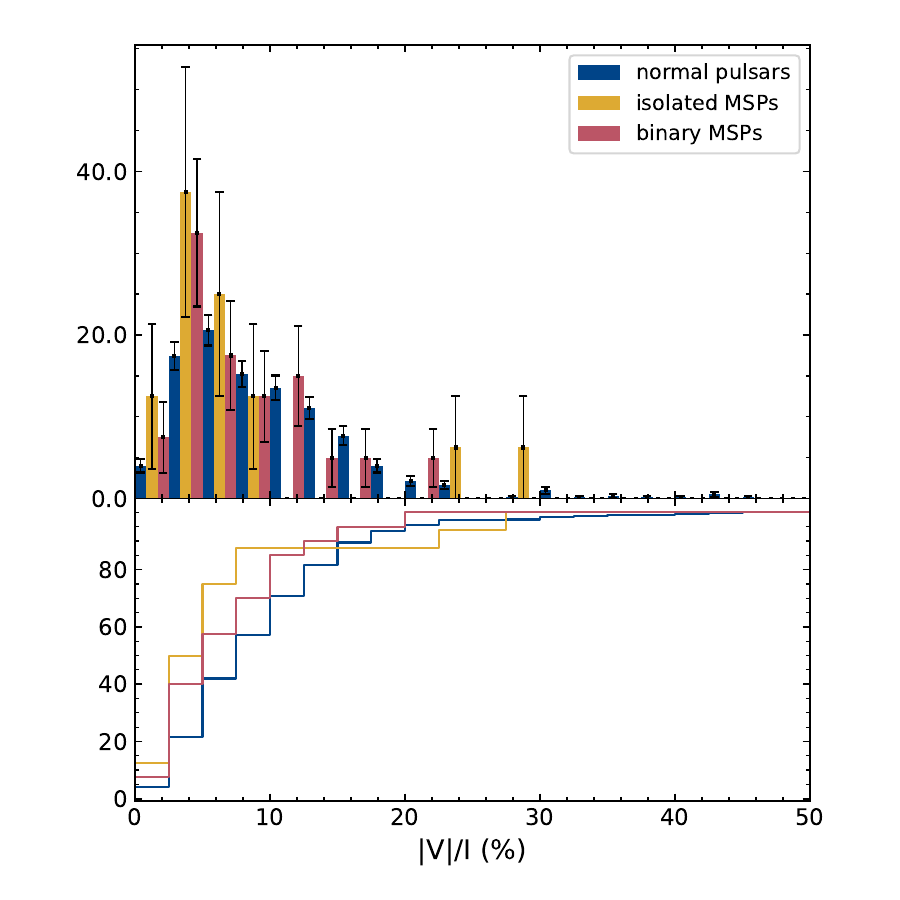}   
\end{minipage}
\caption{Distribution of polarization degree for 56 CPTA pulsars. The panels from left to right show the distribution for linear polarization, circular polarization, and the absolute value of circular polarization, respectively. The top row of panels shows the binned distribution function, while the bottom row is for the cumulative distribution function. Blue, yellow, and red colors denote the distribution for normal pulsars, isolated MSPs, and binary MSPs, respectively. Error bars are for the 68\% confidence level computed from $1/\sqrt{N}$, with $N$ being the counts per each bin. The data for normal pulsars ($P_0 > 50\,{\rm ms}$) is from \cite{WangPengfei2023}(original from \cite{2025A&A_Xu}). \label{fig:cpta_dis}}
\end{figure*}

 
Compared with normal pulsars, MSPs have similar profile complexity \cite{1998ApJ_Kramer} and spectral indexes\cite{2024MNRAS_Karastergiou}. Based on the results of CPTA and GPPS survey, no difference is found on the distribution of fractional linear and circular polarization between normal pulsars and MSPs, as presented in Fig.\ref{fig:cpta_dis}. It indicates that the recycling history has no effect on polarization properties.  Furthermore, the polarization profiles of MSPs, although the magnetosphere is several orders of magnitude smaller than normal pulsars, exhibit same characteristics such as linear polarization angle swing, inter-pulse, OPMs, and sense reversal of circular polarization\cite{1998ApJ_Xilouris,2004ApJ_Manchester}. These observational facts indicate that the radiation mechanisms are same for both populations.

\par 
Different form normal pulsars, the profile widths and separations between pulse components of MSPs remain nearly constant across frequencies, reflecting more compact emission regions\cite{1998ApJ_Kramer,2015MNRAS_Dai}. As shown in Fig.\ref{fig:cpta_pol}, the PA slopes of MSPs are much shallower than normal pulsars, although with higher duty cycles. The emission sites of MSPs may locate at relatively higher emission height, where the magnetic field is azimuthally bent due to the sweepback effect\cite{2004ApJ_Dyks}. For the CPTA sample, only five pulsars present PA curves consistent with RVM model\cite{2025A&A_Xu}, among which four pulsars are found to be aligned rotators. The sole exception is PSR J2214$+$3000 whose magnetic inclination angle and sight line angle are 31.6$^\circ$ and 30.3$^\circ$ respectively. It is similar to the case of B1929$+$10, since it also contains roughly 180$^\circ$ separated inter-pulse. 

\subsubsection{~Magnetars.}

\par
Magnetars are a subclass of isolated neutron stars characterized by their ultra-strong magnetic fields ($B \sim 10^{13} - 10^{15}\ \text{G}$). These objects sometimes exhibit extreme high-energy transient outbursts, the energy of which exceeds the rotational energy loss rate ($\dot{E}_{rot}$), it is generally concluded that these phenomena are powered by the decay of magnetic fields \cite{dt1992}. Magnetars are rare, with only 32 confirmed sources and candidates documented \cite{ok2014,cbi2021,enh2021}. Furthermore, only six have been observed to emit transient pulsed radio emission. Owing to their extreme magnetic environments and episodic energy release, magnetars have long been regarded as the most promising neutron-star candidates for powering FRBs.

\par Prior to the direct identification of a magnetar as an FRB host, the field was divided on whether the highly coherent FRB radiation originated from the compact object's magnetosphere (such as magnetar) \cite{klb2017,z2017,yz2018} or from a synchrotron maser mechanism operating in a relativistic shock far from the source \cite{l2014,mms2019,b2020}. This distinction hinges on the geometry and stability of the emission mechanism. The FAST provided critical early evidence favoring the magnetospheric model through its high-precision polarimetry. Utilizing FAST, Luo et al. (2020) observed 15 bursts from the repeating source FRB~180301 and reported the detection of various PA swings in seven of them \cite{lwm2020}. While both models can account for strong linear polarization and even non-varying PAs (as seen in some other repeaters), their predictions diverge sharply on PA variability. Magnetospheric models, which invoke ordered field lines sweeping the line of sight as the source rotates or as the magnetic configuration is reshaped, are naturally capable of producing diverse PA swing patterns depending on the specific geometry. In contrast, synchrotron maser models require the ordered magnetic field in the shock plane to satisfy the coherent condition, which fundamentally predicts a constant PA during each pulse. The observed diversity and complexity of the PA features from FRB~180301 established a strong observational foundation for the magnetospheric origin hypothesis.

\begin{figure*}
\centering
\includegraphics[width=0.8\textwidth]{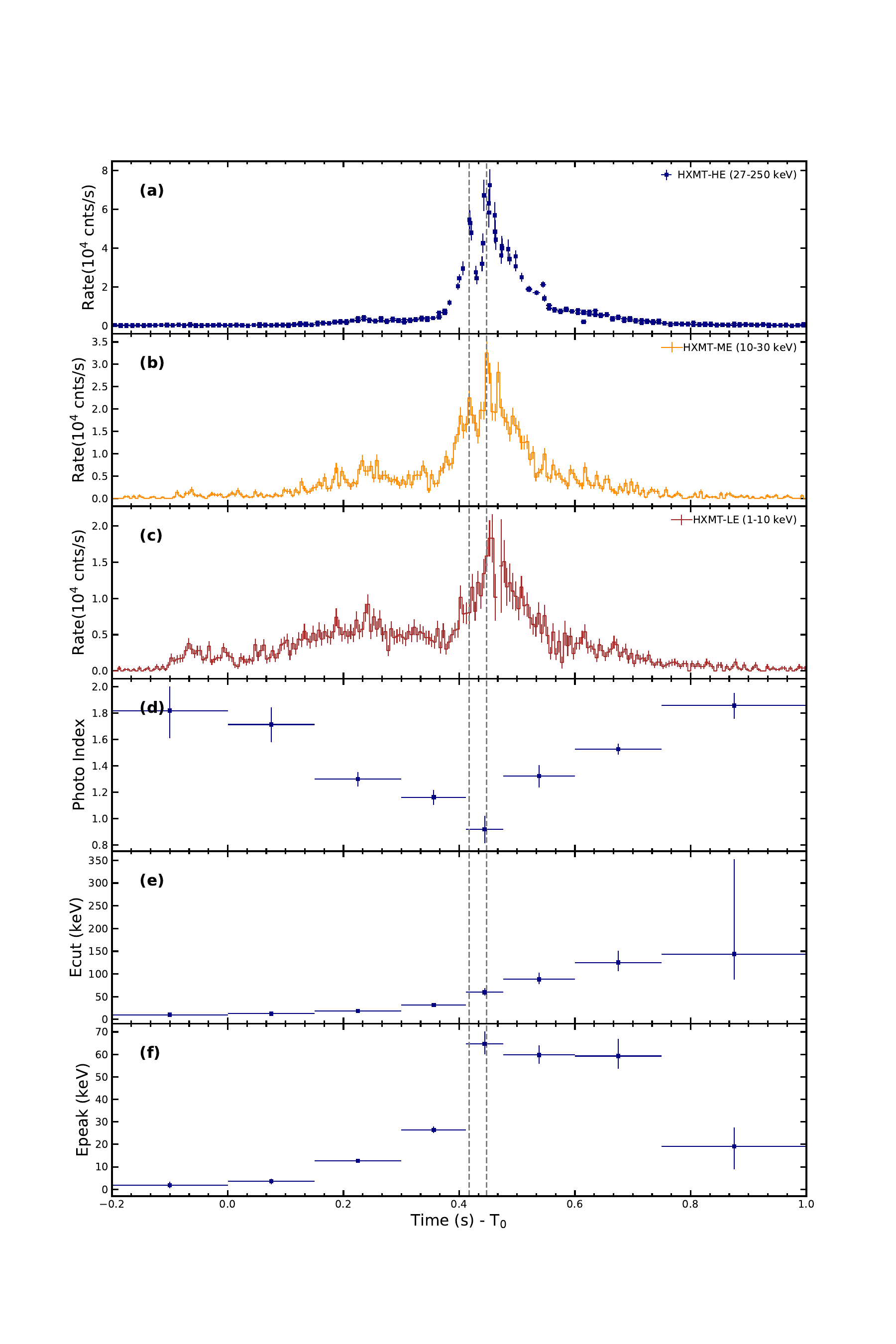}
\caption{Hard X-ray light curves and spectral evolution of the burst from SGR~J1935+2154 associated with FRB~200428, as observed by Insight-HXMT.
Panels (a)–(c) show the light curves in the HE (27–250 keV), ME (10–30 keV), and LE (1–10 keV) bands, respectively. The vertical dashed lines mark the two hard X-ray peaks, whose double-peaked structure is consistent with that of the associated radio bursts. Panels (d)–(f) present the time evolution of the photon index $\gamma$, cutoff energy $E_{\rm cut}$, and peak energy $E_{\rm peak}$ derived from CPL spectral fits (original from Li et al. (2021) \cite{llx2021}).}
\label{fig:HXMT-1935}
\end{figure*}

\par The magnetar hypothesis was observationally confirmed in 2020 with FRB~200428, based on the Insight-HXMT detection of a double-peaked non-thermal hard X-ray burst from SGR~J1935+2154, whose two-peak structure and dispersion-corrected timing were both consistent with those of the associated radio bursts \cite{Bochenek2020, CHIME2020,llx2021}. The corresponding hard X-ray light curves and spectral evolution observed by Insight-HXMT are shown in Fig.~\ref{fig:HXMT-1935}. In a contemporaneous monitoring campaign during SGR~J1935+2154's active phase, FAST performed a critical check on the universality of the FRB-burst link. Lin et al. (2020) reported on an extensive eight-hour targeted radio observational campaign, comprising four sessions, which coincided with intense activity from the magnetar \cite{lzw2020}. During the campaign, a total of 29 Soft SGR bursts were detected at high energies. Throughout the entire observing period, FAST detected no single dispersed pulsed emission coincident with the arrivals of these 29 SGR bursts. This stringent non-detection established the most rigorous upper limit on the radio fluence of typical magnetar bursts to date, with the limit being approximately eight orders of magnitude fainter than the fluence of FRB~200428. These results imply that the generation of coherent radio emission during an SGR burst requires highly specific and extreme physical conditions. These constraints strongly suggest that the FRB emission mechanism may involve either strong geometric beaming and a highly relativistic boost (leading to a narrow viewing cone), or that FRB-like emission associated with common SGR bursts possesses a narrow intrinsic frequency spectrum that typically falls outside the radio band surveyed.

\par Following FAST's continued monitoring of SGR~J1935+2154, Zhu et al. (2023) reported the emergence of a stable radio pulsar phase of the magnetar, detected in 16.5 hours of observation over 13 days, approximately five months after FRB~200428 \cite{zxz2023}. These periodic pulses were approximately eight decades fainter in luminosity than the FRB-like bursts. This study revealed a profound emission dichotomy: the stable pulses were emitted in a narrow phase window anti-aligned with the X-ray pulsation profile, while the transient FRB-like bursts appeared in random rotational phases. This finding suggested a clear separation of emission modes: stable pulses originate from a fixed region within the magnetosphere, whereas bursts occur in random locations, possibly associated with explosive events in a dynamically evolving magnetosphere. This picture successfully reconciles the observed lack of periodicity in cosmological repeating FRBs with the magnetar engine model. Wang et al. (2024) discovered a clear link between the magnetar’s X-ray and radio states, reporting the detection of X-ray spectral hardening associated with the emergence of the periodic radio pulsations \cite{wlj2024}. Their analysis revealed a double-branch association pattern between the X-ray hardness ratio and the radio activity. This evidence supports a unified physical picture where the radio emission originates from the outer magnetosphere of the magnetar, perhaps triggered by a crustal disturbance. Crucially, the surface heating due to the bombardment of inward-going particles from this outer radio emission region is proposed to be responsible for the observed X-ray spectral hardening. 

\subsubsection{~Long-period radio transients.}

\par Long-period radio transients (LPTs) are a newly recognized class of radio sources that emit pulsed radiation with periods ranging from minutes to hours. To date, a total of 15 LPTs have been reported (see Table \ref{tab:1}). In this paper, we define LPTs as sources with periods longer than 1 minute, with the longest known period reaching 6.45 hours. 
They generally exhibit steep spectra with spectral indices $< -2$. The pulse duty cycle varies widely, from as little as 0.5\% to as much as 70\%. The radio pulse profiles can be complex (e.g., GPM\,J1839$-$10 \cite{Men2025}) or remarkably stable (e.g., ASKAP\,J1755$-$2527 \cite{McSweeney2025}). 
Most LPTs show a high degree of polarization, both linear (LP) and circular (CP), with some sources reaching nearly 100\% LP or CP. Multi-wavelength observations have revealed X-ray counterparts for AR\,Scorpii, J1912$-$4410, DART/ASKAP\,J1832$-$0911, and ASKAP\,J144834$-$685644, with DART/ASKAP\,J1832$-$0911 showing pulsed X-ray emission phase-aligned with the radio pulses \cite{Wang2025}. Four LPTs, AR Scorpii, J1912$-$4410, GLEAM-X J0704$-$37, and ILT J1101+5521, are identified as white dwarf (WD)–M dwarf binary systems \cite{Marsh2016, Pelisoli2023, Hurley-Walker2024, deRuiter2025}, where the latter two are synchronized binaries while the former two are not. PSR\,J0901$-$4046 and CHIME\,J0630+25 are likely highly magnetized neutron stars ($B > 10^{14}$ G), with measured period derivatives suggesting that their radio luminosities could be lower than their spin-down power \cite{Dong2025}. In contrast, the emission luminosities of ASKAP\,J1839$-$0756, ASKAP\,J1935+2148, GPM\,J1839$-$10, and GLEAM-X\,J162752$-$35 cannot be explained by a simple spin-down-powered neutron star scenario. In summary, current observations suggest that LPTs comprise at least two distinct populations: highly magnetized neutron stars and WD–M dwarf binaries.

\par Most LPTs have inferred brightness temperatures well above $10^{12}$ K, indicating that their emission mechanisms must be coherent. Their high degrees of polarization further suggest the presence of strong magnetic fields. In particular, GPM\,J1839$-$10 shows possible signatures of Faraday conversion and cyclotron absorption, providing strong evidence for a highly magnetized environment \cite{Men2025}. Several theoretical models have been proposed to explain the radio emission of LPTs: (1) WD–M dwarf binaries, where the emission can be explained either by unipolar inductor–type magnetic interactions \cite{Qu2025, Yang2025} or by the interaction between the WD magnetosphere and the companion’s stellar wind \cite{Horvath2025}; (2) Long-period magnetars, which require a revision of the classical pulsar death line, possibly through mechanisms such as twisted magnetic fields \cite{Tong2023} or fallback disks \cite{Ronchi2022}; (3) Highly magnetized WD pulsars, where the emission is powered by WD spin-down \cite{Katz2022}. However, it has also been argued that isolated WD pulsars face significant challenges in producing pairs within a purely dipolar magnetosphere \cite{Rea2024}.

\par
\begin{figure}
    \centering
    \includegraphics[width=0.8\textwidth]{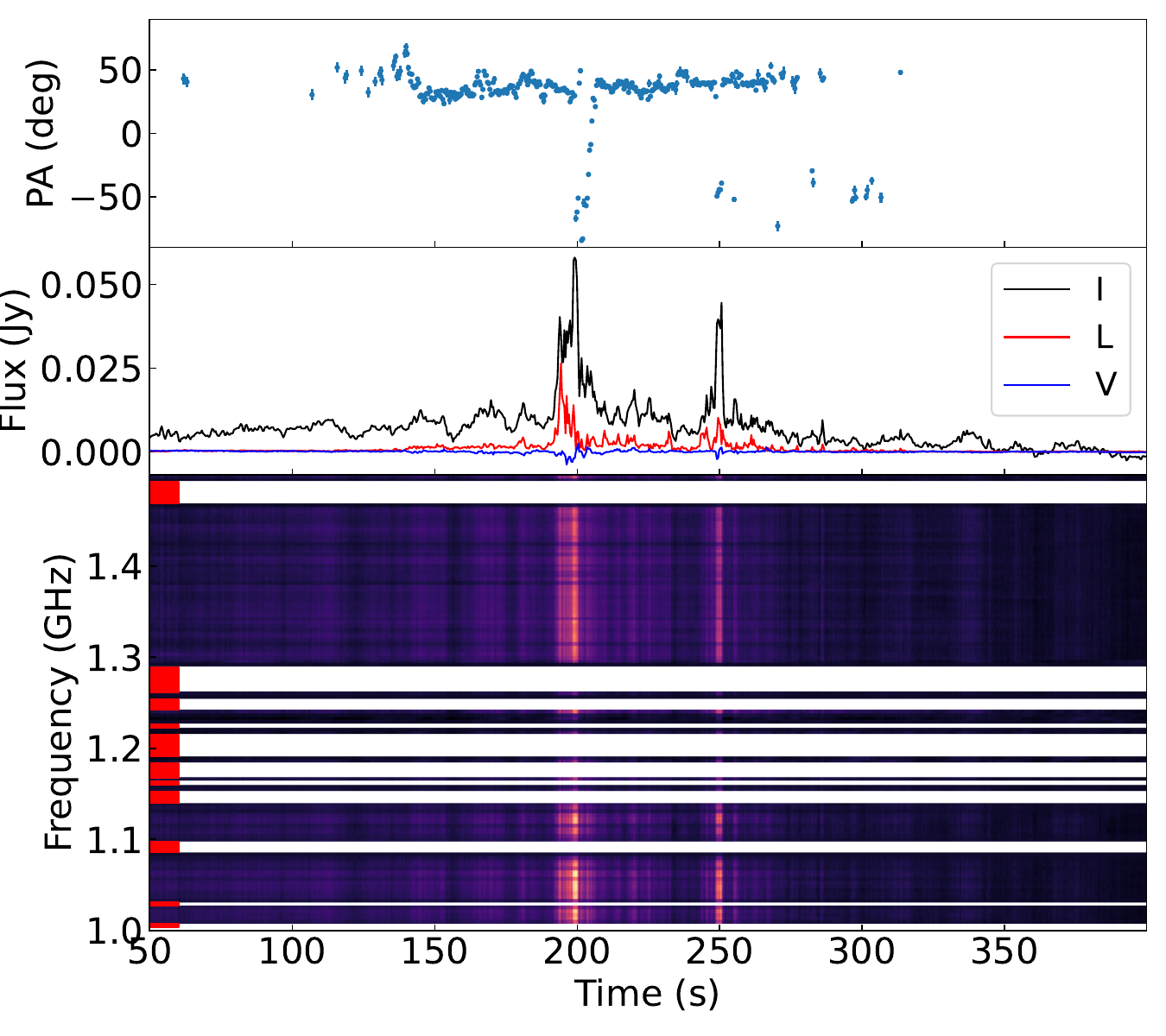}
    \caption{Polarization profiles and dynamic spectra of one burst from GPM\,1839-10. Top panel: PA of linear polarization at infinite frequency. Middle panel: Polarization pulse profile; black, red and blue curves denote total intensity, linear polarization and circular polarization, respectively. Bottom panel: Dynamic spectra of the total intensity as a function of frequency and time. The color bars denote the intensity in the unit of Jy. The observational data were obtained under the open time proposal PT2024\_0124, and a detailed analysis will be presented in a forthcoming work.}
    \label{fig:pa_swings1}
\end{figure}

 FAST is currently the most sensitive radio telescope in the world, capable of detecting much weaker bursts than other facilities. For example, we observed the LPT GPM\,1839$-$10 using the 19-beam L-band receiver at FAST under the open time proposal PT$2024\_0124$. The polarization spectrum of one burst is shown in Figure \ref{fig:pa_swings1}. FAST provides clear advantages for studying fine temporal structures and polarization properties. However, as a single-dish telescope, it is more susceptible to baseline variations compared to interferometric arrays. Nevertheless, some LPTs, such as CHIME J0630+25 and ILT/CHIME\,J1634+44, exhibit bursts that are typically narrow ($<1$ s), making them particularly suitable targets for FAST observations.

\vskip 2mm


\begin{sidewaystable}
\centering
\caption{Observational properties of the published LPTs }

\vskip 2mm \tabcolsep 4.5pt

\centerline{\footnotesize
\begin{tabular}{cccccccccccc}
\hline\hline\hline
\multicolumn{1}{c}{Source}
& \multicolumn{1}{c}{DM}
& \multicolumn{1}{c}{$P$}
& \multicolumn{1}{c}{$\dot{P}$}
& \multicolumn{1}{c}{Duty cycle}
& \multicolumn{1}{c}{$S_\nu$}
& \multicolumn{1}{c}{RM}
& \multicolumn{1}{c}{$\Pi_{L}$}
& \multicolumn{1}{c}{$\Pi_{C}$}
& \multicolumn{1}{c}{Band}
& \multicolumn{1}{c}{Type}
& \multicolumn{1}{c}{Ref.}\\

& \multicolumn{1}{c}{($\mathrm{cm^{-3}\,pc}$)}
& \multicolumn{1}{c}{(minutes)}
& \multicolumn{1}{c}{($\mathrm{s\,s^{-1}}$)}
& \multicolumn{1}{c}{(\%)}
& \multicolumn{1}{c}{(Jy)}
& \multicolumn{1}{c}{($\mathrm{rad\,m^{-2}}$)}
& \multicolumn{1}{c}{(\%)}
& \multicolumn{1}{c}{(\%)}
& 
& 
& \\
\hline
ASKAP\,J1448$-$6856 & \multicolumn{1}{c}{$<720$} & 91.2(9) & $<2.2\times10^{-8}$ & up to 70 & 0.00076-0.01248 & --- & up to 75 & up to 100 & Radio/Optical/X-ray & WD binary? & \cite{Anumarlapudi2025}\\
ASKAP\,J1755$-$2527 & \multicolumn{1}{c}{733$\pm$22} & 69.772141(3) & $-10(92)\times10^{-12}$ & $\sim$2 & $0.003-4$ & 962$\pm$1 & $\sim$60 & $\sim$40 & Radio & --- & \cite{Dobie2024, McSweeney2025}\\
CHIME\,J0630$+$25 & \multicolumn{1}{c}{22$\pm$1} & 7.0225973(13) & $52(11)\times10^{-13}$ & $0.4-0.8$ & $0.4-1.9$ & $-347.8\pm0.6$ & --- & --- & Radio & NS & \cite{Dong2025}\\
ILT/CHIME\,J1634$+$44 & \multicolumn{1}{c}{25$\pm$0.2} & 14.0207649(1) & $-9.03(11)\times10^{-12}$ & $\sim$1.2 & 0.4--9 & 6.27$\pm$0.04 & up to 100 & up to 100 & Radio/UV? & WD binary? & \cite{Bloot2025, Dong2025b}\\
DART/ASKAP\,J1832$-$0911 & \multicolumn{1}{c}{464.5$\pm$0.7} & 44.270783(16) & $<9.8\times10^{-10}$ & 5--10 & 0.03--20 & 90.5$\pm$0.1 & up to 100 & up to 64 & Radio/X-ray & --- & \cite{Li2024, Wang2025}\\
ILT\,J1101$+$5521 & \multicolumn{1}{c}{16$\pm$6} & 125.51950(4) & $3.04\times10^{-11}$ & $\sim2$ & 0.041--0.256 & 4.72$\pm$0.14 & $\sim51$ & $<1.6$ & Radio/Optical & WD+M-drawf & \cite{deRuiter2025}\\
ASKAP\,J1839$-$0756 & \multicolumn{1}{c}{188.4$\pm$5.3} & 387.0290(55) & $<1.6\times10^{-7}$ & 1.4--3.1 & 0.01--1.4 & 214$\pm$1.3 & 60--90 & 30--60 & Radio & --- & \cite{Lee2025}\\
GLEAM-X\,J0704$-$37 & \multicolumn{1}{c}{36.541$\pm$0.005} & 174.9426250(83) & $<1.3\times10^{-11}$ & 0.3--0.9 & $\sim$0.01 & $-7\pm1$ & 20 -- 50 & 10--30 & Radio/Optical & WD+M-drawf & \cite{Hurley-Walker2024}\\
ASKAP\,J1935$+$2148 & \multicolumn{1}{c}{145.8$\pm$3.5} & 53.755150(33) & $<1.2(1.5)\times10^{-10}$ & 0.01--15.5 & 0.009--0.2347 & 159.3$\pm$0.3 & up to 90 & up to 70 & Radio & --- & \cite{Caleb2024}\\
J1912$-$4410 & --- & 5.3224838(13) & --- & $<1.2$ & $\sim$0.012 & --- & --- & --- & Radio/Optical/X-ray & WD+M-drawf & \cite{Pelisoli2023}\\
GPM\,1839$-$10 & \multicolumn{1}{c}{273.5$\pm$2.5} & 21.9699283(33) & $<3.6\times10^{-13}$ & $2.3-23$ & $0.1-10$ & 531.83$\pm$0.14 & up to 100 & up to 95 & Radio/Optical? & --- & \cite{Hurley-Walker2023}\\
PSR\,J0901$-$4046 & \multicolumn{1}{c}{52$\pm$1} & 1.264759118(1) & $2.25(10)\times10^{-13}$ & 0.4 & 0.089--0.169 & $-62\pm2$ & 12.2 & 21.0 & Radio & NS & \cite{Caleb2022}\\
GLEAM-X\,J1627$-$5235 & \multicolumn{1}{c}{57$\pm$1} & 18.1861500(83) & $<1.2\times10^{-9}$ & $2.7-5.5$ & $5-40$ & $-61\pm1$ & $\sim88$ & $\sim$ 0 & Radio & --- & \cite{Hurley-Walker2022}\\
AR Sco & --- & 1.97 & $4.0(3)\times10^{-13}$ & 10 & 0.008 & --- & --- & --- & Radio/Optical/X-ray & WD+M-drawf & \cite{Marsh2016}\\
GCRT\,J1745$-$3009 & --- & 77 & --- & $\sim1$ & $\sim1$ & --- & --- & --- & Radio & --- & \cite{Hyman2005}\\
\hline\hline\hline
\end{tabular}}
\label{tab:1}
\end{sidewaystable}

\medskip
\subsection{~Pulsar Timing.}



\par 
Pulsar timing is a technique that analyses the time of arrival (ToA) of each pulse from a pulsar. This technique provides highly precise measurements of pulsar's astrophysical parameters, including spin parameters, astrometric parameters and binary parameters if the pulsar is in a binary system. The measurements can be done by using timing programs, wherein the most popular programs are \texttt{TEMPO} \cite{2015ascl.soft09002N}, \texttt{TEMPO2} \cite{2006ChJAS...6b.189H} and ``\texttt{PINT} is not TEMPO3'' \cite{2021ApJ...911...45L}. By measuring the spin parameters precisely, one can plot the spin period and its first derivative for all the radio pulsars that have such measurements, deriving $P-\dot{P}$ diagram. From $P-\dot{P}$ diagram, the characteristic age, magnetic field and spin-down luminosity can be inferred under some specific assumptions. These information for different types of pulsars can promote our understanding to the evolution of pulsars.

\par 
Timing pulsar binary systems leads to the measurements of the Keplerian parameters and post-Keplerian parameters, which are particularly useful for testing gravity theories, especially general relativity (GR). It started from the discovery of the first pulsar binary system 52 years ago \cite{1975ApJ...195L..51H}, which unlocked an ideal laboratory for gravity theories, enabling us to investigate the radiation property of the gravity for the first time. The pulsar binary system that represents one of the current frontiers in testing gravitational theories is the Double Pulsar system; it provides the most stringent test for GR in strong fields \cite{2021PhRvX..11d1050K}. In recent years, an increasing number of relativistic pulsar binary systems have been discovered, and FAST is expected to make further significant contributions to tests of gravitational theories.

\par 
Another important application of pulsar timing is pulsar timing array (PTA), which is a galaxy-scale gravitational-wave detector that exploits the exceptional rotational stability of millisecond pulsars. By regularly monitoring the pulse arrival times of a network of precisely timed pulsars, PTAs search for correlated timing deviations induced by passing low-frequency gravitational waves, typically in the nanohertz band. This technique is particularly sensitive to gravitational waves generated by supermassive black hole binaries and other processes in the early Universe. Ongoing PTA experiments, such as Chinese pulsar timing array (CPTA, \cite{2023RAA....23g5024X}) basing on FAST, provide a unique and complementary probe of gravitational wave inaccessible to ground-based and space-based gravitational-wave detectors.

\begin{figure}
    \centering
    \includegraphics[width=0.7\linewidth]{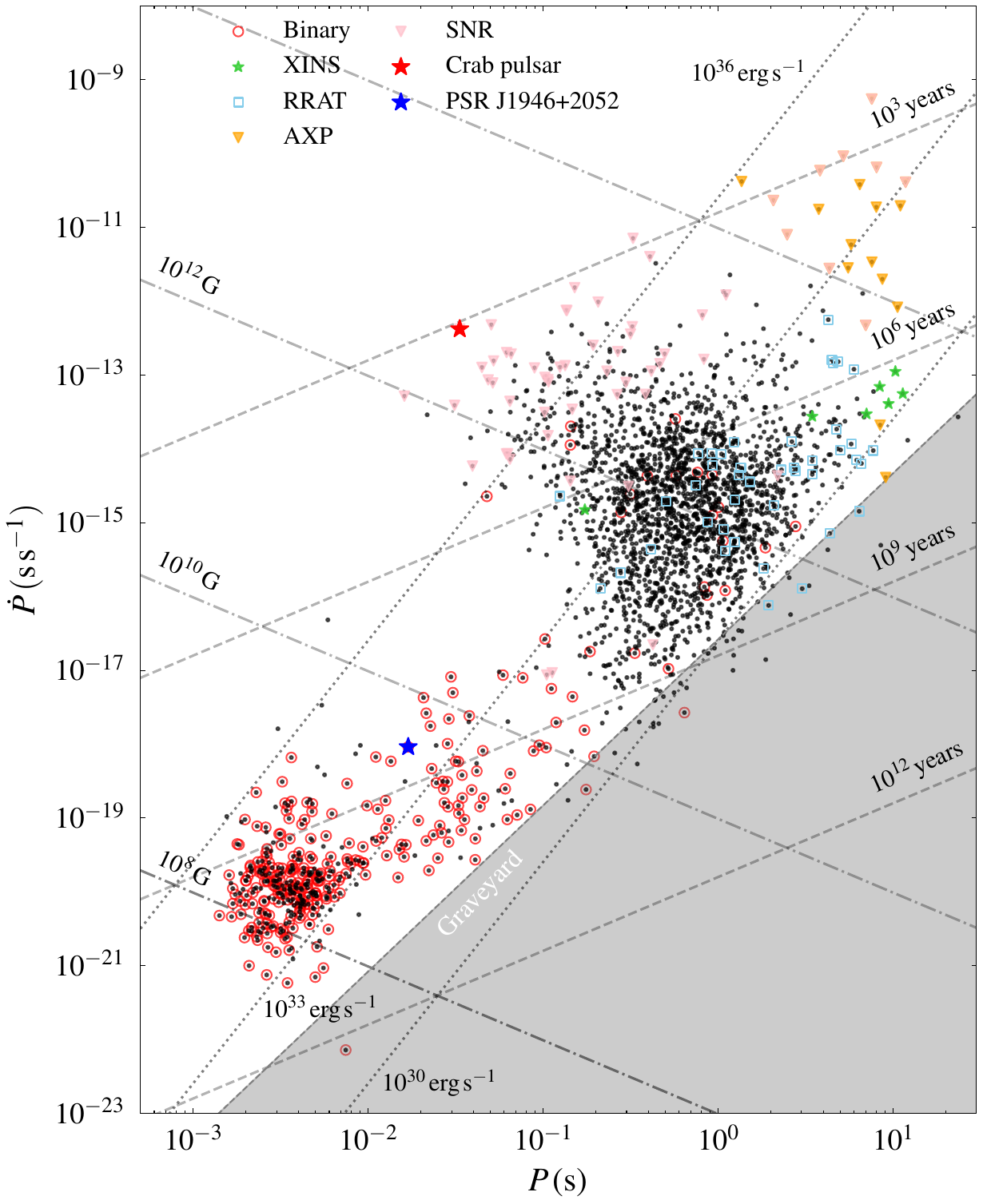}
    \caption{The $P–\dot{P}$ diagram of currently known pulsars. The horizontal axis shows the pulsar spin period, while the vertical axis represents the spin-period derivative. All pulsars are indicated by black dots. Pulsars in binary systems are highlighted by red circles, isolated neutron stars with X-ray emission by green pentagrams, RRATs by light-blue squares, pulsars with anomalous X-ray emission or soft $\gamma$-ray emission by yellow inverted triangles, and pulsars associated with supernova remnants by pink inverted triangles. The Crab pulsar is marked by a red pentagram, while PSR~J1946+2052 is indicated by a blue pentagram. Lines corresponding to constant surface magnetic field strength, spin-down luminosity, and characteristic age are also shown.}
    \label{fig:ppdotdiagram}
\end{figure}

\subsubsection{~Timing Model.}

\par 
When a pulsar signal is received at a ground-based telescope, its time of arrival (ToA) is first recorded in UTC and converted to Terrestrial Time (TT), and then to Barycentric Dynamical Time (BDT) to remove variations caused by Earth’s motion and gravitational potential, yielding the barycentric ToA ($t_{\rm SSB}$). This transformation includes a series of corrections: clock offsets ($t_{\rm corr}$), dispersive delay ($\Delta D$) from the interstellar medium, and relativistic delays such as the Rømer ($\Delta_{\rm R}$), Shapiro ($\Delta_{\rm S}$), and Einstein ($\Delta_{\rm E}$) delays from both the Solar System and the binary system, as well as aberration effects ($\Delta_{\rm AB}$); accurate evaluation of these terms relies on high-precision Solar-System ephemerides (e.g., JPL DE440), enabling ToAs from different epochs and observatories to be coherently combined. Pulsar timing then models the pulse phase evolution via a Taylor expansion of the rotation number $N(t)$ about a reference epoch $t_0$, incorporating the spin frequency and its higher-order derivatives, and determines the astrophysical timing parameters by fitting this model to observed ToAs, such that the predicted pulse numbers accurately reproduce the measurements.

\subsubsection{~Testing GR in Strong Fields.}

\par 
A binary system can be described by the Keplerian parameters, which can be directly measured through pulsar timing observations. However, for relativistic binary systems, classical Newtonian gravity is no longer sufficient to describe the orbital dynamics, and the Keplerian parameters are no longer constant but evolve with time as relativistic effects become significant. These deviations from Keplerian motion are commonly described within the Parameterized Post-Newtonian (PPN) formalism, in which relativistic effects appear as small perturbations to the Newtonian orbit \cite{2014LRR....17....4W}. For binary pulsars, such relativistic corrections can be further reformulated into a set of observable and theory-independent quantities known as post-Keplerian (PK) parameters, which encode relativistic effects in both orbital dynamics and signal propagation and can be directly measured through high-precision pulsar timing \cite{1985AIHPA..43..107D,1986AIHPA..44..263D}.

\par
Within GR, the PK parameters can be expressed as functions of the Keplerian parameters and the masses of the pulsar and its companion. Consequently, the measurement of any two PK parameters determines the component masses. If more than two PK parameters are measured, the additional constraints can be used to test the consistency of GR in strong fields. Among the many relativistic binary pulsars discovered to date, a few representative systems have been particularly influential in testing GR. 
One is the Hulse-Taylor pulsar PSR~B1913+16 \cite{1975ApJ...195L..51H}, which provided the first evidence of the existence of gravitational waves \cite{1989ApJ...345..434T}.
Another is the Double Pulsar system PSR~J0373-3039A/B where the most stringent test for GR in the strong field regime is conducted \cite{2021PhRvX..11d1050K}.
In the following, we highlight a relativistic binary pulsar that has been extensively studied with FAST observations.

\subsubsection{~The Most Compact DNS: PSR J1946+2052}

\leavevmode\par 
PSR~J1946+2052 was discovered by Arecibo telescope in 2017 \cite{2018ApJ...854L..22S}, with the shortest orbital period (1.88\,hours) and spin period (17\,ms) among all DNS discovered by now in the Galaxy. The coalescence timescale of PSR~J1946+2052 is only 46\,Myr, which is almost halfway through the merger of the Double Pulsar (85\,Myr). Furthermore, only one PK parameter was detected in their work, which is the periastron advance ($\dot{\omega}=25.6^\circ/\rm yr$, the largest among all the discovered DNS). In the Arecibo observations, there's no detection of polarization and profile evolution in this pulsar. While FAST has started observing PSR~J1946+2052 since 2019, and detected polarization and significant profile evolution for the first time. These new information can be used to determine the entire geometry of the binary system by assuming GR and the relativistic spin precession \cite{2024ApJ...966...46M}. 

\par
\begin{figure}
    \centering
    \includegraphics[width=\linewidth]{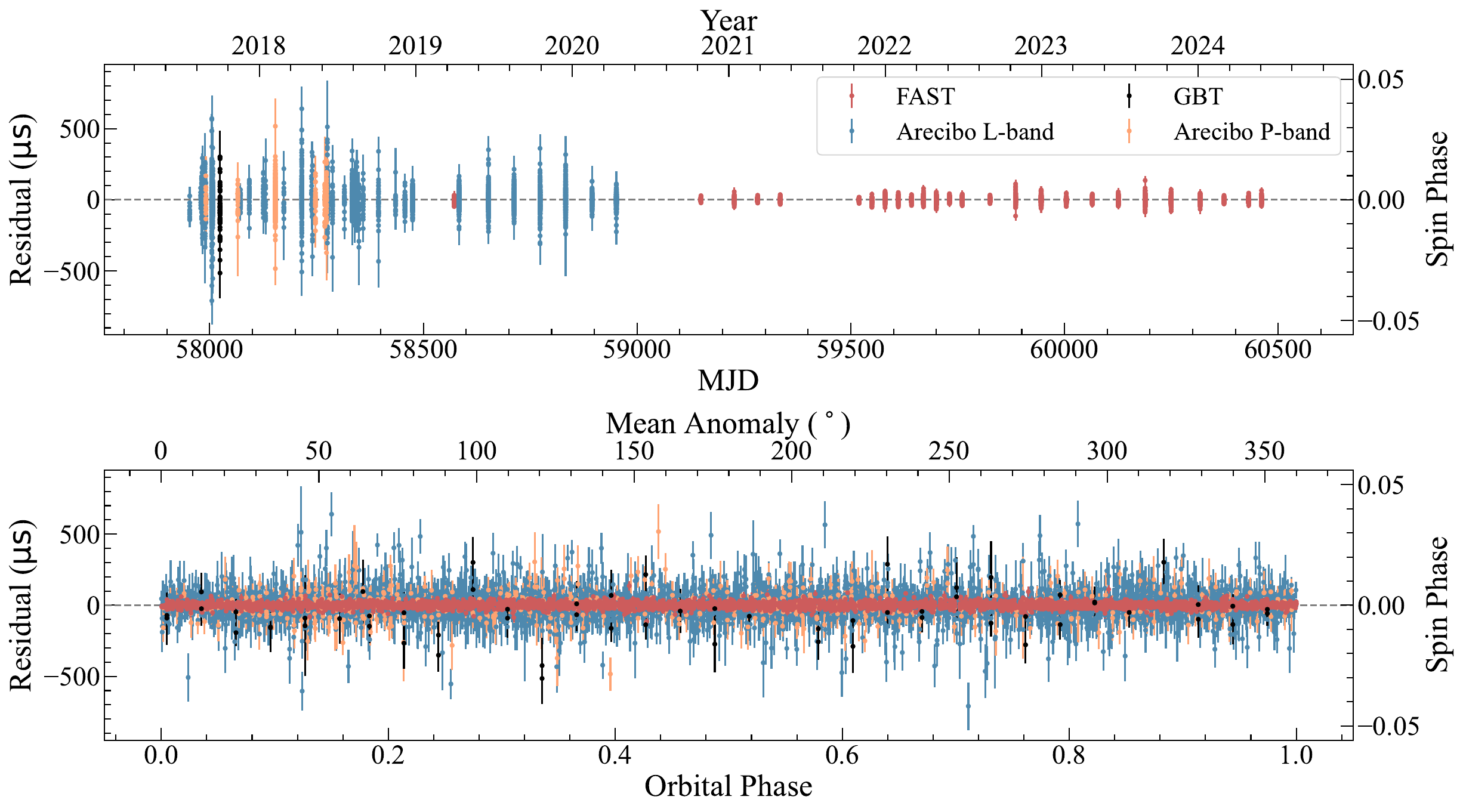}
    \caption{The timing residuals of PSR~J1946+2052. The blue, orange, green, and red residuals correspond to the L-band/PUPPI data, P-band/PUPPI data, a single GBT observation, and FAST data, respectively. The upper panel shows the timing residuals as a function of time, while the lower panel presents the residuals as a function of orbital phase. The root-mean-square (RMS) of the post-fit residuals is consistent with the ToA uncertainties, with RMS values of 66.088 $\mu s$, 87.709 $\mu s$, 138.936 $\mu s$, and 13.581 $\mu s$ for the Arecibo L-band, Arecibo P-band, GBT, and FAST data sets, respectively. (original from \cite{Meng2025})}
    \label{fig:timingresidual}
\end{figure}

\begin{figure}
    \centering
    \includegraphics[width=0.8\linewidth]{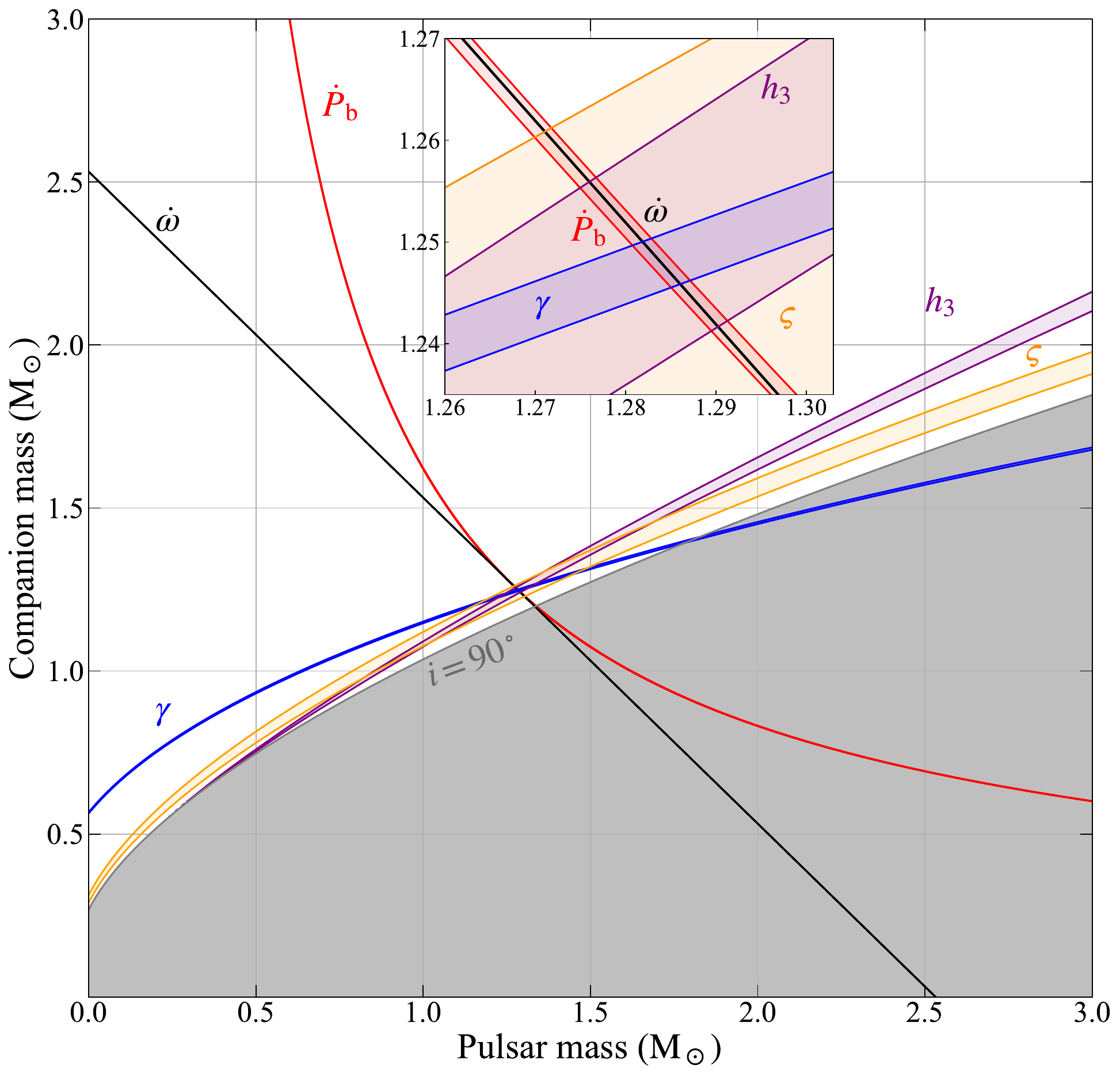}
    \caption{Mass-mass diagram for PSR~J1946+2052. The regions consistent with the measured orbital decay $\dot{P}_{\rm b},\ {\rm periastron\ advance}\ \dot{\omega},\ {\rm Einstein\ delay}\ \gamma,\ {{\rm Shapiro\ delay\ parameters}}\ \ h_3\ {\rm and}\  \varsigma$, and their 1-$\sigma$ uncertainties are displayed in red, green, blue, purple, and orange. Note that $\dot{P}_{\rm b}$ in this diagram has been corrected by removing the external contributions. The grey area is excluded by the mass function and $\sin i \leq 1$.
    The inset is an expanded view, showing in more detail the intersection of all PK parameters and the self-consistency of GR. 
    The $\dot\omega$-line is based on the leading-order.}
    \label{fig:mmdiagram}
\end{figure}

 
After more than 5 years' timing observations with FAST, combined with the data from Arecibo and Green Bank telescope (GBT) since the discovery of PSR~J1946+2052, the measurements of timing parameters has been improved significantly, and can fully describe the ToAs of this pulsar, with no other trend in the timing residual, as shown in Figure~\ref{fig:timingresidual}. The timing results render the precise measurements of 5 PK parameters. By plotting each PK parameter in the mass-mass diagram, one can determine if GR has passed the test through the intersection of the PK parameter curves. The mass-mass diagram of PSR~J1946+2052 is displayed in Figure~\ref{fig:mmdiagram}, where the five bands that represent five PK parameters meet at a small common area, indicating GR has passed the test in PSR~J1946+2052. The measured 5 PK parameters enable 3 independent tests for GR, including the second most precise test for GR's quadrupole formula of the gravitational wave emission: the intrinsic orbital decay can represent 1.00005(91) of the prediction of GR, surpassed the test in the Hulse-Taylor pulsar. In addition, the higher order contributions in $\dot{\omega}$ are already larger than the current uncertainty of $\dot{\omega}$, which leads to the correction of the total mass and shows the potential of measuring the moment of inertia in the future. Note that the present analysis only included five years of FAST observations, resulting in a total timing baseline of approximately seven years. Continued FAST observations in the coming years are therefore expected to substantially improve our understanding of PSR~J1946+2052 \cite{Meng2025}.

\section{~Fast Radio Bursts.}
\par
\begin{figure}[!thp]
    \centering
    \includegraphics[width=0.4\textwidth]{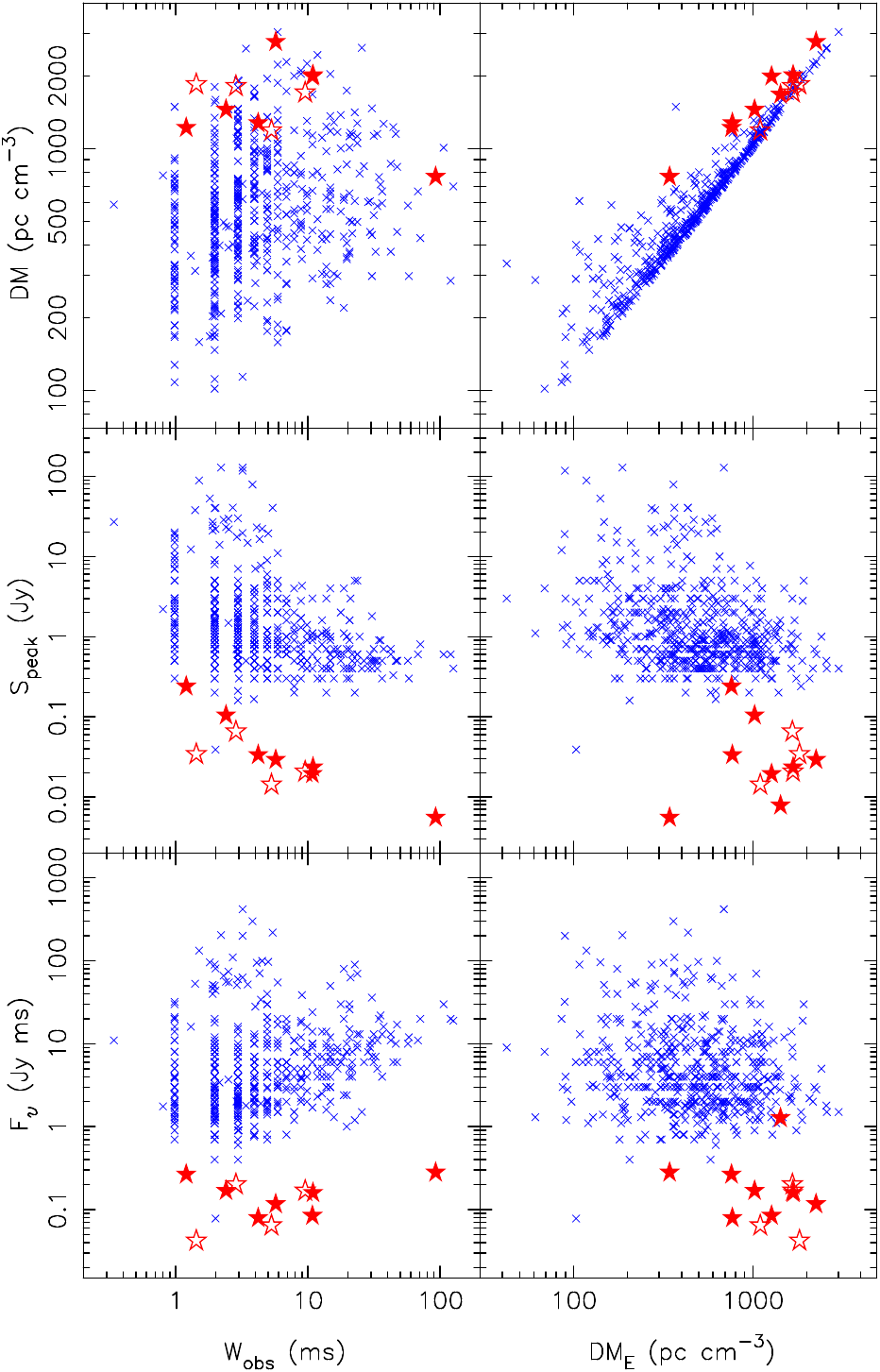}
    \caption{ Basic parameters for 8 one-off FRBs discovered by FAST in the GPPS survey (solid star) \cite{ZhouDJ2023MNRAS.526.2657Z,HanJL2025RAA....25a4001H} and 4 by the CRAFTS (hollow star)~\cite{Zhu2020,Niu2021} and for other known one-off events (cross) detected above 400 MHz, mostly by CHIME \cite{CHIME2021ApJS}. The $DM_{\rm E}$ is estimated by using the YMW16 model~\cite{YMW2016}.}
    \label{fig:sfdm-frb-fast}
\end{figure}

 
Five new FRB sources has been reported during the CRAFTS project. The sample comprises four apparently one-off (non-repeating) FRBs and one repeater, FRB 20190520B \cite{Zhu2020,Niu2021,NiuCH2022}. And eight one-off FRBs the FAST GPPS survey \cite{ZhouDJ2023MNRAS.526.2657Z,HanJL2025RAA....25a4001H}. As show the distributions of burst width, DM, fluence and flux of FAST one-off events Compared those of CHIME in Figure \ref{fig:sfdm-frb-fast}, there is no significant difference of the distributions between these sources, but all these FAST FRBs show lower flux densities and higher DMs, extending the DM-fluence relation into a more distant and fainter region. Identifying such low-luminosity events is pivotal for constraining the low-luminosity end of the FRB luminosity function.

Among the remarkable achievements of the CRAFTS, FRB 20190520B shows a repeating fast radio burst that has unveiled a wealth of intriguing properties distinguishing it from all other known sources.
High-precision localization placed the source within a dwarf galaxy at a redshit of 0.233, and it has an exceptionally high DM contribution from the host galaxy, far exceeding that of other known FRBs \cite{NiuCH2022}.
The source is currently only persistently active repeating FRB known to exhibit continuous activity across multi-year timescales, in stark contrast to most FRBs that are detected only once or display significant quiescent windows.
It provides an invaluable natural laboratory for studying FRB evolution, with long-term monitoring revealing a steady decline in DM at a rate of approximately $12\,\rm pc\,cm^{-3}$ per year, orders of magnitude higher than typical Galactic pulsars, suggesting that FRB 20190520B originates from a young magnetar embedded within an expanding supernova remnant only about 10 to 100 years old \cite{niu2026}.

\subsection{~Morphology.} 

\par The burst morphology of FRBs as revealed in the time–frequency domain, has emerged as a key observational window into their radiative processes. Statistical studies based on large samples, most notably the CHIME/FRB Catalog 1 \cite{CHIME2021ApJS}, have established that repeating FRBs commonly exhibit complex burst structures, including multiple sub-bursts, narrow-band emission, wide burst duration, and systematic frequency drifts, in contrast to the generally simpler morphologies observed in most apparent non-repeaters \cite{PleunisZ2021ApJ...923....1P}. These results suggest that burst morphology is closely linked to the underlying emission mechanism rather than being dominated by propagation effects.

\par
\begin{figure*}[!thp]
\centering
\includegraphics[width=0.3\columnwidth]{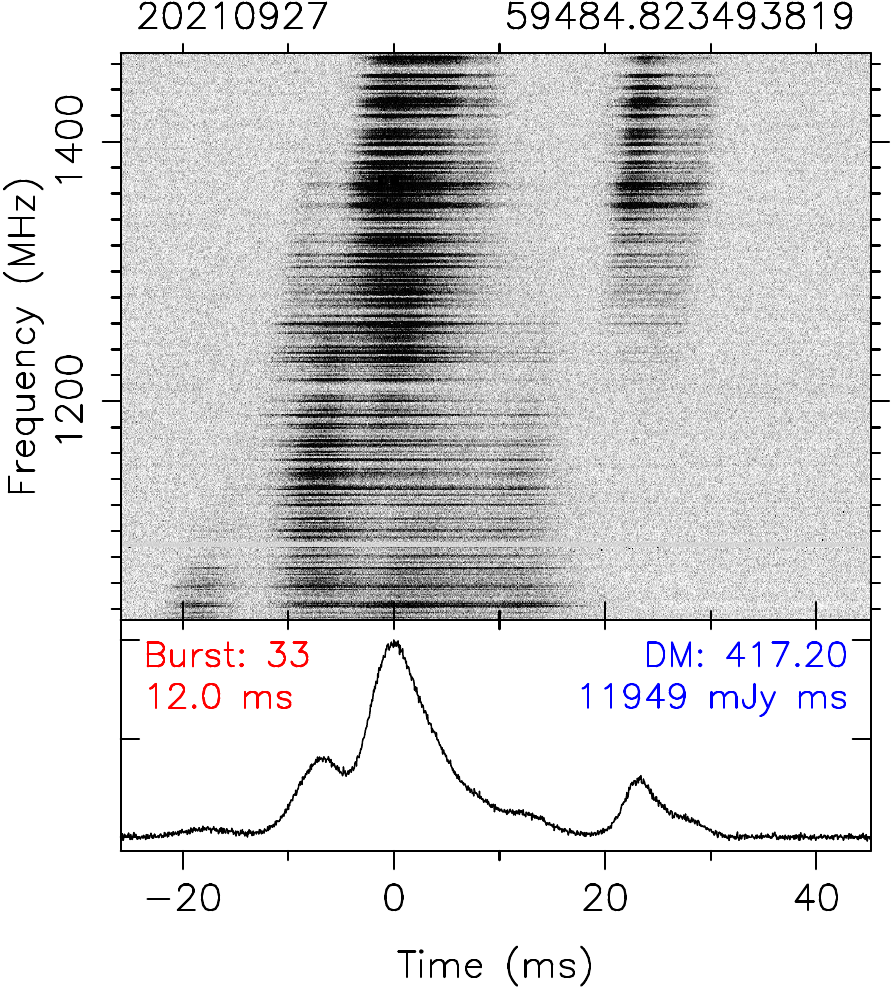}
\includegraphics[width=0.3\columnwidth]{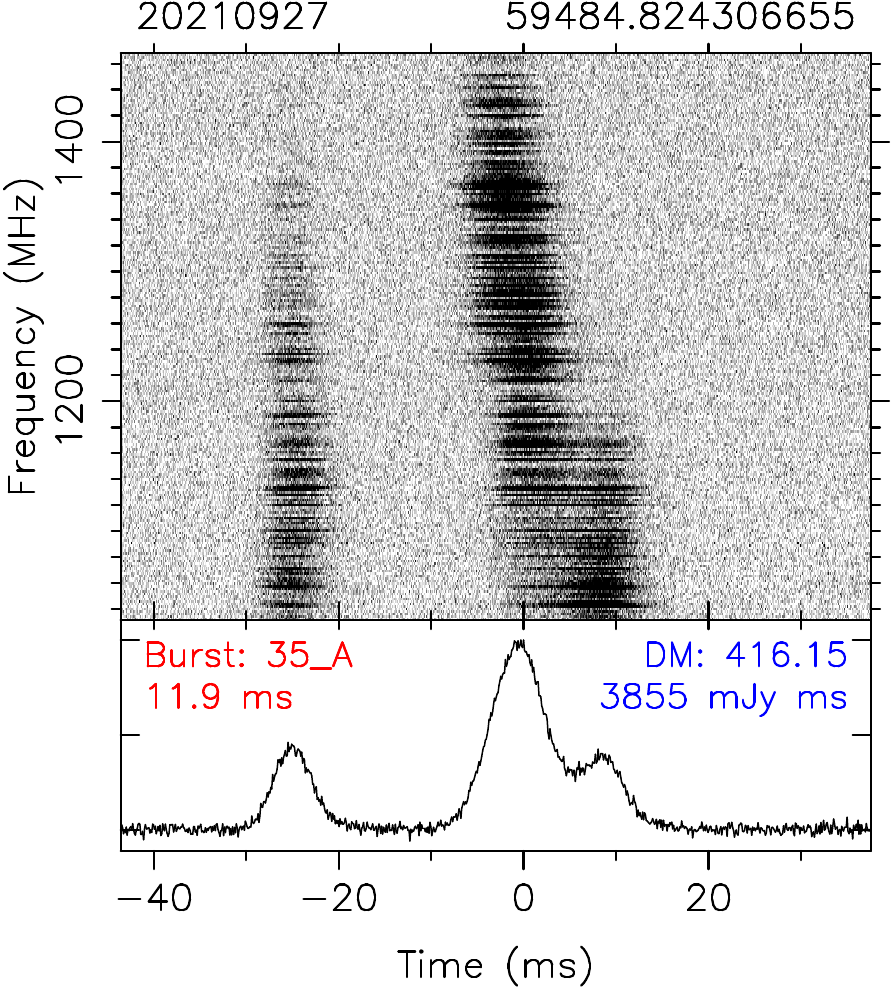}
\includegraphics[width=0.3\columnwidth]{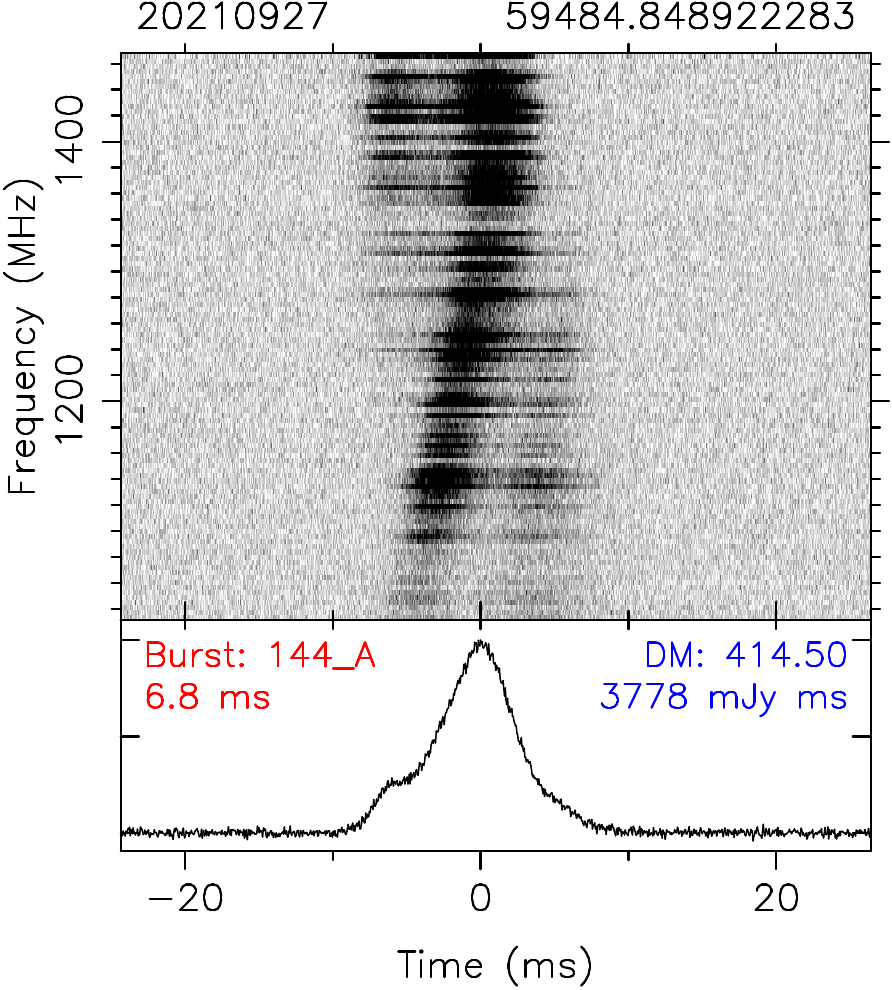}\\
\includegraphics[width=0.3\columnwidth]{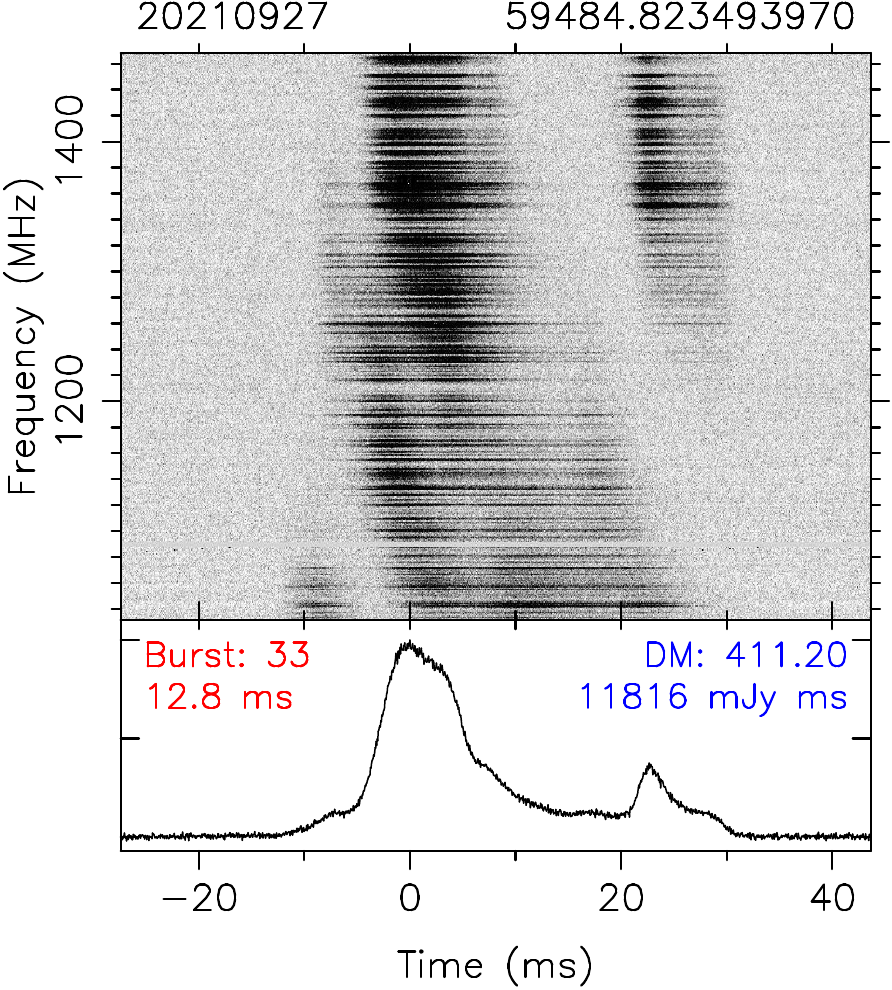}
\includegraphics[width=0.3\columnwidth]{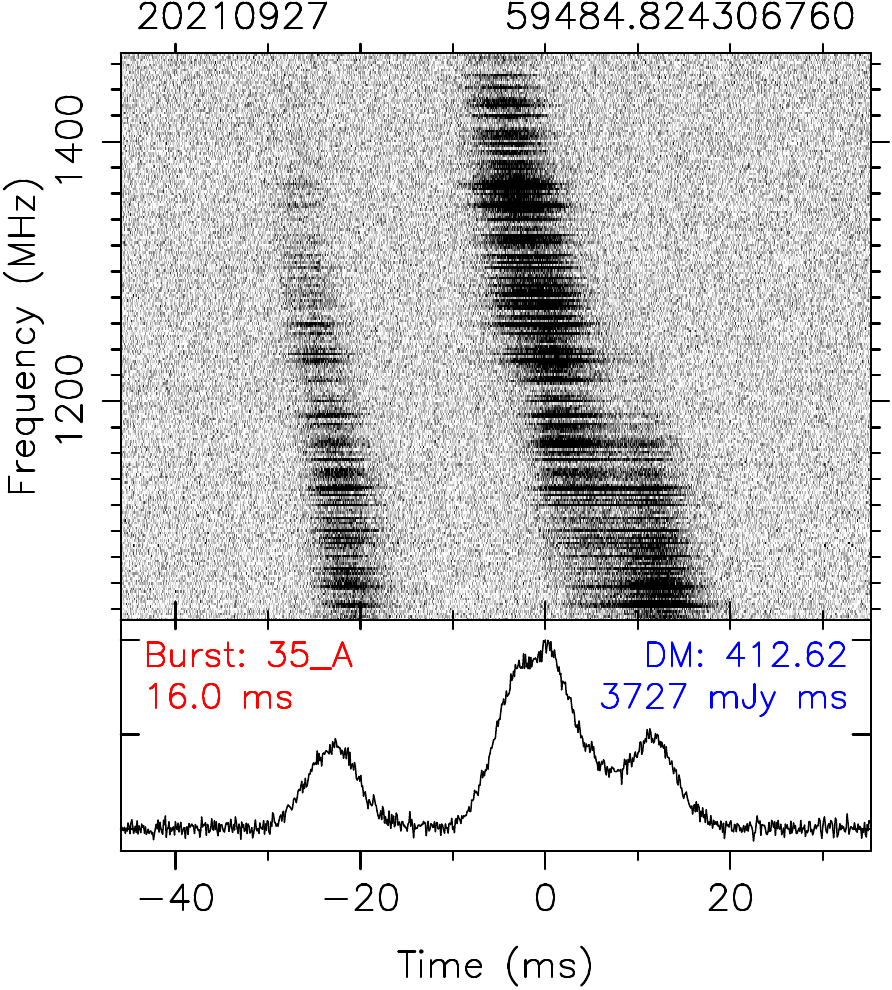} 
\includegraphics[width=0.3\columnwidth]{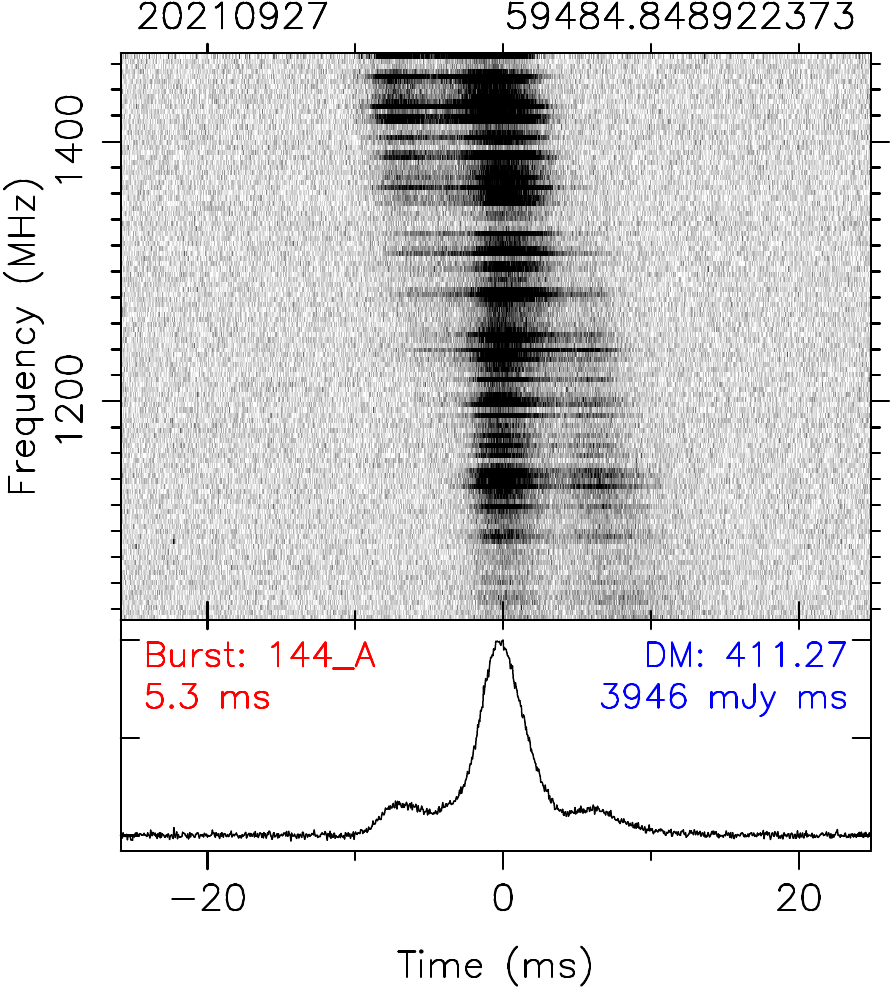}\\
\includegraphics[width=0.9\columnwidth]{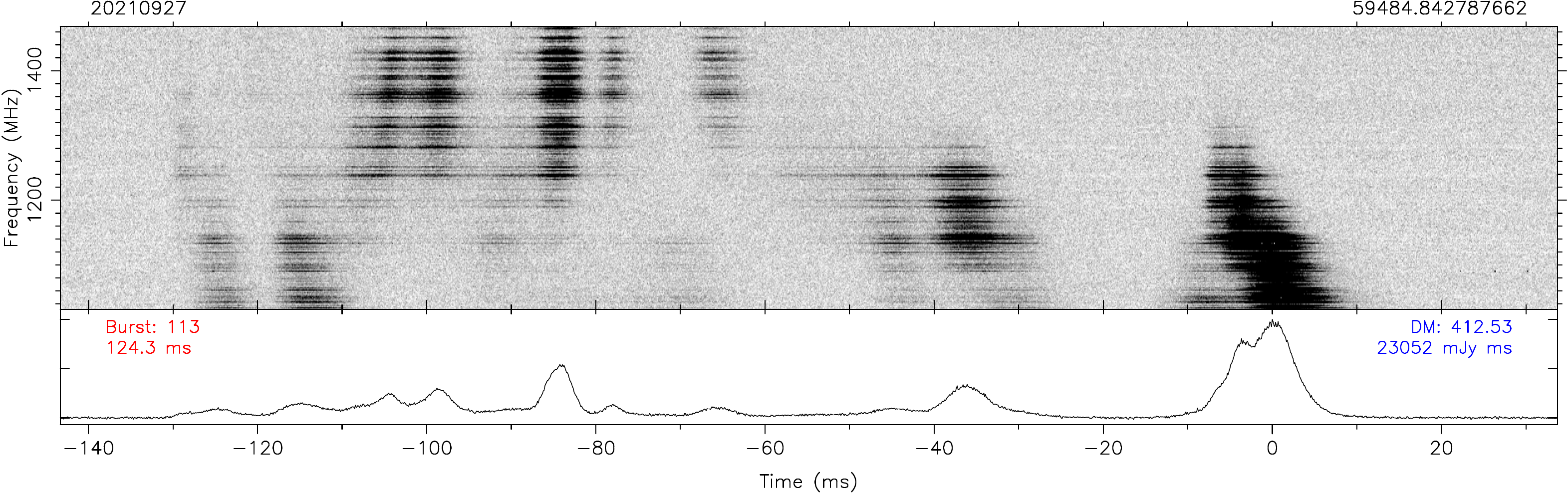}
  \caption{The dynamic spectra of four bursts illuminating the difficulty to determine DMs. In the upper sub-panel of each plot is the frequency-time water-fall plot, and the average burst profile over all frequency channels is shown in the bottom sub-panel. The observational date and the TOA in MJD are marked on the top of each plot. The burst number on that day and burst width in ms, together with the DM value used and the total fluence of this burst are marked in the lower-sub-panel. In the top three panels, the DM values of bursts No.~33 and No.~35$\_$A and No.~144$\_$A on 20210927 are determined from burst emission at the upper half band of the FAST observations, and the middle panels are the same data but with DMs determined from the emission at the lower half band. The wide panel in the bottom shows a complex, No.~113 on 20210927, which have many sub-bursts so that it is difficult to obtain one DM to align with all the burst emission. We then have to compromise by using the averaged DM of the day to generate this plot (original from \cite{ZhouDJ2022RAA....22l4001Z}).
}
\label{fig:difmorphology}
\end{figure*}

 High-sensitivity observations with FAST provide a crucial complementary perspective for repeaters by resolving fine morphological details that are often inaccessible in survey-mode observations. FAST observations of the extremely active episode of FRB 20201124A have enabled one of the most comprehensive morphological studies to date, based on more than 600 detected bursts within a few days. The bursts show a remarkable diversity in their dynamic spectra, including single- and multi-component structures, clustered emission episodes, and predominantly narrow-band radiation within the 1.0–1.5 GHz band (shown as in Figure \ref{fig:difmorphology}). A key result is the prevalence of systematic downward frequency drifting, observed not only in multi-component bursts but also in apparently single-component events when examined with sufficient signal-to-noise ratio. Such drifting behavior, together with characteristic emission bandwidths of a few hundred MHz and sub-burst durations of several milliseconds, appears to be an intrinsic property of the source rather than a consequence of interstellar scattering.

\par
\begin{figure*}[!thp]
  \centering
  \includegraphics[width=0.9\textwidth]{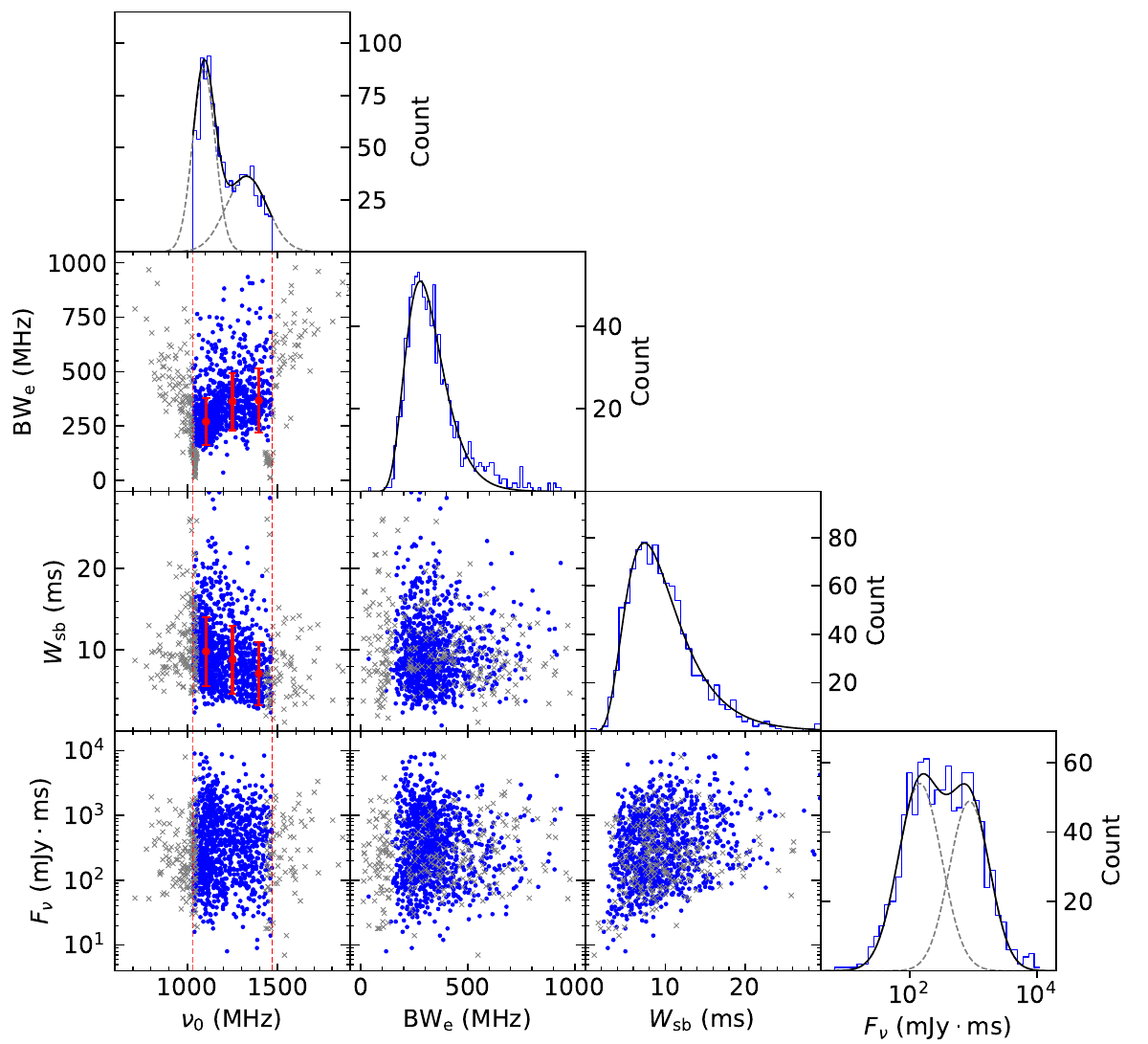}
  \caption{Data distributions of emission peak frequency ($\nu_{\rm 0}$), burst emission bandwidth ($\rm BW_{e}$), sub-burst width $W_{\rm sb}$ and specific fluence $F_{\nu}$ of all sub-bursts of FAST detected bursts. The values are estimated via Gaussian fittings. The histogram of $\nu_{\rm 0}$ distributions shows two peaks, which are fitted with two Gaussian functions peaking at 1901.9~MHz and at 1327.9~MHz. The histogram of $\rm BW_{e}$ and $W_{\rm sb}$ distributions can be fitted by a log-normal function, and they peak at 277~MHz and 7.4~ms, respectively. The histogram of the log~$F_{\nu}$ distribution has a dip near the peak but can be fitted with two Gaussian functions with peaks at log~${F_\nu= 2.1\pm0.3}$ and log~${F_\nu= 2.9\pm0.4}$, respectively. It is noticed that the sub-burst width $W_{\rm sb}$ tends to be larger for the bursts emerging at the lower part of the band, while the emission band widths  $\rm BW_{e}$ of these bursts tend to be smaller than the bursts emerging at high frequency part of the band, as illustrated by the median and the standard deviation for the three sub-bands in the first column of data distribution of $\nu_{\rm 0}$ vs $\rm BW_{e}$ and distribution of $\nu_{\rm 0}$ vs $W_{\rm sb}$ (original from \cite{ZhouDJ2022RAA....22l4001Z}). }
\label{fig:FRB_freq_w_flux}
\end{figure*}

 Hundreds of bursts allowing robust statistical measurements of several key morphological parameters (see Figure \ref{fig:FRB_freq_w_flux}). The emission peak frequency distribution within the FAST L-band shows two preferred frequency ranges, indicating that burst emission is not uniformly distributed across the observing band. Individual sub-bursts are typically narrow-band, with characteristic emission bandwidths of a few hundred megahertz, substantially smaller than the total instrumental bandwidth. Temporally, the sub-burst durations cluster around several milliseconds, with longer durations preferentially associated with lower emission frequencies. The burst fluence distribution spans more than an order of magnitude and exhibits a non-trivial structure, reflecting significant diversity in burst energetics even within a single active episode of one source.

\par Based on their time–frequency characteristics, the bursts from FRB 20201124A can be classified into multiple morphological categories, including downward-drifting, upward-drifting, non-drifting, and complex bursts with mixed behaviors. This systematic taxonomy demonstrates that FRB emission is highly structured on millisecond timescales and likely involves rapidly evolving emission regions or radiation conditions. Similar morphological features have since been reported in other repeating FRBs, suggesting that such complexity is a generic aspect of the coherent radio emission mechanism. In this context, FAST plays a crucial role by bridging sensitivity and time–frequency resolution, allowing weak substructures to be resolved and enabling morphology to be used as a key diagnostic of FRB radiative processes.

\subsection{~Polarization.} 
\par 
FAST has played a pivotal role in advancing our knowledge of FRB polarization, revealing unprecedented details about the magnetic environments and emission mechanisms of these enigmatic cosmic signals. Through its exceptional sensitivity, FAST has enabled detailed polarimetric studies of numerous FRBs, with its most significant contributions centering on the intricate and often surprising polarization characteristics of repeating sources.
\par 
The polarization of FRBs is crucial to understanding the emission mechanism. Normally, one can measure the polarization properties of burst profile and possible Faraday Rotation caused by magnetized environment. To calibrate the polarization, the prerequisite is to obtain a sample of FRBs with significant signal-to-noise ratio. Thus, FAST would be one of the best facilities to study FRB polarization in the world. 

\par
Polarimetry offers a vector-based probe of the emission geometry and the magneto-ionic material along the line of sight. Unlike total intensity (Stokes $I$), which traces the energetics of the burst, the polarization state—defined by the linear polarization (Stokes $Q, U$) and circular polarization (Stokes $V$)—encodes the orientation of magnetic fields at the source, the turbulence of the circum-burst medium, and the relativistic electrodynamics of the emitting plasma. The detection of complex polarization phenomena, including rapid PA swings\cite{Luo2020Natur}, extreme Rotation Measures (RM)\cite{liye2025}, and unexpectedly high fractions of circular polarization\cite{jiang2025ninety}, has forced a re-evaluation of theoretical models. The simplistic view of FRBs as generic shocks is being replaced by nuanced models involving magnetospheric coherent emission, relativistic winds, and binary interactions.

\par
The polarization state of electromagnetic radiation is fully described by the four Stokes parameters: $I$(Total intensity); $Q$ and $U$(Linear polarization intensity), defining the orientation of the electric vector position angle; $V$(Circular polarization intensity), representing the degree of ellipticity. The total polarized intensity is $P = \sqrt{Q^2 + U^2 + V^2}$, and the degree of polarization is $\Pi = P/I$. The linear polarization fraction is $L/I = \sqrt{Q^2 + U^2}/I$. Because $L = \sqrt{Q^2 + U^2}$ is a positive-definite quantity, any noise in the Stokes $Q$ and $U$ parameters will always add positively to $L$, causing the measured value ($L_{\text{meas}}$) to systematically overestimate the true linear polarization ($L_{\text{true}}$) \cite{everett2001}. $$L_{\text{true}} = \begin{cases} \sqrt{L_{\text{meas}}^2 - \sigma_{\text{noise}}^2} & \text{if } L_{\text{meas}} > \sigma_{\text{noise}} \\ 0 & \text{otherwise} \end{cases}$$
$L_{\text{meas}} = \sqrt{Q^2 + U^2}$ is the measured linear polarization intensity at a specific bin.$\sigma_{\text{noise}}$ is the root-mean-square (RMS) noise level of the Stokes parameters.
The PA, $\Psi$, is derived as:$$\Psi = \frac{1}{2} \arctan\left(\frac{U}{Q}\right)$$
In the context of FRBs, the evolution of $\Psi$ and $V$ as a function of time ($t$) and frequency ($\nu$) provides the primary constraints on the emission mechanism. 

\par
\begin{figure}
    \centering
    \includegraphics[width=0.8\textwidth]{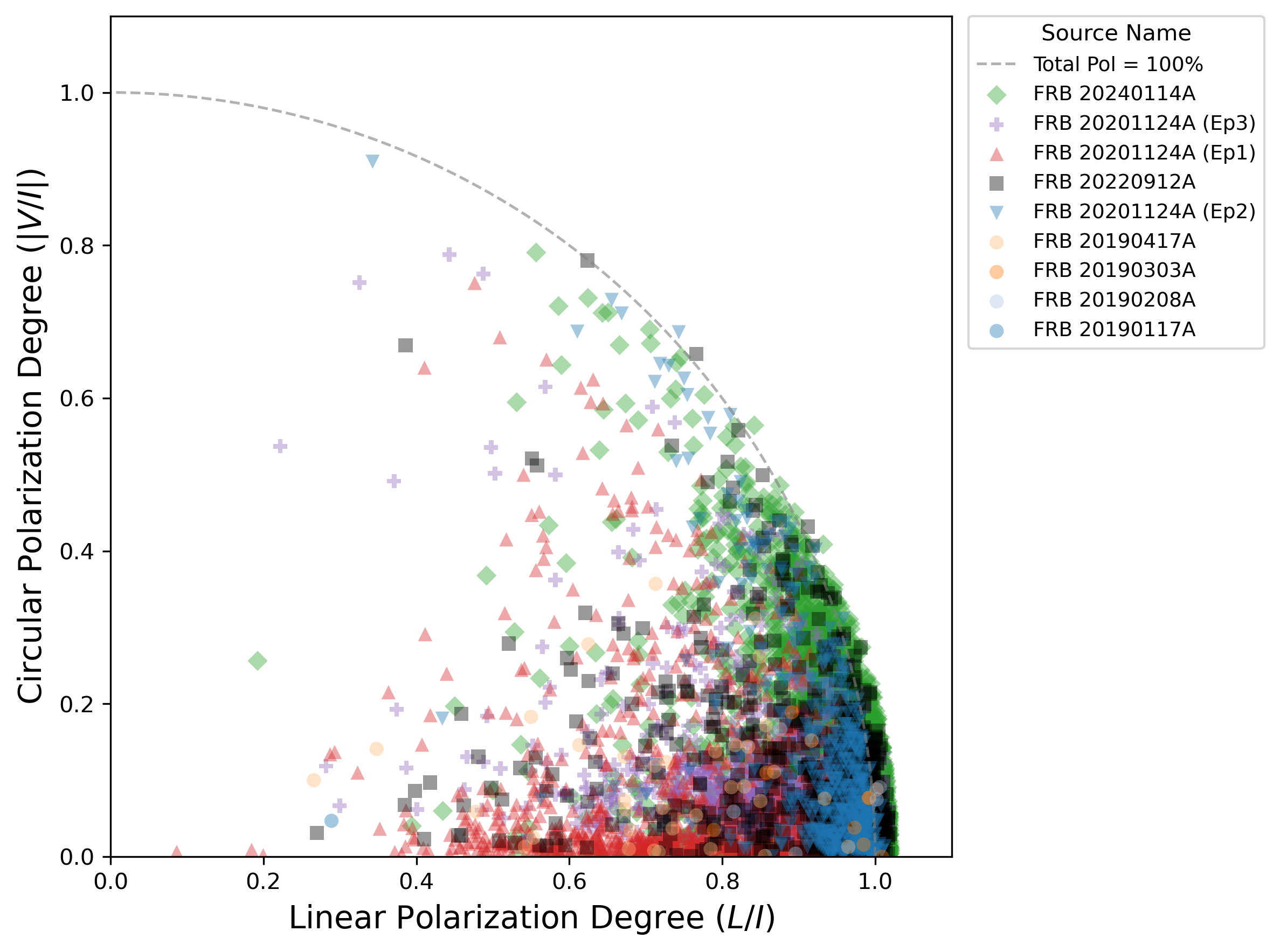}
    \caption{Polarization degree statistics of repeating bursts observed by FAST. The X-axis represents the degree of linear polarization, the Y-axis represents the degree of circular polarization, and the dotted line indicates that the total degree of polarization is 100\%.}
    \label{fig:fastLV}
\end{figure}

 
To date, FAST has published a series of scientific papers on the polarization of repeating FRBs, including FRB~20180301A\cite{Luo2020Natur}, FRB~20121102A\cite{li2021bimodal}, FRB~20201124A, FRB~20220529A\cite{liye2025}, FRB~20220912A\cite{zhang23}, FRB~20240114A etc. Figure\ref{fig:fastLV} shows the statistics of the linear and circular polarization degrees of the repeating bursts observed by FAST. 

\par 
As the number of detection increases, the presence of circular polarization becomes common in FRB repeaters. The initial polarimetry result of the first FRB repeater 20121102A showed 100\% linear polarization at C band \cite{2018Natur.553..182M,2018ApJ...863....2G}. However, FAST detected multiple bursts from it with significant circular polarization at L-band \cite{2022SciBu..67.2398F}. As the detection accumulates, circular polarization has become a common feature among FRB repeaters, including FRB 20190417A \cite{2025SCPMA..6889511F}, 20190520B \cite{2022SciBu..67.2398F}, 20201124A \cite{xuheng2021,2022RAA....22l4003J,jiang2025ninety}, and 20240114A \cite{2025ApJS..278...49X}. The presence of circular polarization in FAST observations, as high as 90\% \cite{jiang2025ninety}, raises challenge for the radiation mechanisms of FRB repeaters.

\par
\begin{figure}
    \centering
    \includegraphics[width=0.8\textwidth]{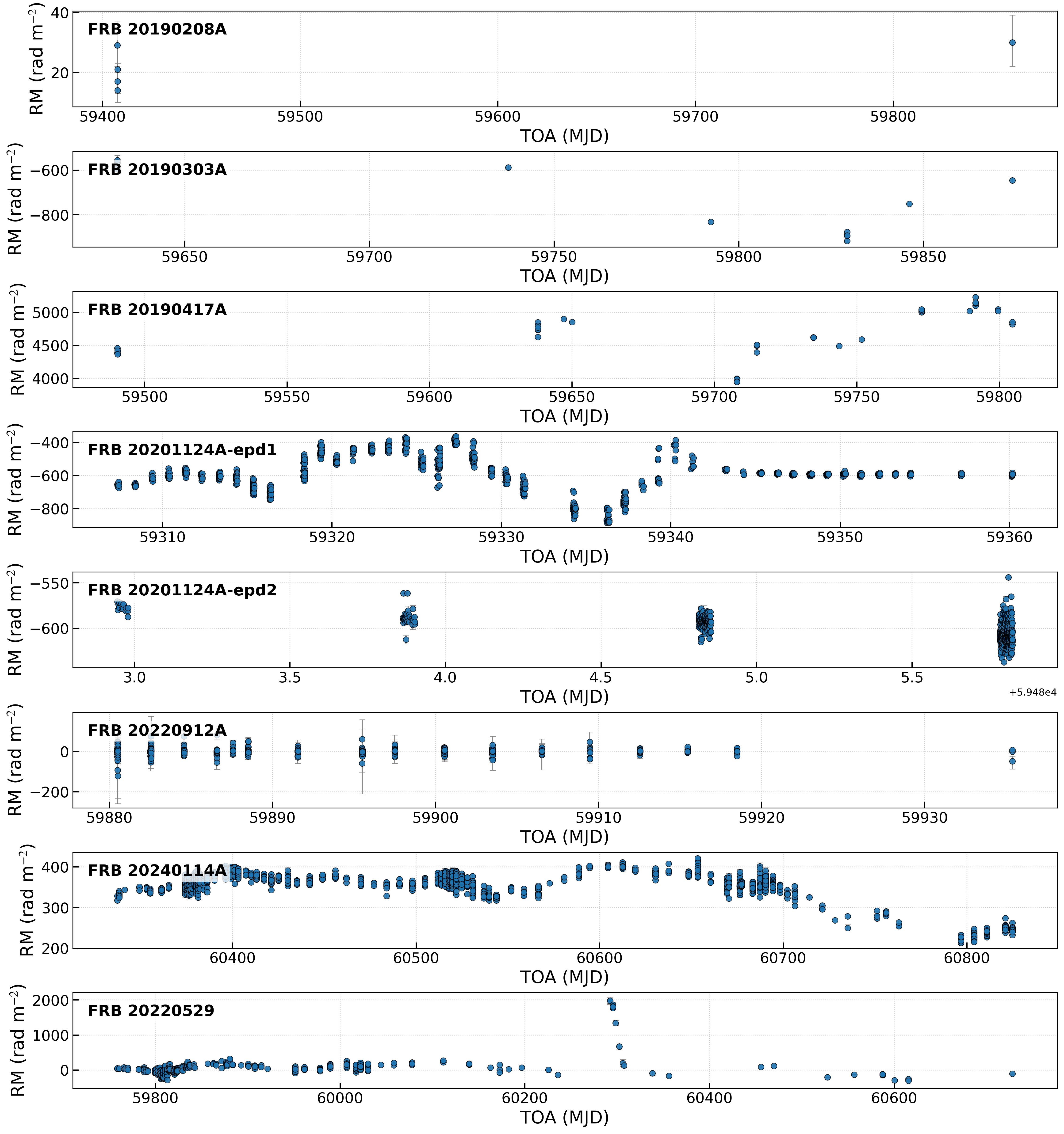}
    \caption{Long time scale evolution of RM in repeating FRBs.}
    \label{fig:fastfrbrm}
\end{figure}

In the ``cold plasma'', relevant for the propagation of radio waves through the ISM and IGM where thermal velocities are non-relativistic, the two natural modes are Right-Hand Circularly Polarized (RCP) and Left-Hand Circularly Polarized (LCP). These modes possess different phase velocities due to the presence of the background magnetic field $B$. As the wave propagates, the phase difference between RCP and LCP accumulates, resulting in a rotation of the linear polarization plane. This is the classical Faraday Rotation effect. The rotation angle $\Delta \Psi$ is proportional to the square of the wavelength ($\lambda^2$):$$\Psi(\lambda) = \Psi_0 + {\rm RM} \cdot \lambda^2$$The Rotation Measure (RM) is the path integral of the electron density ($n_e$) weighted by the parallel magnetic field component ($B_{||}$):$$RM \approx 0.81 \int_{source}^{observer} n_e [\text{cm}^{-3}] B_{||} [\mu\text{G}] dl [\text{pc}] \quad \text{rad m}^{-2}$$
Figure\ref{fig:fastfrbrm} shows long time scale evolution of RM in repeating FRBs. This maps the FRB's immediate environment.  Steady decay implies an expanding supernova remnant around a young magnetar. Fluctuations may suggest interaction with binary winds or turbulent nebulae.  Thus, RM changes reveal the source's evolutionary stage and the dynamics of its local magnetic field.

\begin{figure}
    \centering
    \includegraphics[width=0.8\textwidth]{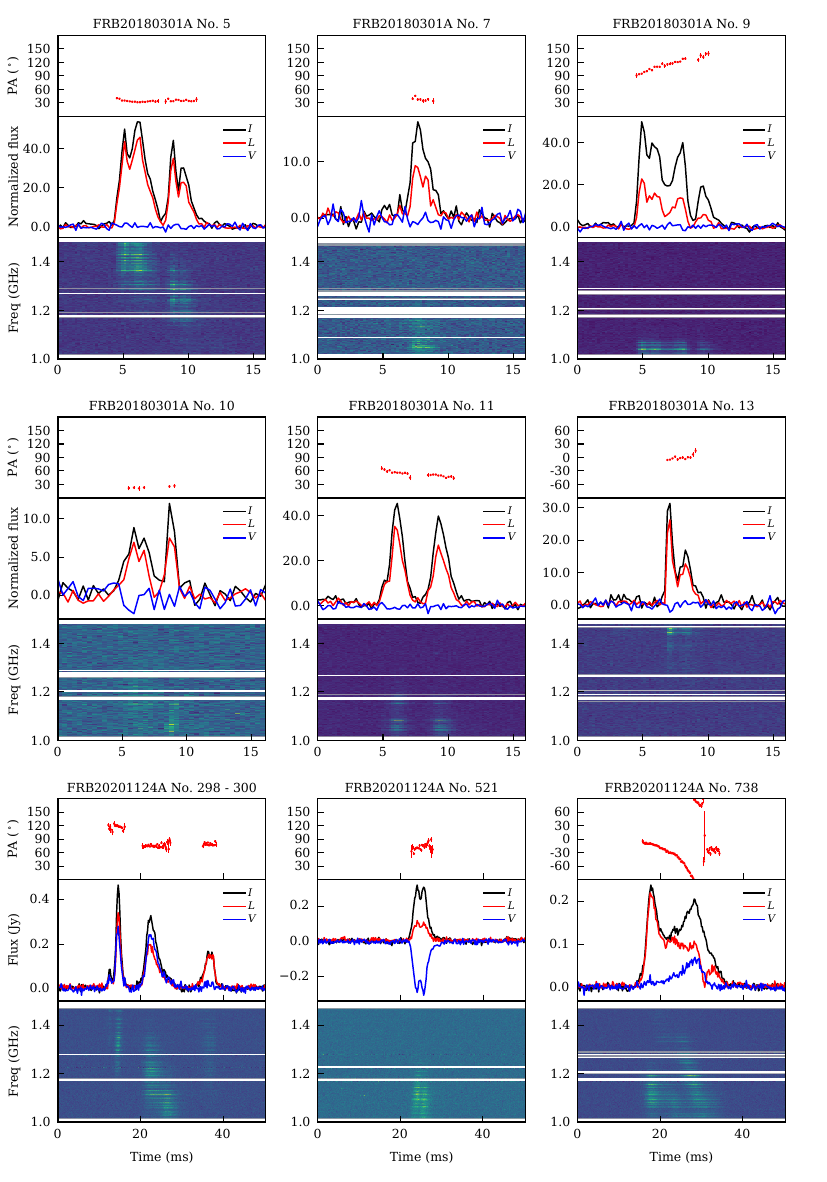}
    \caption{Polarization profiles and dynamic spectra from FRB 20180301A\cite{Luo2020Natur}(Top two panels) and FRB 20201124A\cite{jiang2025ninety}(Bottom panel). Diverse PA features are observed, including PA swings, PA orthogonal jumps, and highly circularly polarized emission in some bursts.} 
    \label{fig:pa_swings}
\end{figure}

Lots of intriguing features were discovered using FAST, for instance, diverse polarization position angle swings \cite{Luo2020Natur}, orthogonal polarization mode jump \cite{niu2024PAjump} and extremely high circular polarization \cite{jiang2025ninety}(see Figure~\ref{fig:pa_swings}). Most bursts are highly linearly polarized. For example, the median linear polarization fractions are 95.5\% for FRB~20201124A and 96.0\% for FRB~20220912A \cite{2022RAA....22l4003J,zhang23}.
Statistically, the linear polarization seems to show a frequency-dependent relation from repeaters, which may be characterized using a parameter caused by multi-path scattering, i.e., $\sigma_\mathrm{RM}$ \cite{feng22}.

\subsection{~Burst activity.}
\par 
Repeating FRBs cannot only exhibit various spectro-polarimetric characteristics, but also show very unpredictable burst activities. Normally, the observed event rates of repeating bursts can be changeable from quiescent to hyper-active states, such as, FRB~20201124A and FRB 20240114A. Among these repeaters, many bursts were detected with clustering feature in time. This can be described as the Weibull distribution, which is a generalized form of the Poissonian process. With both “Pincus Index” and the “Lyapunov Exponent”, repeating FRBs are far more random and less chaotic than events like earthquakes on the randomness-chaos two-dimensional plane, closely resembling Brownian motion\cite{zhang24}. While a couple of repeaters can be observed with periodic activities, for example, 16.35-day activity window from FRB~20180916B \cite{2020Natur.582..351C}, we have not obtained a clear evidence on both short-term and long-term periodicity from any repeaters using FAST \cite{2022RAA....22l4004N}. 

\section{~Hints at Radiative Mechanism.}

\par The high-precision timing enables detailed studies of pulse microstructure, nulling, and mode switching, which are direct manifestations of plasma processes near the stellar surface.
Understanding where and how these particles originate is therefore central to solving the emission problem for both pulsars and FRB sources.
Polarization measurements allow us to probe the distribution of currents and particle acceleration regions just above the surface, thereby linking observed radio emission to the physical conditions at the neutron star surface.

\par The observations of individual pulses, which are the basic radiation units likely, can reveal the coherence characteristics. These radiation units, whether individual bunches, charge clouds, or localized plasma instabilities, must be coherent. These radiation must be 100\% polarized in observations, and that their special behaviors provide important evidence for understanding the coherence of radiation.

\par FAST contributes to understanding how diverse observational phenomena emerge from magnetospheric dynamics.
By detecting faint emission from a large sample of pulsars, it can help map the beam geometry and constrain the configuration of surface magnetic fields, whether they are predominantly dipolar or contain higher-order multipole components.
The long-term monitoring campaigns capture the evolution of pulsar emission states, revealing quasi-periodic modulations, intermittent activity, and correlations between radio emission and changes in spin-down rates.
For FRBs, some repeaters exhibit complex activity cycles and variations in burst morphology, hinting at magnetospheric processes analogous to those seen in young, active pulsars but scaled to extreme energy budgets.
By enabling direct comparisons between pulsars and FRBs, it fosters a unified understanding of how magnetospheric dynamics.

\subsection{~The Geometry of Beamed Radiation.}


Thanks to the polarization observations of pulsed radio emission, we can determine the geometry of pulsar radiation, including the structure of the magnetosphere, the configuration and location of the emission beam, as well as their corresponding characteristics.
~In fact, the radiation geometry is usually used in describing the quasi-steady process in radiation, instead of the dynamic process. Although the motion and distribution of radiative particles are also aspects of radiation geometry, they are always abstracted to be a static or quasi-static emission beam. Thus the radiation geometry is often related to the time invariant characteristics of radiation, for example, the mean pulse profile, the PA curve, and their corresponding evolution with radiation frequency. 

\par
\begin{figure}
    \centering
    \includegraphics[width=0.8\textwidth]{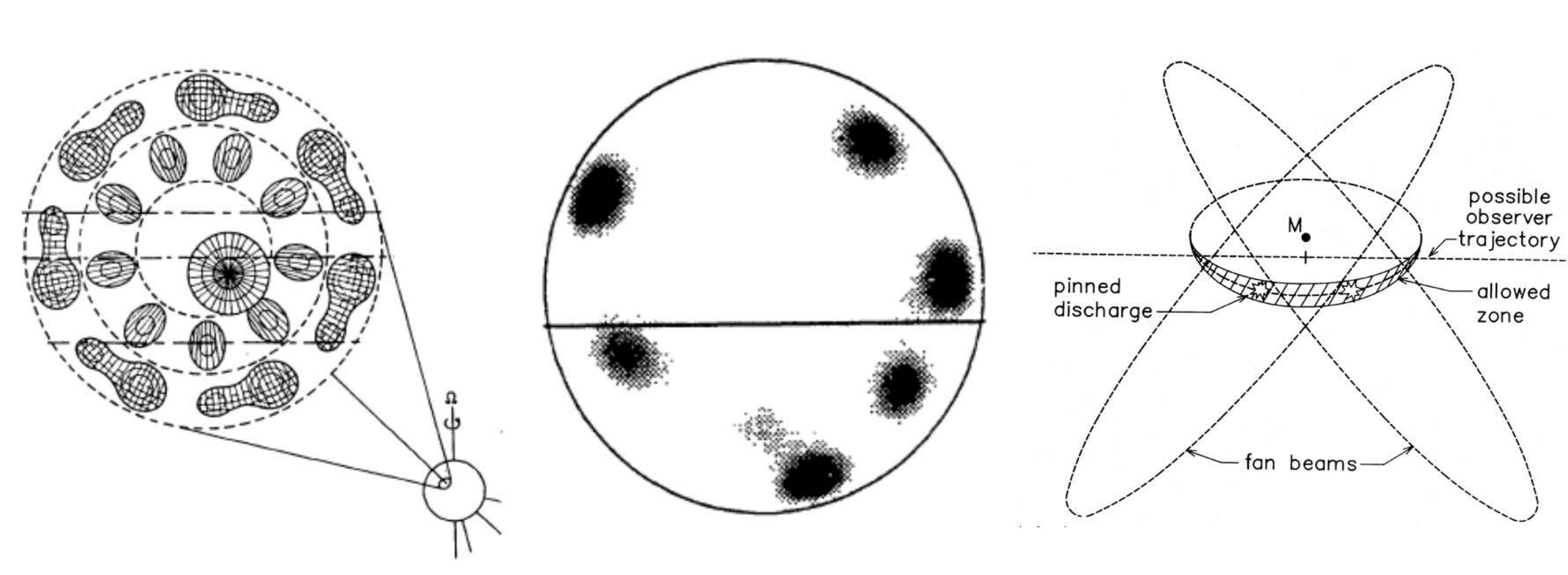}
    \caption{Phenomenological models of pulsar radio emission beams. Left: the conal beam model~\cite{1983Rankin}, middle: the patch beam model~\cite{1995Manchester}, right: the fan beam model~\cite{1987Michel}.
    \label{fig:threebeams}}
\end{figure}

 
In the normal pulsars with period of $\sim$1\,s, the radiation geometry is comparatively simple. On one hand, the radiation altitude of radio emission is always far lower than the light cylinder, and the induced magnetic field and the distortion of magnetic field around radiation position are therefore weak. On the other hand, the radiation altitude is always much larger than the pulsar radius in the normal pulsar, which leads to the weak multipolar components of magnetic field. Hence the magnetic field near radiation point can be approximated to static dipole field induced by the compact star.
~In this case, there is little overlapping between the radiation from different regions, and the emission beam at specified frequency can be treated as a combination of several patches. In some models, the radiative patches are like rings, which are circular and symmetric about the magnetic pole without considering the asymmetry induced by the obliqueness \cite{1983Rankin}. While in some other models, the radiation beams are shaped as fan-like structures extending along magnetic flux tubes \cite{1987Michel, 2010Dyks, 2014ApJ...789...73W}, or appear as discrete and randomly distributed patches \cite{1995Manchester} . Figure~\ref{fig:threebeams} shows the cartoons of three kinds of emission beam geometry. 

\begin{figure}[!tbh]
    \centering
    \includegraphics[width=1\linewidth]{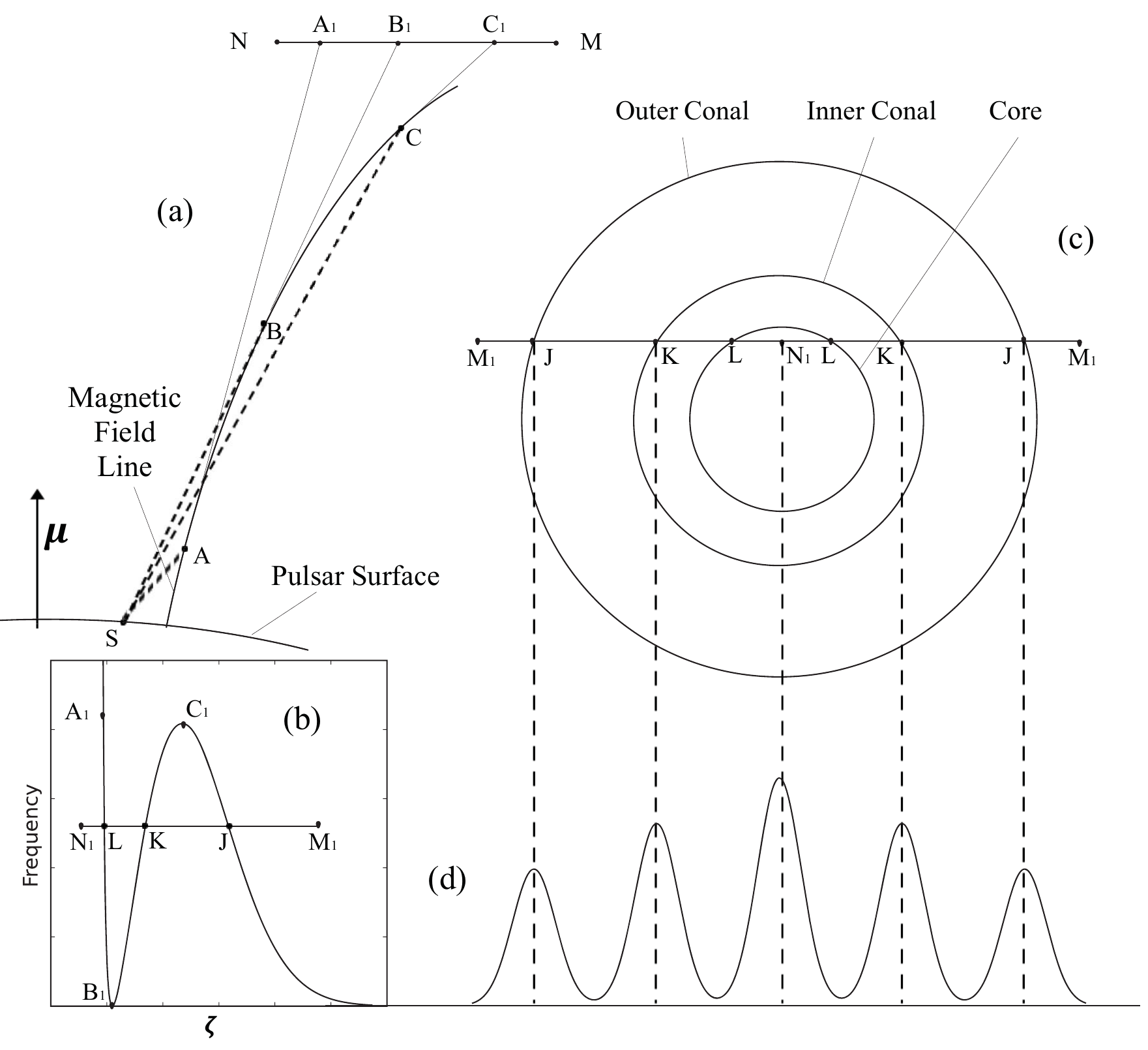}
    \caption{The radiation geometry (a), radiation frequency (b), emission beam (c) and resulted pulse profile (d) of ICS model.
    \label{fig:ics}}
\end{figure}

\par 
In addition to phenomenological beam models, radiation geometry is also related to the underlying physical mechanism of radiation. 
In fact, radiation under various mechanisms can be understood as resulting from particle acceleration. 
For example, Curvature Radiation (CR) arises from charged particles moving along curved paths, while Inverse Compton Scattering (ICS) involves particles accelerated by background fields to produce radiation. 
Although the radiation direction of ultra-relativistic particles is highly aligned with their motion, different radiation mechanisms produce distinct radiation spectra.
These radaiative mechanisms are also used to understand FRBs.

\par The RS model~\cite{RS75} posits that pulsar radiation is primarily caused by curvature radiation and, by combining it with the polar gap picture, yields a single-hollow-cone radiation geometry. However, considering other mechanisms like ICS~\cite{1998A&A...333..172Q} leads to drastically different results. The radiation geometry of ICS model is shown in Figure~\ref{fig:ics}. In panel (a), the radiation process is shown. The low frequency radiation is emitted from the point S on the stellar surface, and is scattered by the charged particles moving along the magnetic field line. In panel (a), 3 points A, B and C are marked, among them the point B is the point of tangency between the emission track of low frequency radiation and the magnetic field line. The scattered radiation from points A, B and C are received at points $\mathrm{A_1}$, $\mathrm{B_1}$ and $\mathrm{C_1}$ on the segment MN, which is the projection of the observer on the observer-magnetic axis plane while pulsar rotates. The curve of radiation frequency versus emergence angle $\zeta$ (the angle between line of scattered radiation and magnetic axis) is shown in panel (b). The radiation frequency corresponding to point B is zero, because the ICS cannot work for the case that low frequency radiation and charged particle motion having same direction. In fact, the practical observation is always taken near a fixed frequency, just like the segment $\mathrm{M_1N_1}$ in panel (b), which intersect the curve on points J, K and L, corresponding to different emergence angles. As a consequence, the emission beam of the ICS model have 3 cones with different emergence angles as shown in panel(c). When the line of sight passes the beam along the segment $\mathrm{M_1N_1M_1}$ which has 6 intersection points with the 3 cones, the observed radiation will have 5 peaks (the pulse phase between two points L is always very small thus there is always one peak remained). Furthermore, if the line of sight passes the beam along other path or the observation is taken on other frequency, the pulse profile would have other shapes.

\par 
The fan-beam models postulates that relativistic secondary particles flowing outward along magnetic flux tubes in pulsar magnetosphere produce radially extended beams. Desvignes et al. (2019) found that the main pulse beam of the precessing pulsar PSR J1906+0746 exhibits an elongated strip-like morphology, providing a direct observational evidence for fan beams\cite{2019Sci...365.1013D}. However, the shape of the interpulse beam of this pulsar aligns neither with the conal beam model nor with existing fan-beam models. It should be noted that the beam shape in fan-beam models depends on the distribution of particles within the magnetosphere. Thus, the challenge posed by the interpulse beam of PSR J1906+0746 likely stems from our incomplete understanding of the magnetosphere structure and particle distributions. Long-term monitoring of precessing pulsars with FAST will help reconstruct the radio beams of more pulsars, thereby providing additional direct samples for studying pulsar beam structures. 

\par 
Some studies have developed methods to test beam models through the pulse width–impact angle relation\cite{2014ApJ...789...73W} or the frequency evolution of many single pulses at different phases\cite{2019MNRAS.489..310O}. Recently, Wang et al. (2023) fitted the RVM to linear polarization position angles from FAST observations for many pulsars and derived magnetic inclination angles and impact angles for 190 pulsars\cite{2023RAA....23j4002W}. Based on this dataset and supplemented by other samples, Deng et al. (2026) confirmed that for impact angles larger than a few degrees, the relationship between the 10\% pulse width and the impact angle remains consistent with the prediction of the fan-beam model\cite{Deng2026}. However, for very small impact angles, better models are needed to explain the observations. Undoubtedly, the strong capability of FAST to resolve single pulses and to build large samples of magnetic inclination and impact angles will also play a crucial role in discriminating among different beam models.

\par 
In the millisecond pulsars, the radiation geometry is much more complicated. The radiation position is always near the stellar surface or light cylinder, thus the distortion and multipolar components of magnetic field cannot be neglected. Hence, the emission from different radiation regions overlaps significantly, and that the aberration and retardation effects even intensify the overlapping of the emission. As a result, it is hardly to differentiate the radiation position with only the time invariant characteristics of radiation, and the features on single pulse, e.g. spectrum structure, PA distribution and microstructure, may indicate more information of radiation geometry.

\subsection{~Boundary Conditions on the pulsar Surface.}
The radiative processes of pulsars, and the interior structures of compact stars, are both related to surface properties. At the early stage of pulsar research, the work function of ions on pulsar surface was a hot topic to debate. Different emission models~\cite{RS75,Arons79} were constructed under different estimations of ion work function. Surface conditions, including crust, atmosphere, and multipolar magnetic field, play an important role in glitches, thermal X-ray emissions, and the coherent radio emission. FAST has provided us with more details in pulsar radio emission, therefore could help us explore surface properties.

\par 
Small ``mountains'' or ``zits'' can formed on the the rugged surface of a pulsar, and then parallel electric field of the inner gap might be enhanced, enabling more efficient positron acceleration \cite{2025arXiv250612305X}.
This peculiar feature reduces the voltage
required to form a spark, thereby explaining the unexpected radio emissions observed from those two dead pulsars, PSR J0250+5854 and PSR J2144--3933, as well as understanding the swoosh phenomenon described in \S2.5.1. 
Small mountains may locate randomly on the pulsar surface, which could result in patch-like emission beams in Fig.~\ref{fig:threebeams}. 
A fan beam could be reproduced by a bunch of particles from point-discharging around a small mountain if curvature-like radiation dominates.

Additionally, these mountains can also be applied to explain radio-X-ray synchronization and some highly asymmetric pulse profiles, by inducing local discharge processes. For some examples, please refer to the analysis of FAST observing PSR~B0950$+$08~\cite{WangZhengli2024} and PSR~B0943$+$10~\cite{CaoShunshun2024}.

\subsection{~FRB Trigger and Radiation.}
Even there are more than thousands of FRB sources been found, the origin of FRB is still unknown, constituting the paramount unresolved question in FRB research. In the early stages of FRB research, catastrophic events such as supernova and binary compact star mergers were once considered potential origins of FRBs. However, the discovery of the repeating FRB 20121102A ruled out purely catastrophic events as the sole origin of such events. Compact objects with strong magnetic fields, i.e., neutron stars/magnetars, have gradually emerged as potential candidates for FRBs. The detection of an FRB 20200428D originating from the Galactic magnetar SGR J1935+2154 suggests that magnetars are the sources for at least some FRBs\cite{Bochenek2020,CHIME2020}. It bridges the critical gap connecting magnetar radio emissions to FRBs when this magnetar emitted radiation similar to FRB 20200428D again in 2022.

\par The leading hypothesis is that these bursts are triggered by sudden, large-scale energy releases within the magnetar's magnetosphere. One possible trigger mechanism is the starqauke caused by sudden and catastrophic fractures in the crust of a highly magnetized neutron star or magnetar\cite{Wang2018}. During a starquake, the cracking of the crust releases the accumulated strain in the internal magnetic field, which unleashes a tremendous amount of energy. The starquake process analogous to an earthquake, so that FRBs may share similar statistical behaviours with earthquakes.

\par Let's briefly mention the merits of solid strangeon stars in relation to starquakes. The energy emitted by FRBs can be released in any form through magnetic energy, or gravitational energy due to the bulk variation\cite{2024RAA....24j5012W},  but the total free magnetic energy for toroidal fields released can be up to $10^{46}$ erg. The energy budget of gravitational energy is much higher with anisotropy when the stellar mass is close to $M_{\rm TOV}$\cite{2024RAA....24b5005C}. For a star with a mass exceeding $M_{\rm TOV}$, it will collapse into a black hole during a relatively short timescale and will then no longer produce FRBs~\cite{2026arXiv260101949X}. In the case where the mass is below the TOV limit, the star will gradually approach its demise. Ultimately, both of the above scenarios lead to the cessation of FRB following extremely high levels of activity.

\par Other proposed mechanism involves an external trigger \cite{Dai2016,Zhang2017}. Falling material is 
injected into the magnetosphere through the interaction with an external source, producing a large population of charged particles. These particles are then accelerated to relativistic speeds within the magnetosphere generates the observed radiation.

\par The bright temperature of FRBs is calculated to
\begin{equation}
T_b \simeq \frac{\mathcal{S}_{\nu, p} D_{\mathrm{A}}^2}{2 \pi k(\nu \Delta t)^2}=\left(1.2 \times 10^{36} \mathrm{~K}\right)\left(\frac{D_{\mathrm{A}}}{10^{28} \mathrm{~cm}}\right)^2\left(\frac{\mathcal{S}_{\nu, p}}{1\,\mathrm{Jy}}\right)\left(\frac{\nu}{1\,\mathrm{GHz}}\right)^{-2}\left(\frac{\Delta t}{1\,\mathrm{~ms}}\right)^{-2},
\end{equation}
where $D_A$ is the luminosity distance, $\mathcal{S}_{\nu, p}$ is the flux, $\nu$ is the frequency, $\Delta t$ is the burst width, exceeds the maximum bright temperature of $\sim10^{12}$ K inferred by the incoherent synchrotron radiation, so that FRBs must extremely coherent radiation. Based on the emission region from the neutron satr/magnetar center, the radiation models can be generally divided into two categories: emission within the magnetosphere, and emission from shock regions far outside the magnetospheres \cite{Zhang2023}. However, the latter models are challenged for both the observational and theoretical aspects. Driven by a particular trigger mechanism, charged particles in the magnetosphere stream along the magnetic field lines and assemble into bunches. These bunches have a length smaller than $\lambda/2$, where $\lambda$ is the wavelength, thereby enabling the coherent radiation of FRBs.

\par
\begin{figure*}
    \centering
    \includegraphics[width=1\linewidth]{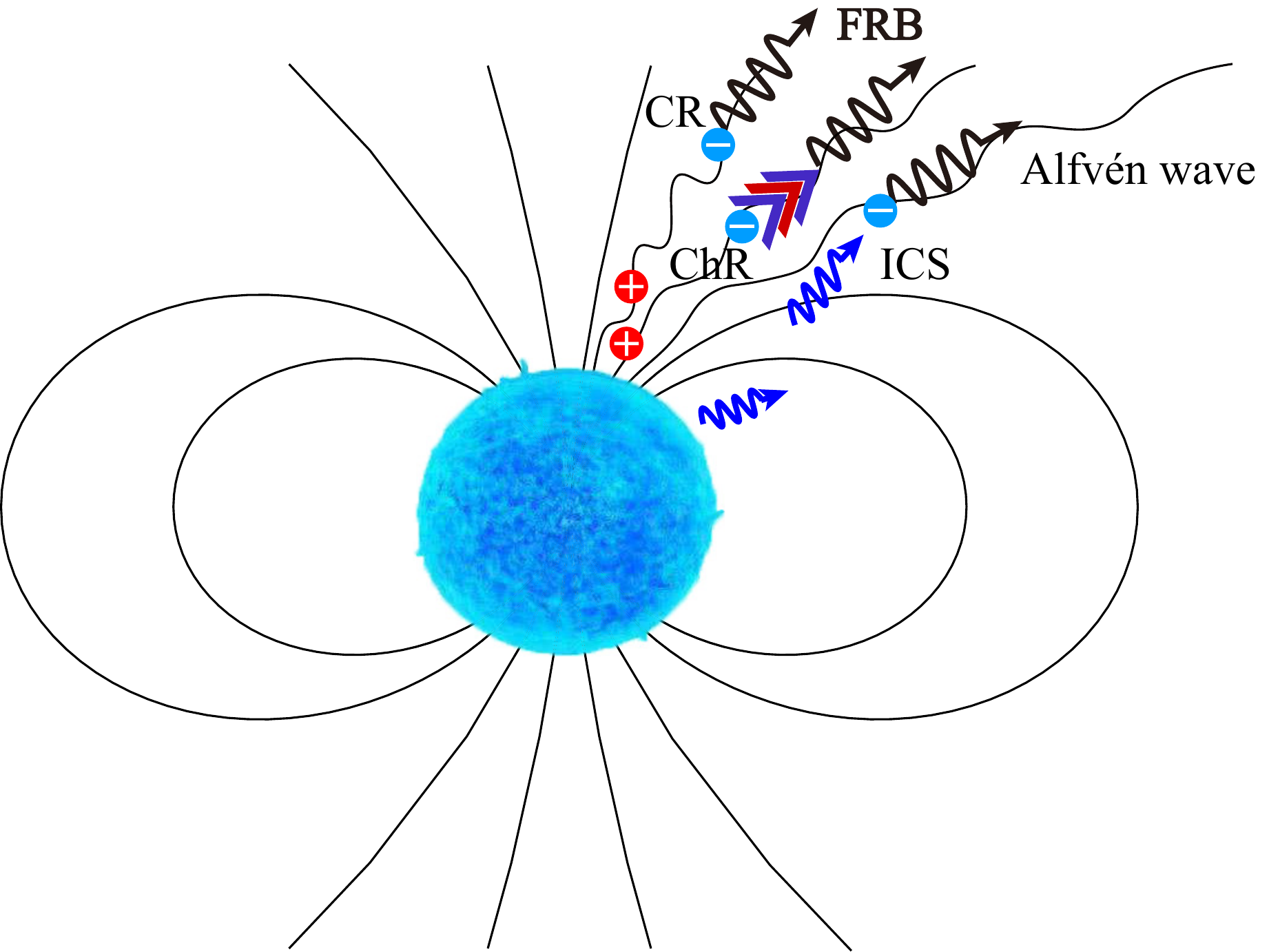}
    \caption{Schematic diagram of magnetospheric radiations. The blue solid lines denote low-frequency waves.}
    \label{FRBemission}
\end{figure*}

 During the charged bunches move along the curved trajectories, they emit coherent CR due to the perpendicular acceleration, shown as Figure \ref{FRBemission} (also for the following two radiation mechanisms). Since the electrons are ultra-relativistic, their emission is confined to a narrow radiation cone. The angular size of this cone is primarily determined by the bunch's half-opening angle and the Lorentz factor $\gamma$ \cite{Wang2022a,Wang2022b}. The emission is highly linearly polarized near the center of the cone. As the line of sight sweeps across the edge of the cone, the degree of circular polarization gradually increases. The characterized frequency of CR is determined by the height, i.e., $\nu_c=3\gamma^3/(4\pi c\rho)$, where $\rho$ is the curvature radius. Consequently, when the line of sight sweeps across two distinct bunches at different heights, the emission from the lower one always arrives earlier than that from the higher one. As the lower bunch typically radiates at higher frequencies, this scenario naturally accounts for the observed downward-drifting pattern of at least some FRBs \cite{Wang2019}. However, the spectrum of CR is inherently broadband, i.e., $\Delta\nu/\nu>0.1$. Generating the observed narrowband emissions therefore requires an additional mechanism, such as quasi-periodic discharges within a gap region \cite{Wang2024,Yang2023}.

\par ICS is also a radiation mechanism that could potentially generate FRBs \cite{Zhang2022}. When incoming low-frequency waves are scattered by high-energy electrons, ICS can occur. The low-frequency waves here might be the fast magnetosonic wave triggered during the starquake. Similar to the bunched CR scenario, the ICS waves are highly linearly polarized for the on-beam case whereas highly circularly polarized for the off-beam case. Note a common property of the bunching antennas coherent radiation is that the sign change of circular polarization can be seen when the LOS sweeps across the bunch center. The difference is that the CR from a single electron can generate circular polarization even when the  
LOS is not confined to the trajectory plane, while the ICS emission is linearly polarized and requires the superposition of electric vectors from different directions to produce circular polarization. The ICS spectrum can be narrow-banded, with its frequency being a twice Doppler-boosted version of the monochromatic low-frequency incoming wave \cite{Qu2024}. Currently, there is no established mechanism of ICS to explain the drifting pattern observed in FRBs.

\par Coherent Cherenkov Radiation (ChR) is also proposed to explain FRBs in which a crucial condition that the index of the plasma medium $n_r>1$ \cite{Liu2023}. The bunched ChR has a narrow spectrum but with $\sim100\%$ linear polarization.

\section{~Future Prospects.}
To overcome inherent limitations in angular resolution and fully exploit the optimal local electromagnetic environment, a significant strategic upgrade, known as FAST Phase II, is envisioned. This project aims to construct an ultra-high sensitivity and high angular resolution hybrid interferometric array comprised of the existing main dish and additional smaller antennas. The ultimate goal, targeted for completion by the end of 2030, involves establishing a large-scale array consisting of 64 new 40-meter diameter antennas distributed within a 30-kilometer radius of FAST, forming a ``1+64" synthesis array. Upon completion, this configuration is projected to yield an unprecedented observational sensitivity of 4000\,m$^2$/K and achieve a spatial resolution of approximately 1\arcsec. These substantial enhancements in detection sensitivity, precise localization, and imaging capabilities across time, frequency, and spatial domains are expected to facilitate major breakthroughs, particularly in probing the radiative processes of pulsars and the precise localization of fast radio bursts, alongside broader applications in neutral hydrogen, active galactic nuclei, and gravitational wave studies.

 The steerable Jingdong 120 m Pulsar Radio Telescope  will be built in Jingdong county of Yunnan Province, China. It will become the world’s largest fully steerable radio telescope with observation frequencies covering 100 MHz–10GHz (300cm–3cm).  JRT’s scientific goals include (1) performing long-term, high-time-density, and high-precision timing observations for more than 90\% of millisecond pulsars in the sky to research the construction of a pulsar-based time scale and the detection of nanohertz low-frequency gravitational waves; and (2) contributing observational studies in the fields of pulsar physics, fast radio bursts, gravitational testing, black hole, and VLBI astrometry; (3) serving the national strategic needs in the fields of pulsar navigation, deep space exploration, and space target monitoring. JRT developed key technologies such as large-scale antenna structure design, ultra-wideband feed, low-noise receiver, and multi-functional digital backend. The design and development of each subsystem of JRT will be completed along with relevant domestic research forces. JRT construction will promote China’s development in engineering and technical fields, such as large-scale machinery design and processing, automatic control, low-temperature electronics, and system integration. The completion of JRT will progress China’s radio astronomy and improve its international influence.
 
 The QTT (Qitai Radio Telescope), located at an elevation of 1,800 m in the eastern Tianshan Mountains of Xinjiang's Qitai County, is a fully steerable, 110-m diameter Gregorian telescope with a standard parabolic primary reflector \cite{2023SCPMA..6689512W}. This telescope works at 150 MHz up to 115 GHz. The QTT will operate in pulsar, spectral line, continuum and Very Long Baseline Interferometer (VLBI) observing modes, and also plays an important role in improving the Chinese and international VLBI networks.
 


\section{~Summary.}

Thanks to the extremely high sensitivity of FAST, we can not only detect faint single pulses but also precisely measure their polarization. 
These diverse observations provide crucial insights into the physics driving pulsars and FRBs.
The remarkable achievements are summarized as following.

\par1) Over 800 pulsars are discovered during the GPPSs. The survey also serendipitously detected some highly dispersed FRBs. FAST's GC pulsar surveys (e.g., GC FANS) have also discovered over 60 new pulsars, more than doubling the known population in its sky coverage. Furthermore, more than 200 MSPs have been detected.

\par2) By utilizing the high sensitivity of FAST, some pulsars can be observed to exhibit a ``valley-like'' structure in their average pulse profiles. Accurately delineating these fine profiles can help us to understand the radiation geometry of pulsars. Meanwhile, this also enables the detection of variety of polarization features in many pulsars. Among the CPTA and GPPS samples, MSPs can share the same distribution of fractional linear and circular polarization with normal pulsars. These features may reflect various aspects of pulsar radiation geometry, radiative mechanism, and propagation effects within the magnetosphere.

\par3) Single pulses, especially those with extremely high polarization, may represent the fundamental emission units of pulsars. FAST detected as numerous sporadic, weak, and narrow pulses during the traditional ``nulling'' state of pulsar B2111+46. All pulsars currently known to exhibit dwarf pulses are located in the ``death valley'' of the $P-\dot{P}$ diagram. The phenomenon has since been confirmed in at least 13 pulsars, including PSR J2323+1214 and PSR B1931+24. The detection of dwarf pulses prompt us to reexamine the radiative mechanisms of pulsars.

\par4) Some pulsars exhibit temporal variability, like pulsars with mode changes and LPTs. These temporal variability reflect the fact that the global magnetospheric structure or local emission regions are not static. The radio emission from magnetars are also time-variable. Magnetars' radio emission only appears following X-ray outbursts, a radio pattern reconfirmed by FAST observations of SGR 1935+2154 which pulsations matches the X-ray observations.

\par 5) FAST pulsar timing can yield precise mass measurements puslar binary system. PSR J1946+2052 as the most compact double neutron star, has the shortest known orbital period among Galactic double neutron stars. Long-term observations, primarily from FAST, have enabled the measurement of five post-Keplerian parameters. These collectively provide three independent, high-precision tests of GR. One test of orbital decay is the second most precise ever, surpassing the Hulse-Taylor result.

\par 6) The radiation geometry relates to stable observational features like pulse profiles and polarization. In typical pulsars, emission occurs far below the light cylinder but well above the surface, allowing a static dipole approximation where beams appear as distinct patches (ring, fan, or random). This geometry is inherently linked to radiation mechanisms (e.g., curvature radiation and ICS), as different acceleration processes dictate emission direction and spectrum—requiring combined study.

\par 7) FAST has great advantages in monitoring known FRB source to detect faint bursts. According to FAST's samples of hundreds of bursts, FRBs with time-frequency structures can be classified into multiple morphological categories, including downward-drifting, upward-drifting, non-drifting, and complex bursts with mixed behaviors. Some polarization observations, such as PA swings and PA jumps, indicate that the radiation from FRBs likely originates within a magnetosphere. Within the magnetosphere, charged particles can generate coherent radiation through mechanisms 
like curvature radiation, inverse Compton scattering, and Cherenkov radiation.

\par FAST observations are expected to advance our understanding of fundamental questions, such as the nature of dense matter inside neutron stars and its impact on magnetospheric dynamics, the coherent emission mechanism of both pulsars and FRBs, and the magnetospheric plasma physics.
In the future, without a doubt, more facilities being commissioned would help us to understand and to solve the problems.

\section*{Acknowledgments}
This work made use of data from the FAST. This work is supported by the NSFC (No. 12403058, 12261141690, 12573104, 12133004) and National SKA Program of China (2020SKA0120100) and the NSFC (Nos. ). Y.F. is supported by National Natural Science Foundation of China grant No. 12522305, and by the Leading Innovation and Entrepreneurship Team of Zhejiang Province of China grant No. 2023R01008. J. R. Niu is supported by the National Natural Science Foundation of China (NSFC, No. 12503055) and the Postdoctoral Fellowship Program of CPSF under Grant Number GZB20250737. J. C. Jiang is supported by the European Research Council (ERC) starting grant ``COMPAC'' (No. 101078094). R.L. is supported by the NSFC (No. 12303042).
Lei Qian were supported by the Youth Innovation Promotion Association of CAS (id.~2018075, Y2022027), and the CAS ``Light of West China'' Program.
Y.J. Wang is supported by the National SKA Program of China No. 2020SKA0120200. This work used resources of China SKA Regional Centre prototype (\cite{2019NatAs...3.1030A}, \cite{2022SCPMA..6529501A}) funded by Ministry of Science and Technology of the People's Republic of China and Chinese Academy of Sciences.
This work is supported by the National Natural Science Foundation of China under Grand No. 11703047, 11773041, U2031119, and 12173052. ZP is supported by the CAS ``Light of West China'' Program and the Youth Innovation Promotion Association of the Chinese Academy of Sciences (ID 2023064), National Key R/\&D Program of China, No. 2022YFC2205202.

\newpage
\bibliographystyle{unsrt}

\end{document}